\documentclass[12pt]{elsarticle}
\usepackage{amsfonts}
\usepackage{graphicx}
\usepackage{amscd}
\usepackage{amsmath}
\usepackage{xcolor}
\usepackage{epsfig}
\usepackage{subfigure}
\usepackage[english]{babel}
\usepackage{algorithm,algpseudocode} 
\usepackage{listings} 
\usepackage{multicol}
\usepackage[colorlinks=true, allcolors=blue]{hyperref}
\usepackage{tikz, amsmath}
\usetikzlibrary{shapes,arrows}

\providecommand{\U}[1]{\protect\rule{.1in}{.1in}}
\providecommand{\U}[1]{\protect\rule{.1in}{.1in}}
\allowdisplaybreaks

\numberwithin{equation}{section}

\numberwithin{table}{section}
\numberwithin{figure}{section}
\usepackage{tikz}
\usetikzlibrary{arrows.meta,positioning}
\usetikzlibrary{arrows}
\usetikzlibrary{fit,shapes.geometric}
\newcommand{\beginsupplement}{%
        \setcounter{table}{0}
        \renewcommand{\thetable}{S\arabic{table}}%
        \setcounter{figure}{0}
        \renewcommand{\thefigure}{S\arabic{figure}}%
     }

\begin{document}
\begin{flushleft}
{\Large
\textbf{Modeling the Effect of Observational Social Learning on Parental Decision-Making for Childhood Vaccination and Diseases Spread over Household Networks}
} \newline
\\
Tamer Oraby$^{*}$, Andras Balogh
\\
School of Mathematical and Statistical Sciences, The University of Texas Rio Grande Valley, \\
1201 W. University Dr., Edinburg, Texas, USA,
\\
 $^*$Corresponding author: tamer.oraby@utrgv.edu 
\end{flushleft}
\section*{Abstract}
\indent 

In this paper, we introduce a new model of parental decision-making concerning vaccines against a childhood disease that spreads over a contact network. We consider a bilayer network composed of two overlapping networks which are either Erd\H{o}s-R\'{e}nyi (random) networks or Barab\'{a}si-Albert networks. The new model uses a Bayesian aggregation rule for observational social learning, occurring over a social network, of which other decision models, like voting and DeGroot models, are special cases. Using our new model, we show how some levels of social learning about
vaccination preferences can lead to the convergence of opinions and affect levels of vaccine uptake and so disease spread. In addition, we study the effect of the existence of two cultures of social learning on the establishment of social norms of vaccination and levels of vaccine uptake. In all cases,
the mutual influence between the dynamics of observational social learning and disease spread is dependent on the network's topology and vaccine safety and availability.


{\bf Keywords:} Social Learning; Children Vaccination; Networks; Disease Model; Social Norms.

\section{Introduction}

Human herding behavior or cascading is a convergence in opinion driven by social learning \cite{Dietz1995,Hirshleifer2003}. Such herding could also be induced by payoff externalities, sanctions, preference interaction, direct communication, and/or observational influence \cite{Hirshleifer2003}. Observational influence results from integrating learned behavior or perception of others' opinions with one's own opinion. Observational social learning causes a subtle pressure on people to conform \cite{Cai2009,Sasaki2018,Goeschl18}. That social learning takes place through different channels of information sharing, and through observation and/or perception of public choices \cite{bandura1977social}. However, boundedly rational observational learning occurs when there is incomplete or insufficient information on the behavior of others, \cite{Mueller-Frank2015}. 

Social learning has been shown to play a vital role in people's decision-making even in the presence of information \cite{Bikhchandani2021}. Parents learn about the vaccination choices/attitudes of other parents and look for consensus vaccination signals \cite{Damnjanovic2018}. When parents share their feelings about the vaccination of their children while others observe these opinions, pressure is placed on other parents \cite{Brunson2013,Ebi2022}. Mixed messages and signals of parents' choices can cause problems for other parents when interpreting them. Incomplete information on the vaccine and the opinions of friends on social networks, creating boundedly social agents, can cause problems in vaccine acceptance and therefore in vaccine uptake and disease spread \cite{Oraby2015}. 

People and households are connected to each other through different types of networks. From a graph theory point of view, there are several types of network models, e.g. Erd\H{o}s-R\'{e}nyi (random) network model (ERN) and Barab\'{a}si-Albert network model (BAN), see \cite{Easley2010}. Different network models represent various real-life systems. Many real-life networks are scale-free networks (SFN) in which the degree distribution is a power law with exponent $2<\gamma \leq 3$, see \cite{Namatame2016}. 

Many studies modeled the spread of diseases on networks, see, e.g., \cite{NdeffoMbah2012a,Pastor-Satorras2004,Boguna2013,Keeling2005}, and on multiplex network models, see, e.g., \cite{Granell2013a,Scata2016,Pan2018,Guo2015,Fatima2017,Wang2015}. Some studies also considered the structure of the home \cite{Liu2004,Ball2010b,Ma2013,Pellis2012}. Few models of spread of diseases considered vaccine decision making on those networks through voter model \cite{Fatima2017}, through DeGroot model of selection \cite{Namatame2016}, through social norms \cite{Phillips2020}, and through social learning \cite{Carrignon22}. Moreover, most of the mathematical models in those articles consider rational agents making decisions to vaccine children with complete information, e.g. \cite{NdeffoMbah2012a,Brunson2013,Fukuda2014,Wells2012}, except for a few like \cite{Wang2015}.

Here, we consider bilayer networks in which two same-type overlapping networks and their mutual influence are of interest. The first network is the physical network through which face-to-face contacts take place and pediatric disease transmission occurs. The second network is a social (bidirectional) weighted network through which information and opinions about the vaccine are shared, shaping parents' decisions. Pediatric disease spreads on the physical network within and between households, while information and opinion sharing and perception transpire on the social network of households, parents in that case. We assume that the two networks overlap with parents linked to a larger number of other parents, relatives, coworkers, and distant friends, whose children might or might not be connected on the physical network. We introduce some adjustments to network models to reflect some important aspects of household networks. First, household networks should account for the number of children in households. Second, households without children must not have physical connections. That is, the expected degree of a household should increase with the number of occupying children, on average.   

In this paper, we present a new model of parental decision-making to protect their children against a measles-like disease that spreads over household networks. This model considers boundedly rational observational social learning using a Bayesian aggregation formula. This Bayesian formula is different from that of Muller-Frank and Neri \cite{Mueller-Frank2015}, who presented a quasi-Bayesian model of boundedly rational observational learning in a general context. We show that our model can give rise to social norms and at the same time encompasses other selection models like voting and DeGroot. Our new model considers socially bounded agents (parents who value children the most) who possess imperfect information about the vaccination choices of their network's neighbors. 

We postulate that those agents can only perceive a correct or wrong message, probably, because of fear of retribution or confusion. There is a chance that the agents will send a correct message of their opinion with probability $q$ and a message with probability $1-q$, see \cite{Easley2010a}. Using that model, we study the cascading opinion of vaccination in a boundedly rational observational social learning and compare it to other models of social pressure \cite{Phillips2020,Oraby2014}, see also \cite{Oraby2015}. Moreover, we study the influence of those types of signal games on the spread of vaccine opinion and disease in social and physical networks in the presence of resource limitations represented in vaccine efficacy and accessibility, as well as vaccine safety. Finally, we study the effect of the presence of two cultures of social learning on the establishment of a social norm and so on vaccine uptake.    

\section{Model and Methods}
\subsubsection*{Networks.}
To model the spread of the disease, we use an agent-based network model whose nodes are households $N$. Households are occupied by a number of children ($C_i$, such that $0\leq C_i \leq n_C$, for $i=1,\ldots,N$) who are connected through a physical network. The number $n_c$ is the maximum number of children per household. Parents are connected through a different (social, internet and physical) bi-directional weighted network via which they exchange opinions and information, and observe choices. Parents in a household without children could still be connected to other parents and shape their opinions. We use two types of networks: Erd\H{o}s-R\'{e}nyi (random) network model (ERN) and Barab\'{a}si-Albert network model (BAN), see \cite{Easley2010}.  In physical networks, we postulate the degree of nodes to be proportional to the number of children in the households. Meanwhile, the parent's network overlaps with the children's network via random rewiring of the children's network but with wiring probabilities that are greater than severance probabilities. We assume that the parent's network is weighted bidirectional with weights given by the (learning) probabilities $q_{j,i}$, for $i,j=1,\ldots,N$ and $(i,j)$ is a social network link. See supplementary material (SI. Model) for complete information about the networks.  An instance of a histogram for each random network is shown in Fig. S1.
\subsubsection*{Birth process.}
A birth process is postulated to depend on the number of children occupying a household. The probability of a new pregnancy is modeled using a logistic function with a median of $C^*$ such that it decreases as the number of children in a household approaches $n_c$. See supplementary material (SI. Model). A pregnancy lasts for a period of 280 days. The delivery of a newborn updates the number of children, but not the number of links in the children's network. We consider miscarriages and children's deaths to be rare and so are not included.
\subsubsection*{Disease spread.}
We assume that a new measles-like disease (vaccine-preventable, pediatric) is spreading between children within and between households. We assume a mean length $m_p$ days of incubation period with a maximum of $\ell$ days. A new infection in a household occurs with a probability $1-(1-\beta_h)^{I(i)} \cdot(1-\beta)^{n_I(i)/C_i}$; where $\beta$ is the probability of infecting a child in another household (through the physical network), $\beta_h$ is the probability of infecting a sibling within the same household, and $I(i)$ is the total number of infected siblings in the same household. The number $n_I(i)$ is the number of infected children connected to the household $i$ through the network.  The number $n_I(i)$ is divided by $C_i$ to approximate the probability of infection formula based on the assumption that the children in a household have the same number of friends on average. We assume that the epidemic starts with $I_0$ infected children in different households and is randomly chosen.

\subsubsection*{Vaccination decision-making.}
Parents are randomly allocated into three types: never-vaccinators, who oppose vaccination at all times, non-vaccinators, and vaccinators. We assume a small percentage of parents are never-vaccinators who are still sharing their opinions. Parents in the household $i$ shape their subjective decision to vaccinate their children based on the reward of vaccination, given by $\pi_i=\alpha_i I -\gamma_i A$, where $I$ and $A$ are the total number of adverse events infected and vaccine, respectively, up to the decision-making.  Parents who experience an adverse event due to vaccinating their children are switched to never-vaccinators in the sequel. The parameter $\alpha_i$ is the degree of relevance of disease infectiousness and $\gamma_i$ is the degree of relevance of vaccine adverse events to the subjective opinion of the household $i$. The probability of accepting vaccination against the disease is given by $pr_i=1/(1+\exp(-\pi_i))$ for the household $i$ and is equal to zero for never-vaccinators. Probabilities $pr_i=1/(1+\exp(-\alpha_i I_0))$ are used to generate the initial stance of parents in the household $i$ toward vaccination.

\subsubsection*{Observational social learning.}

Let the learning probability $q_{i,j}$ that the household $i$ has the correct perception or learning about the household's $j$ opinion or stance about vaccination. Meanwhile, $1-q_{i,j}$ is the probability that the household $i$ will make the wrong perception of the household opinion $j$'. Such social learning does not have to be symmetric, that is, $q_{i,j}\neq q_{j,i}$. For instance, followers of a celebrity learn from the celebrity more than in the other way around. Reciprocal or symmetric social learning involves $q_{j,i}=q_{i,j}$. Let household $i$ have a set of vaccinator neighbors in the social network $N_V(i)$ with cardinality $n_V(i)$ and have a set of non-vaccinator's social network neighbors $N_N(i)$ with cardinality $n_N(i)$. Let also the total number of neighbors be $n_S(i)=n_V(i)+n_N(i)$, where $N_S(i)=N_V(i)\cup N_N(i)$. Then parents in the household $i$ make the decision to vaccinate their children based on the following posterior probability:
\begin{equation}\label{qeq0}
P_S(i)=\dfrac{pr_i \cdot \prod_{j\in N_V(i)}q_{j,i}\cdot \prod_{k\in N_N(i)}(1-q_{k,i})}{pr_i \cdot \prod_{j\in N_V(i)}q_{j,i}\cdot \prod_{k\in N_N(i)}(1-q_{k,i})+(1-pr_i) \cdot \prod_{j\in N_V(i)}(1-q_{j,i})\cdot \prod_{k\in N_N(i)}q_{k,i}}
\end{equation}
or what we call the Bayesian aggregation rule in observational social learning. The rationale of the formula in equation \eqref{qeq0} is that the prior probability of vaccination $pr_i$ is updated by independent information collected or perceived from neighbors in the network. A vaccinator is perceived to have that opinion with probability $q_{i,j}$, and a non-vaccinator is perceived to have the opinion to vaccinate with probability $1-q_{i,j}$.


Nondirectional social learning, in the sense of outward uniformity, means $q_{j,i}=q_{j,k}=:q_j$ for all $i\neq k$ and for all $j$.  

In case that $q_j=q$ for all $j$, then
\begin{equation}\label{qeq}
P_S(i)=\dfrac{pr_i \cdot q^{n_V(i)}\cdot (1-q)^{n_N(i)}}{pr_i \cdot q^{n_V(i)}\cdot (1-q)^{n_N(i)}+(1-pr_i) \cdot (1-q)^{n_V(i)}\cdot q^{n_N(i)}}
\end{equation}
for $0<q<1$. 
An uninformative probability $q=.5$ results in no social influence on the parent, since $P_S(i)=pr_i$. 

The model in equation \eqref{qeq0} can be re-written as
\begin{equation}\label{seq0}
P_S(i)=\dfrac{1}{1+\exp\left(-\Delta_i-\pi_i\right)}
\end{equation}
where $\pi_i=\text{logit}(pr_i)$ and $\Delta_i=
\sum_{j\in N_V(i)}\text{logit}(q_{j,i})-\sum_{k\in N_N(i)}\text{logit}(q_{k,i})$. (Note: $\text{logit}(p)=\log(\dfrac{p}{1-p})$.)

It is also a model of bounded rationality. Putting $PS(i)$ in the Boltzmann distribution form
$$
P_S\left(i\right)=\frac{\exp{\left(.5{(\Delta}_i+\pi_i)\right)}}{\exp{\left(.5{(\Delta}_i+\pi_i)\right)}+\exp{\left(.5{(\bar{\Delta}}_i+{\bar{\pi}}_i)\right)}}$$
with temperature $\tau=.5$, reveals a bounded rational model supported by physics \cite{Ortega2013}. In that case, the reward for vaccination is given by $\pi_i=\alpha_iI-\gamma_iA=-{\bar{\pi}}_i$. Meanwhile, $\Delta_i=\sum_{j\in N_V\left(i\right)}\text{logit}\left(q_{j,i}\right)-\sum_{k\in N_N\left(i\right)}\text{logit}\left(q_{k,i}\right)={-\bar{\Delta}}_i$ is the induced social pressure to vaccinate. 

DeGroot model of selection \cite{Namatame2016}, can also be seen in the term $\Delta_i$ if the neighbor $j\in N_S(i)$ is given a weight of $\text{logit}(q_{j,i})$ and using the discrete opinions to be valued as $+1$ for vaccination and $-1$ for no vaccination. This makes a stochastic DeGroot model of selection a special case of the Bayesian aggregation rule in equation \eqref{qeq0}.

Similarly, when $q_i=q$ and $pr_i=pr$ for all $i$ as in the model of equation \eqref{qeq}, the model can be re-written as
\begin{equation}\label{seq00}
P_S(i)=\dfrac{1}{1+\exp\left(-\delta_i \dfrac{n_V(i)-n_N(i)}{n_S(i)}-\pi\right)}
\end{equation}
where $\pi=\text{logit}(pr)$ and $\delta_i=n_S(i)\,\text{logit}(q)$ is the degree of injunctive social norm practiced by household $i$. See \cite{Phillips2020,Oraby2014,Oraby2015} for the case $\delta_i=:\delta$ for all $i$, in which case

\begin{equation}\label{seq}
P_S(i)=\dfrac{1}{1+\exp\left(-\delta(2\frac{n_V(i)}{n_S(i)}-1)-\pi\right)}
\end{equation}
which leads to the voting model of selection using $G:=2\frac{n_V(i)}{n_S(i)}-1$ to decide the winner strategy when $G>0$ and loser when $G<0$. In other words, the latter model is also a special case of \eqref{qeq0} in case of homogeneous observational learning. But since $\delta$ is nonnegative, the equation \eqref{seq} can only reveal the human behavior when $q\geq 0.5$ and so the model in \eqref{seq} is less attractive than that in \eqref{qeq0}.

On each day, the parental position on vaccination is updated randomly based on the probabilities $P_S$ and vaccinators are chosen with the probability $\rho$ of vaccinating all their children where $\rho$ is the probability of getting access to vaccination based on the resources. 

\subsubsection*{Epidemiological measures.}
To analyze the effect of the probability of learning $q_{j,i}$ on the opinion of parents and on the spread of disease, we use a number of epidemiological measures: the size of epidemics, the peak of the epidemic, the uptake of vaccines, the number of vaccinators, and the basic reproduction number $R_0$ that we use only for the calibration of ERN. We use $R_0$ as an epidemiological measure and not as a threshold, see \cite{Pastor-Satorras2001}.
The size of the epidemic is the total number of children infected at the end of the epidemic. The uptake of the vaccine is the total number of children who are vaccinated. The ending number of vaccinators is used to measure whether the opinion of vaccination becomes a consensus at the end of the epidemic. The basic reproduction number $R_0$ is defined to be the average number of secondary cases in a completely susceptible population. We use that definition to build an algorithm to estimate the value of $R_0$ (see Algorithm 1 in Supplementary Material SIII, Methods). In that algorithm, we use Bayes' theorem to calculate the probability that infection happens due to contact with the index case, which is then used to calculate the mean number of infections. To find the grand mean, we averaged over several simulations of disease transmission and then over the $N$ households, which can include the index case, then finally over simulations of the various networks.

\subsubsection*{Model simulation.}
The model is implemented using stochastic simulation for 100 runs to examine the effect of the probability of social learning $q_{j,i}$ on the uptake of vaccines and the spread of pediatric diseases. We assume that $q_{j,i}$ are uniformly distributed on $q\pm 0.05$ for a prespecified value of $q$ where $0<q<1$. The stance towards vaccination and the disease states of infected children are updated at the beginning of each time step (day). Multiple infections can occur on the same day in the network and the numbers $n_I(i)$ and $I(i)$ are updated every day for all $i$, $i=1,2,\ldots,N$. An infected child on the $j^{th}$ day after infection jumps either to the end of the incubation period (recovers) or remains infected with a transition to the following day with probabilities given by the truncated exponential distribution (see supplementary material SI, Model).

Our simulation codes rely on the NumPy-compatible CuPy Python library \cite{cupy}, accelerated with NVIDIA CUDA \cite{cuda} for parallel calculations in Graphical Processing Units (GPUs). Most of the calculations were performed on a CentOS workstation with $8$ NVIDIA Tesla (Kepler) K80 GPU cards, each of which has $2496$ CUDA cores and 12GB memory. For more information, see Supplementary Material SV, Codes of Simulations.


The eight GPU cards were used to distribute the 100 runs of stochastic
simulations, with each card running one network simulation at a time.
While the epidemic process can be run immediately after the network
generations, we typically run the two processes separately. After
network generation, the networks were saved to a hard drive, and at the beginning of the epidemic process, these networks were read from the hard drive.

\begin{itemize}

\item Average generation time of one network, including the savings of the networks to hard drive:

\begin{itemize}

\item Erd\H{o} s-R'enyi Network (ERN): $21$ seconds of wall clock time.

\item Barab\textbackslash'asi-Albert Network (BAN): $116$ seconds of wall clock time. This network has more serial steps than the ERN,
since the households are added one by one to the network.

\end{itemize}

\item Average epidemic run time over one network:

\begin{itemize}

\item Erd\H{o} s-R'enyi Network (ERN): $13$ seconds of wall clock time.

\item Barab\textbackslash Asi-Albert Network (BAN): $11$ seconds of wall clock time. The shorter time is due to the fact that the sizes of the epidemics are smaller than those on the ERNs.

\end{itemize}

\end{itemize}
\subsubsection*{Parameter values.}
Model parameterization is done using the literature, calibration, and guesstimation. We use a number of $N=100,000$ households in a moderate-sized city with a random number of children in each household of a mean equal to two and half children. We assume that the mean degree in the children's network is $40$ and the mean degree of the parents' network is $60$ in the ERN. We assume that the epidemic starts with $I_0=10$ initially infected children that are randomly dispersed among $N$ households. We postulate a disease of a mean incubation period of $11$ days and a maximum of $16$ days, \cite[p.8]{Vynnycky2010}. We find the values for $\beta$ by calibration using the $R_0$ values between 12 and 18, \cite[p.8]{Vynnycky2010}. We assume that a fraction of $5\%$ of the population will refuse to vaccinate at all (never-vaccinators), due to medical or ideological reasons. A full table of the definition of parameters and their values can be found in the supplementary material SII and Table S1.

\section{Results}
Using voting models of selection, the effect of the degree of the injunctive social norm or peer pressure $\delta$ on epidemic sizes and their peaks as well as the vaccine uptake is almost not noticeable for selected values of $\delta$ in $[0.025,0.225]$, Fig.  1. That is true for the case of Erd\H{o}s-R\'{e}nyi (random) network model (ERN) in Fig. 1 (a), (b) and (c), and clearer for Barab\'{a}si-Albert network model in Fig. S1 (d), (e) and (f).The degree of injunctive social norm works differently on the two types of networks on vaccine uptake and on the sizes and peaks of epidemics as indicated by the results of the simulations shown in top panels versus bottom panels of Fig. 1. Irrelevant to the group pressure, vaccine uptake on Barab\'{a}si-Albert networks is larger than on Erd\H{o}s-R\'{e}nyi networks.     

\begin{figure}[H]
\subfigure[]{\includegraphics[width=5.5cm]{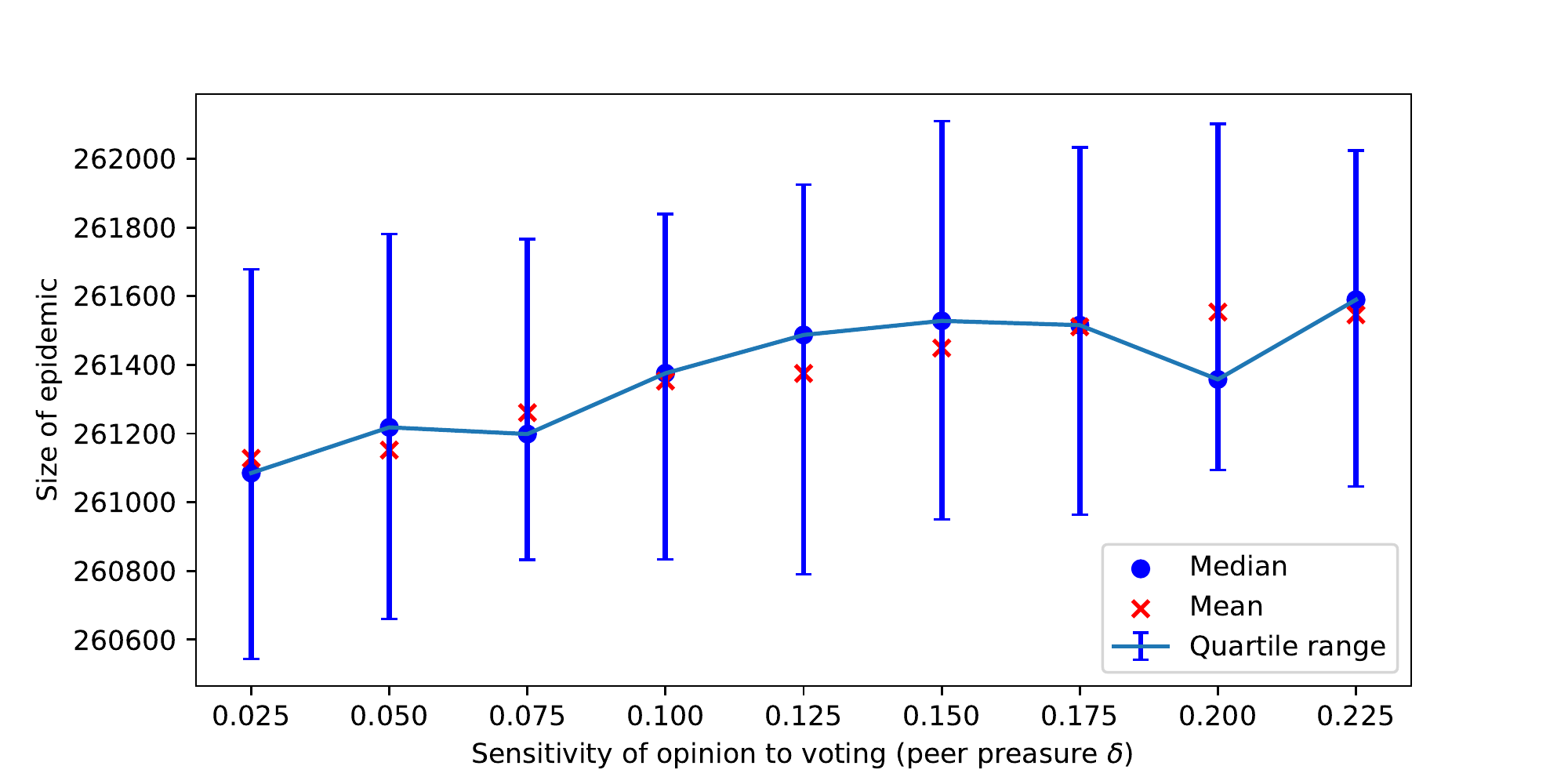}} 
\subfigure[]{\includegraphics[width=5.5cm]{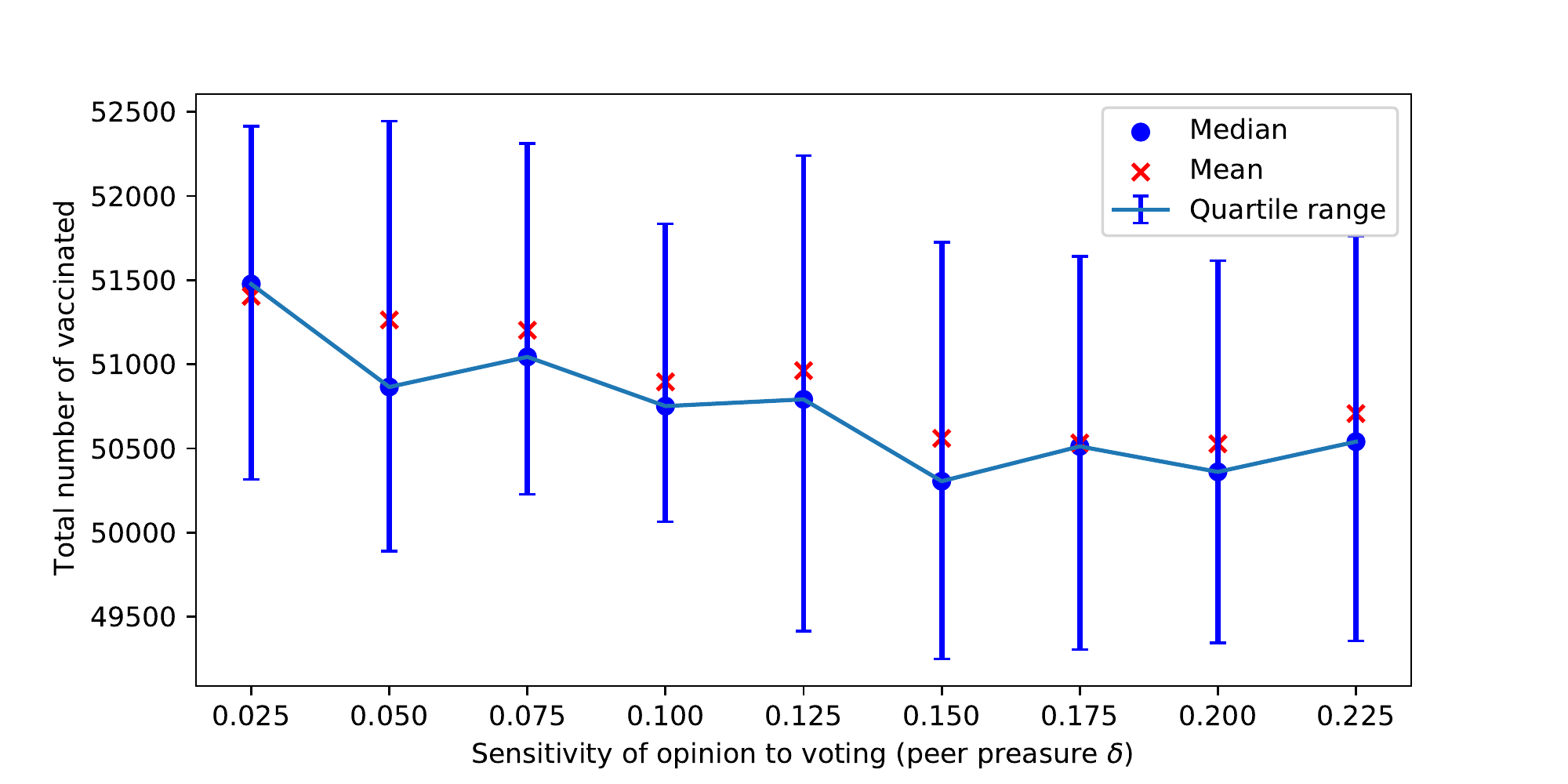}}
\subfigure[]{\includegraphics[width=5.5cm]{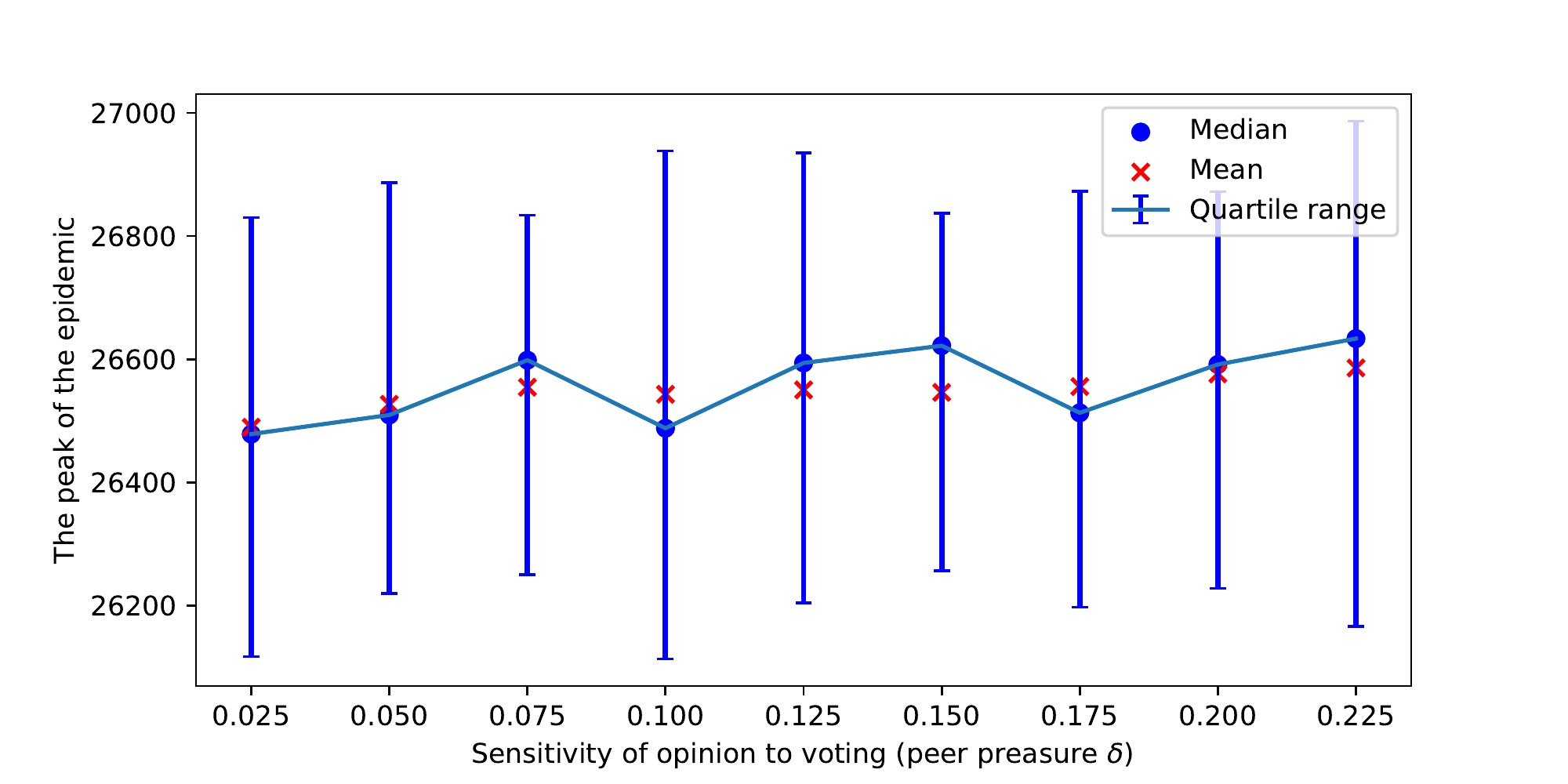}} 
\subfigure[]{\includegraphics[width=5.5cm]{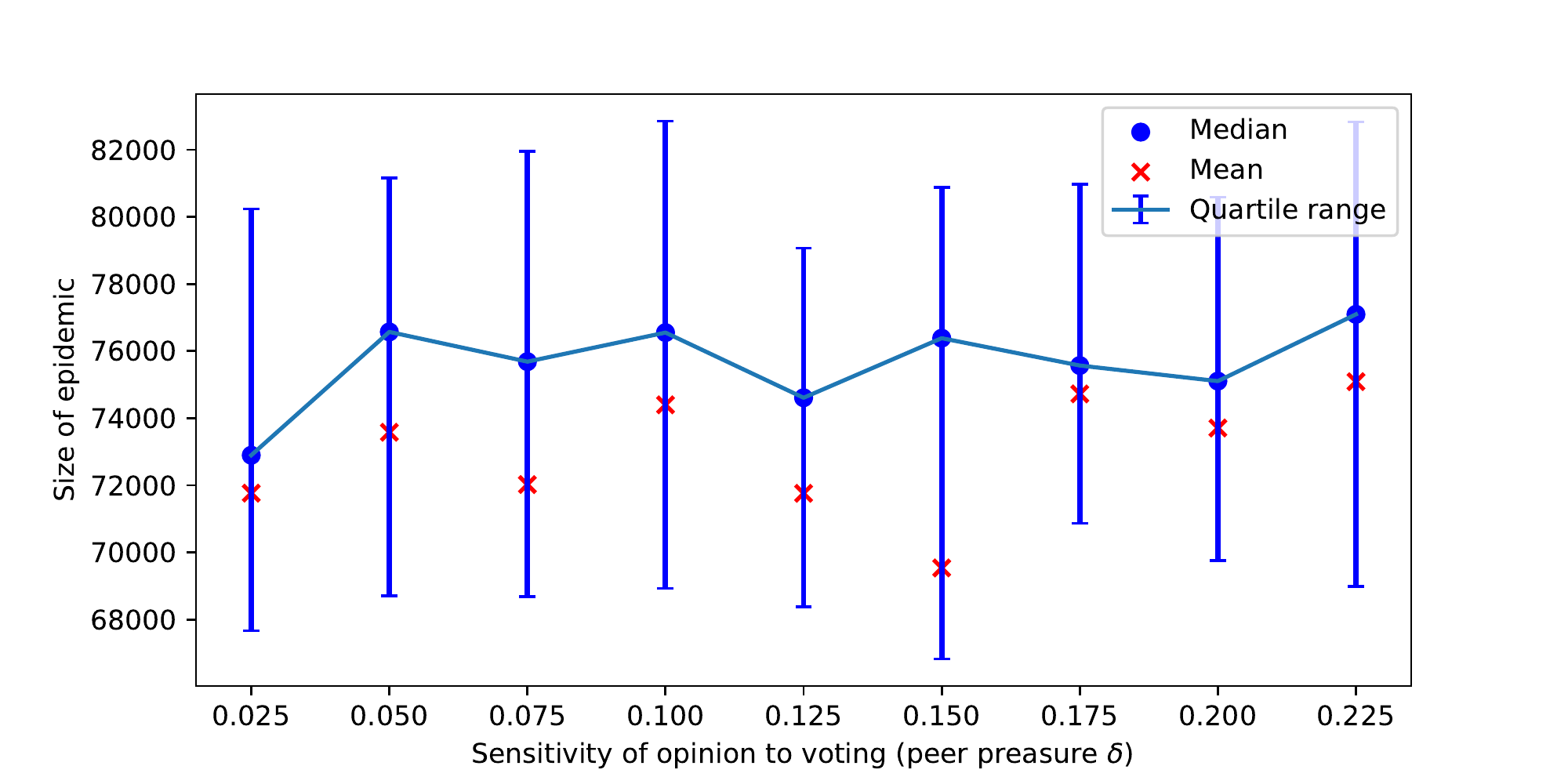}}
\subfigure[]{\includegraphics[width=5.5cm]{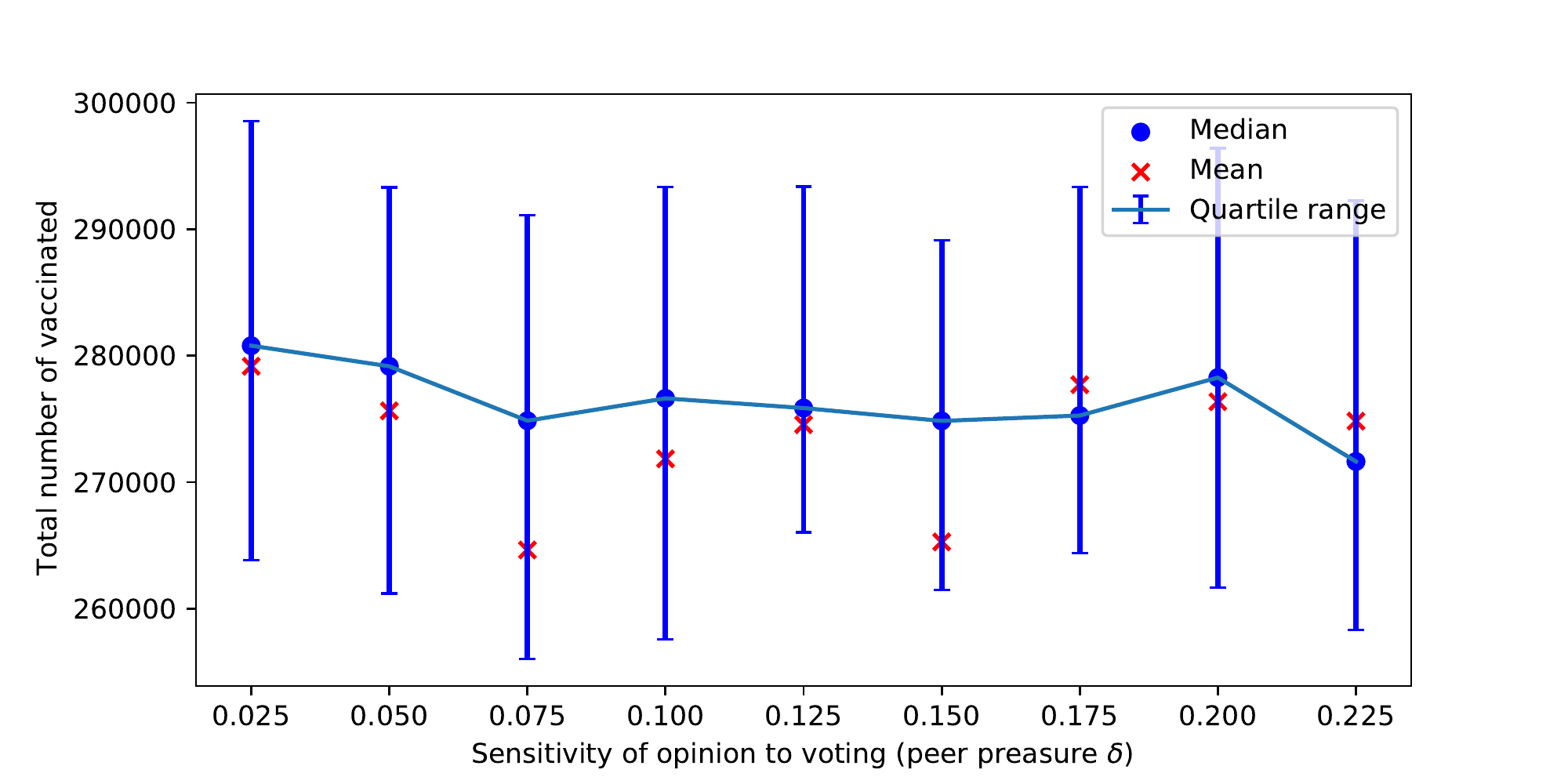}} 
\subfigure[]{\includegraphics[width=5.5cm]{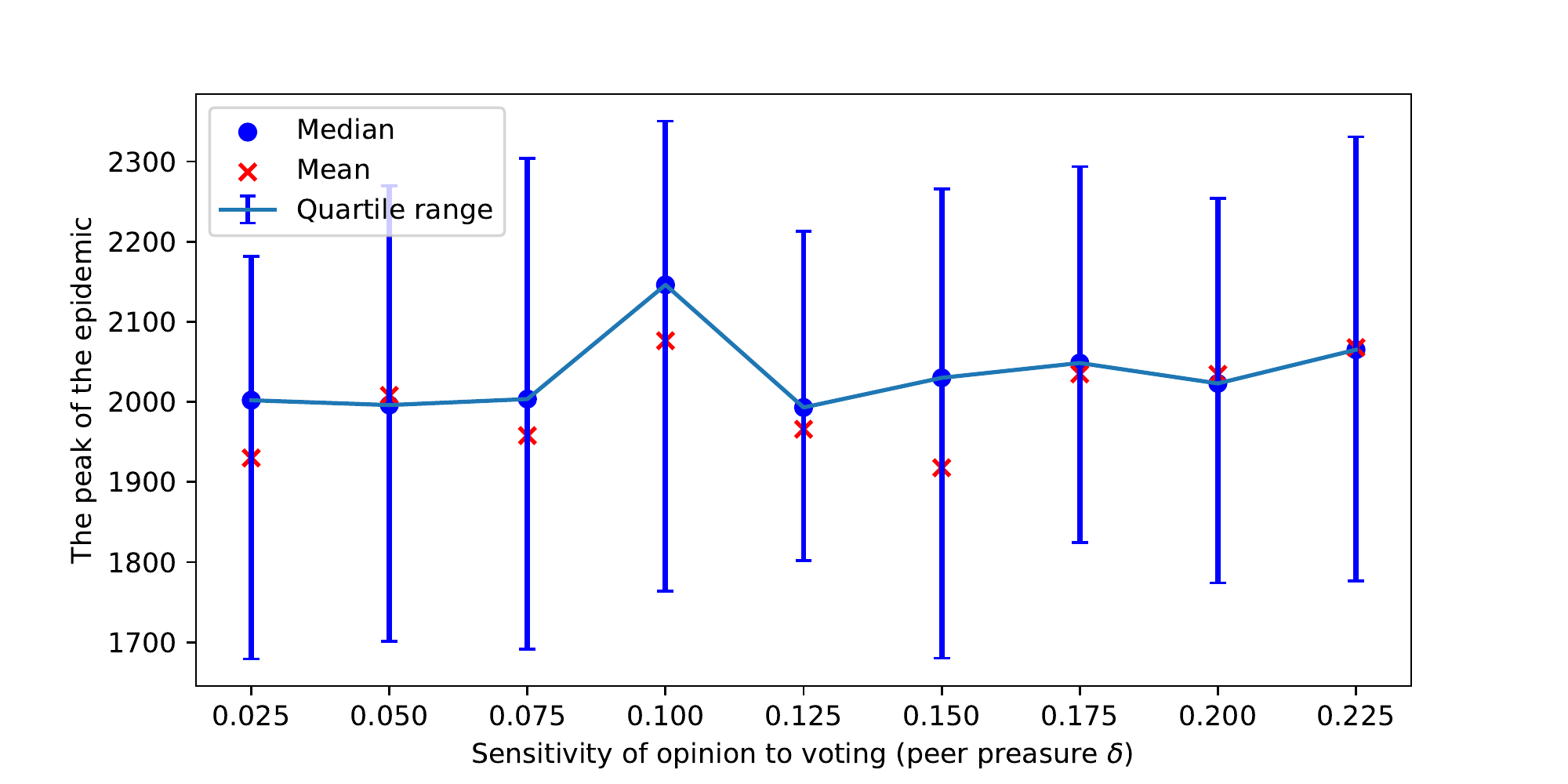}}
\caption{{\bf Fig. 1. Simulations of sizes of the epidemic, the total number of vaccinated children, and the peak of the epidemic for different values of $\delta$ in equation \eqref{seq}.} Simulations are done on Erd\H{o}s-R\'{e}nyi (random) network model (ERN) in  (a), (b), and (c), and on Barab\'{a}si-Albert network model (BAN) in (d), (e) and (f). In all of the simulations $P_{adv}=.0001$ and $\rho=.01$.}\label{fig0}
\end{figure}
In case of Erd\H{o}s-R\'{e}nyi (random) network model (ERN) and using the general Bayesian aggregation rule in equation \eqref{qeq0}, we find richer dynamical behaviors than those when voter model with an injunctive social norm or peer pressure $\delta$ is used. First, let us assume that there are enough vaccines to vaccinate one child in every 100 children every day. When $P_{adv}=.0001$, as $q$ increases, the pressure imposed on parents increases and leads to higher uptake of vaccines and, consequently, to a smaller size and peak of epidemics, Figs. 2 (a), (b), and (c). But when the probability of an adverse event increases, and so more adverse events occur, as the probability of perceiving the correct position increases, vaccine uptake levels drop. When $P_{adv}=.001$, the uptake of the vaccine will continue to increase with the values of $q<.5$, a pattern that changes for $q>.5$, Figs. 2 (d), (e), and (f). A higher probability of adverse events of $P_{adv}=.01$, results in a decrease in vaccine uptake as the value of $q$ increases beyond $q\sim .2$. In that case, both the size and the peak of the epidemics increase with the increase in the probability $q$, Fig. 2 (g), (h) and (i). 

\begin{figure}[H]
\centering
\subfigure[]{\includegraphics[width=5.5cm]{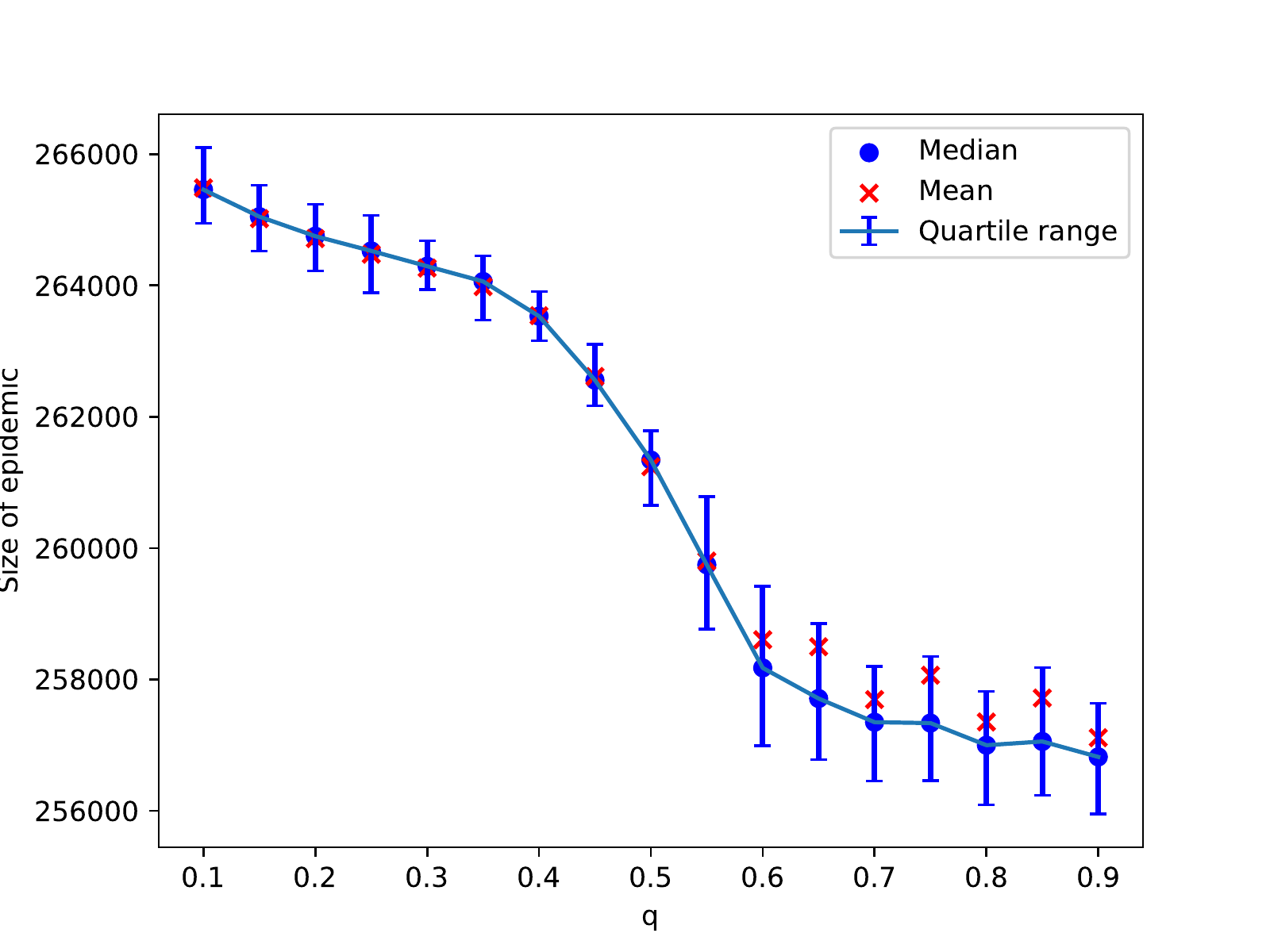}} 
\subfigure[]{\includegraphics[width=5.5cm]{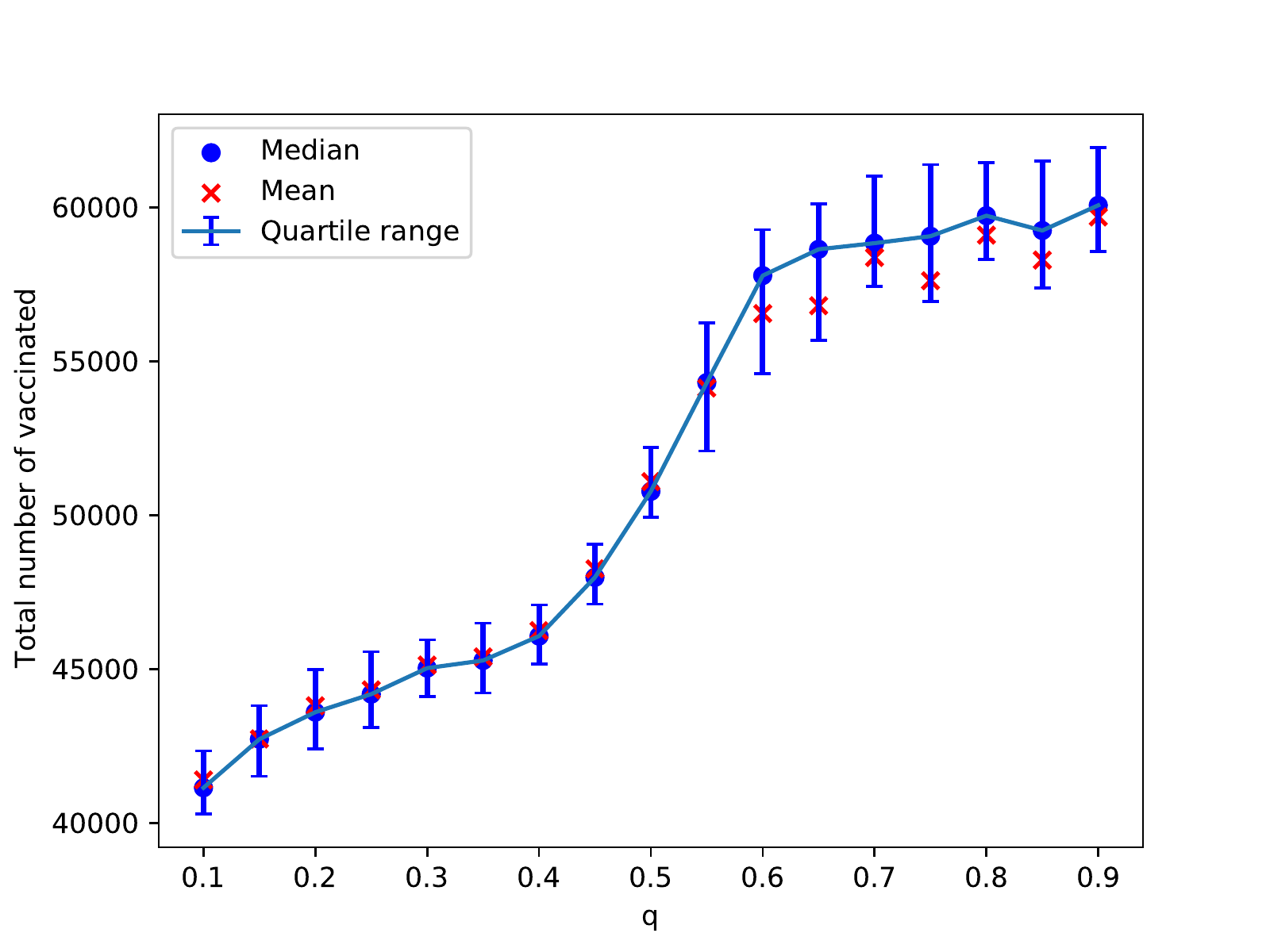}}
\subfigure[]{\includegraphics[width=5.5cm]{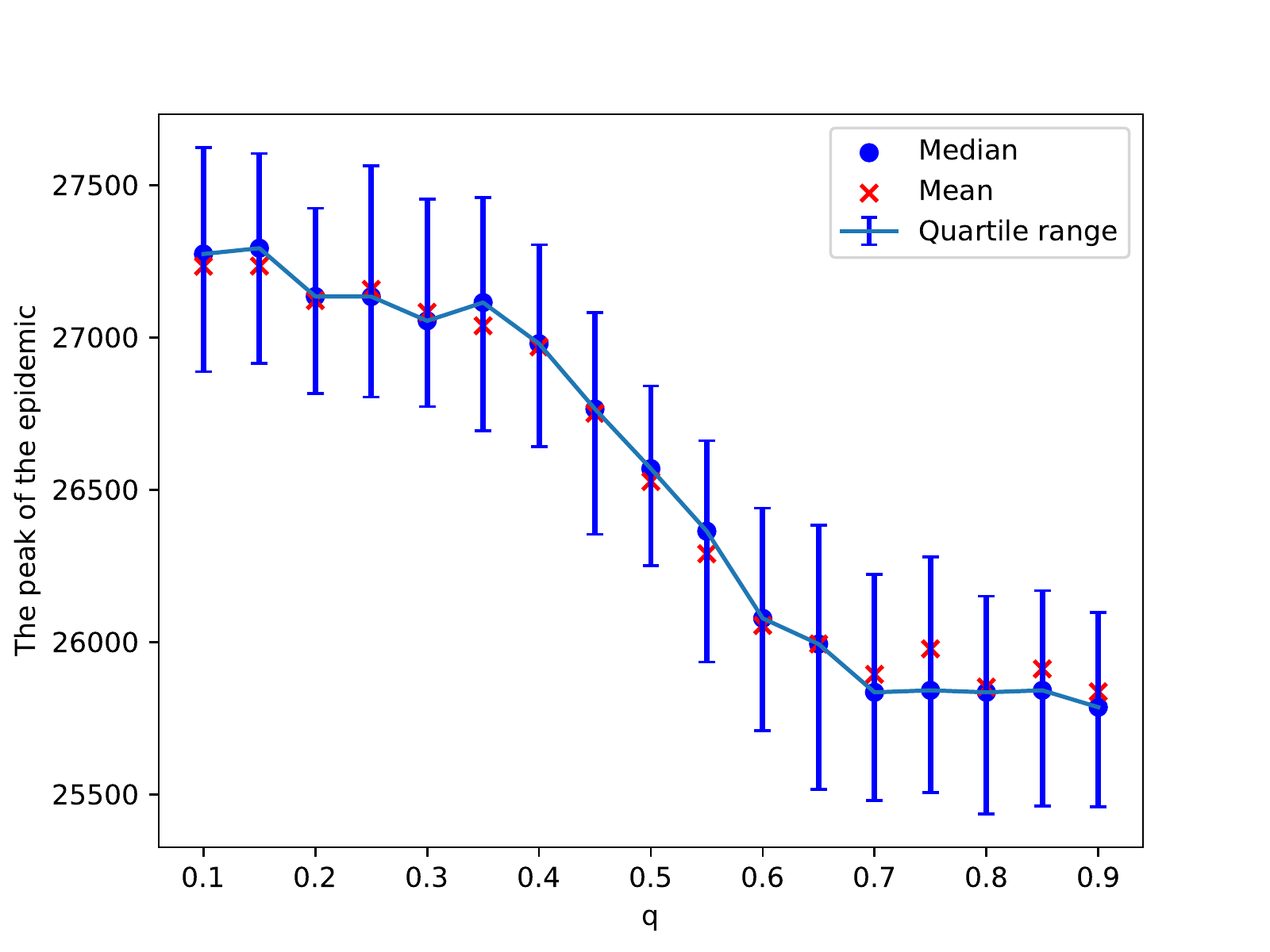}}
\subfigure[]{\includegraphics[width=5.5cm]{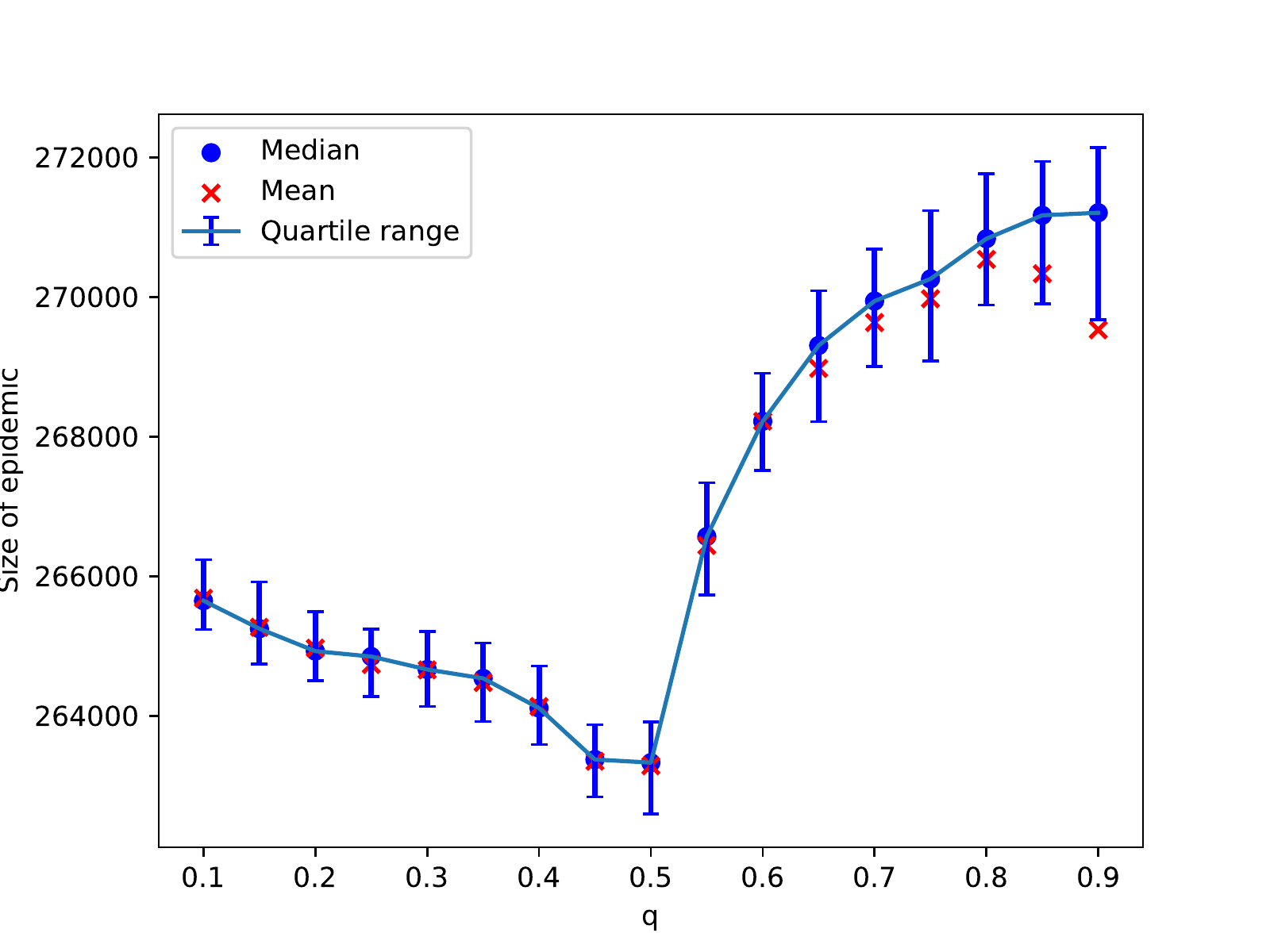}} 
\subfigure[]{\includegraphics[width=5.5cm]{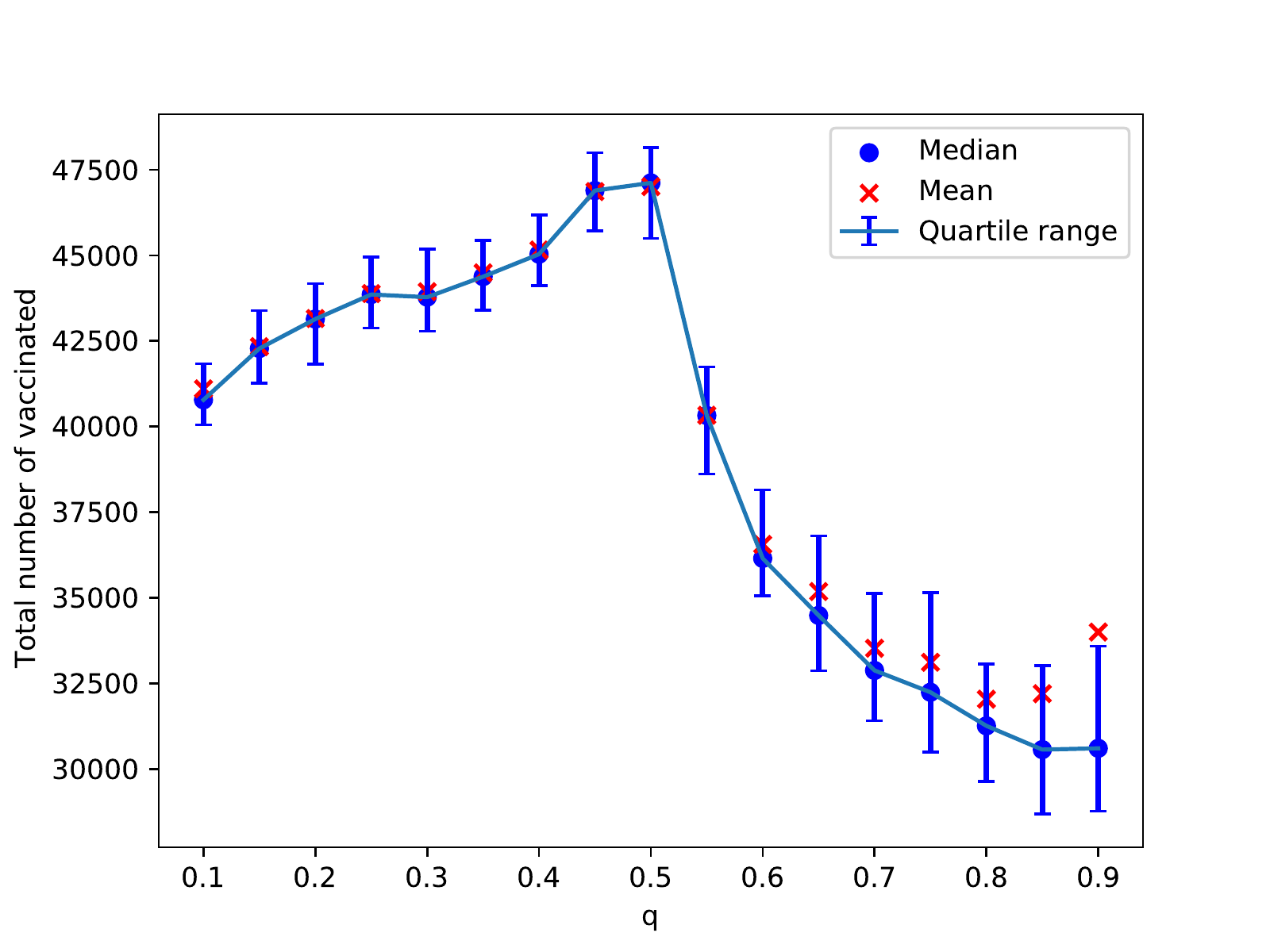}}
\subfigure[]{\includegraphics[width=5.5cm]{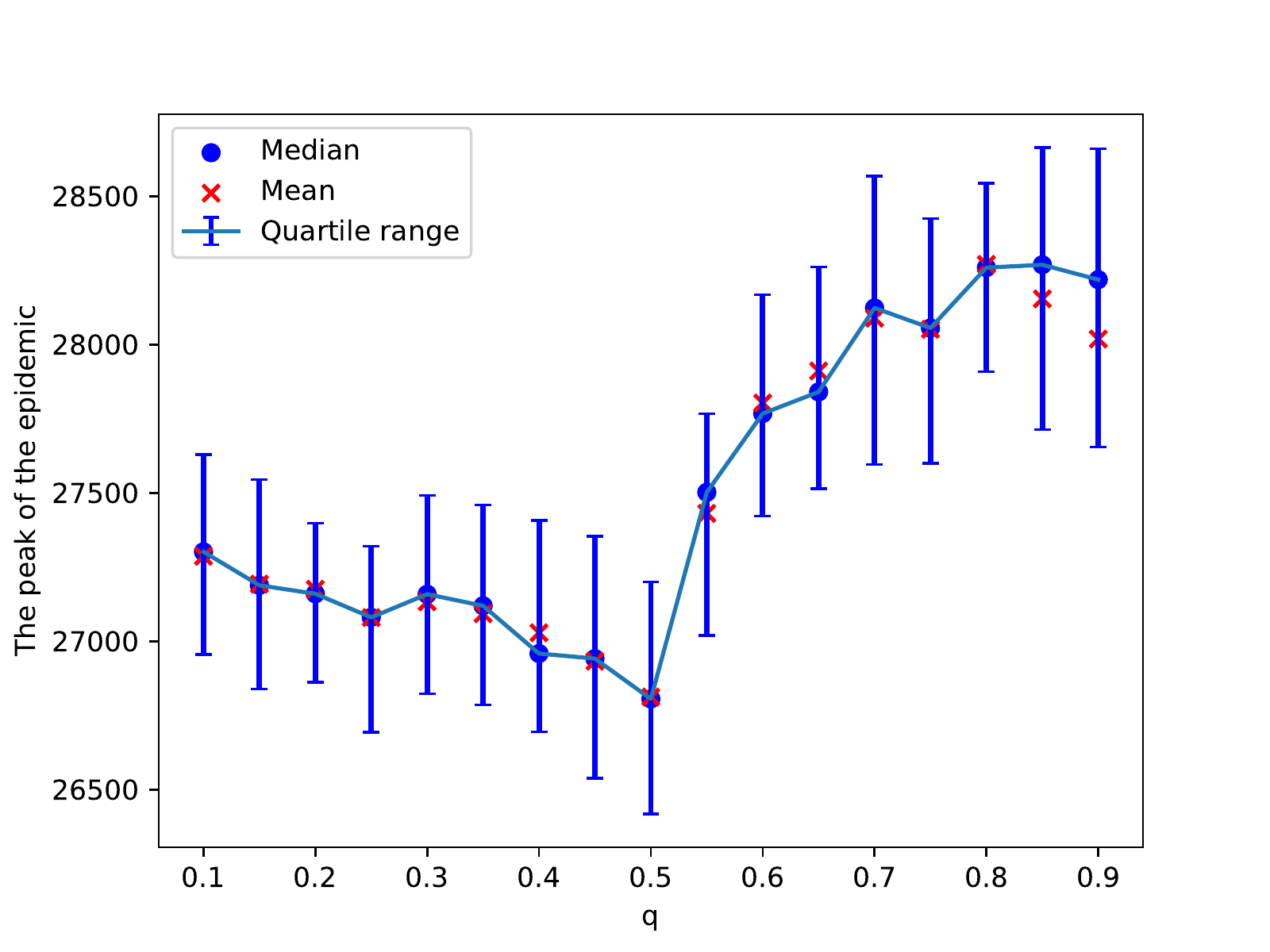}}
\subfigure[]{\includegraphics[width=5.5cm]{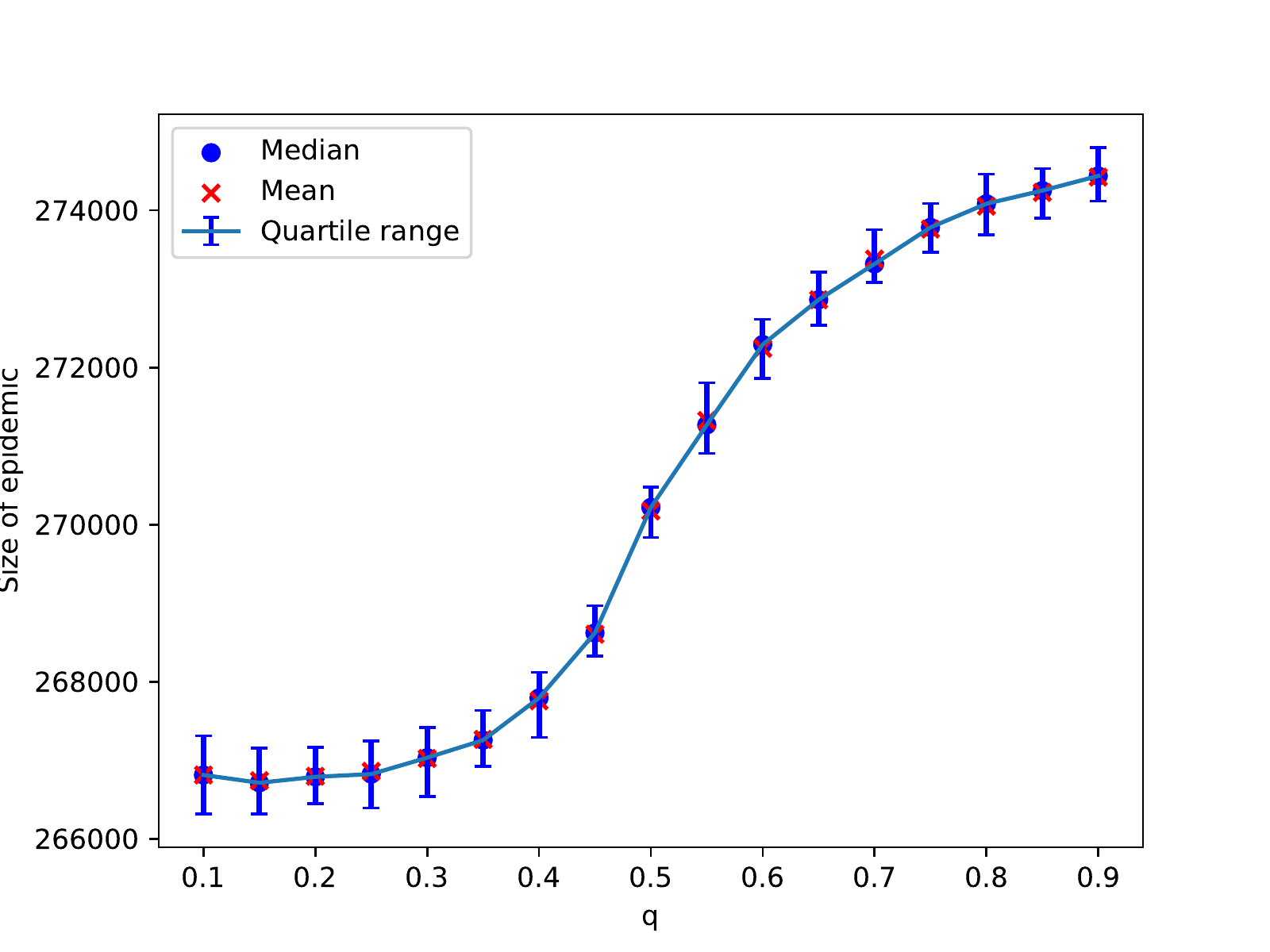}} 
\subfigure[]{\includegraphics[width=5.5cm]{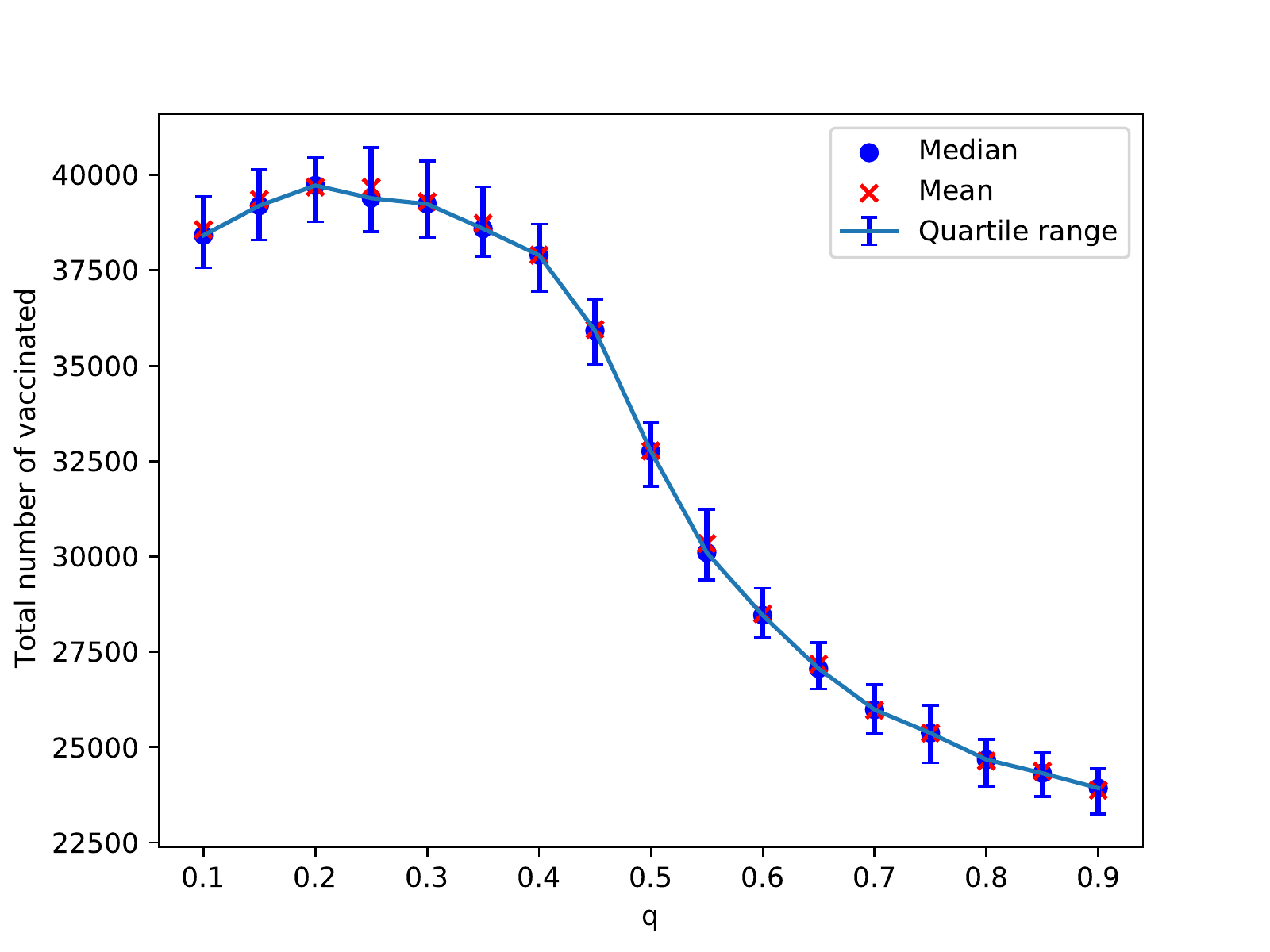}}
\subfigure[]{\includegraphics[width=5.5cm]{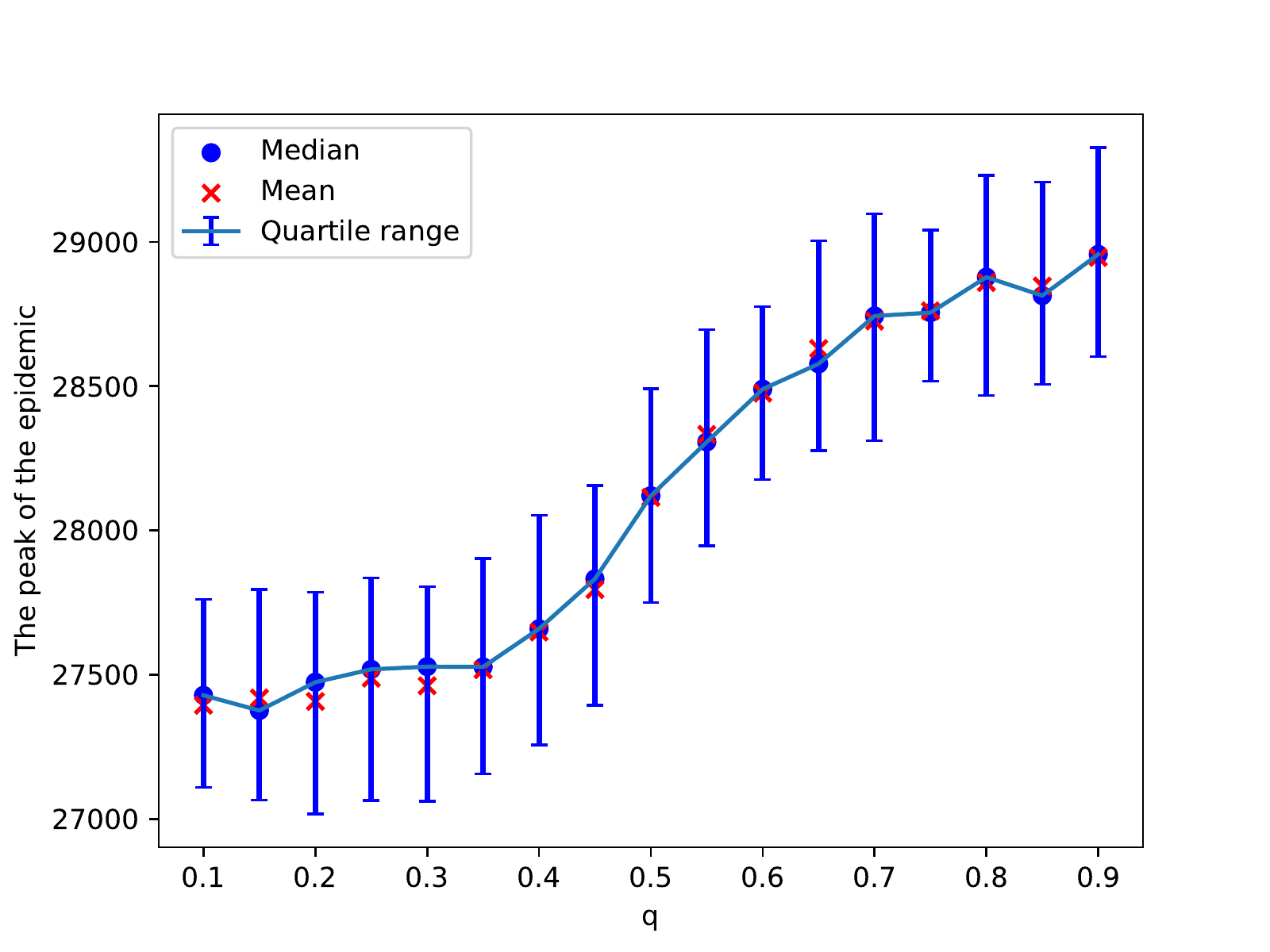}} 
\caption{{\bf Fig. 2. Simulations of sizes of the epidemic, the total number of vaccinated children, and the peak of the epidemic on Erd\H{o}s-R\'{e}nyi (random) network model (ERN) for different values of $q$.} Simulations are done using $P_{adv}=.0001$ in  (a), (b), and (c), $P_{adv}=.001$ in (d), (e), and (f), and $P_{adv}=.01$ in (g), (h) and (i). In all of the simulations $\rho=.01$.}\label{fig1}
\end{figure}

Those patterns change when we assume that vaccines are scarce, for instance, when it is only possible to vaccinate one child in every 1000 children every day. First, the uncertainty of the outcomes increases, compare panels of Fig. 2 to Fig. 3. The less vaccine available, the less number of adverse cases will appear. It takes a higher probability of an adverse event to effectively motivate parents to refuse vaccination of their children, Fig. 3 (b), (e) and (h). It is noticeable that the peak of the epidemics in that case do not change significantly with the change in the learning probability.

\begin{figure}[H]
\centering
\subfigure[]{\includegraphics[width=5.5cm]{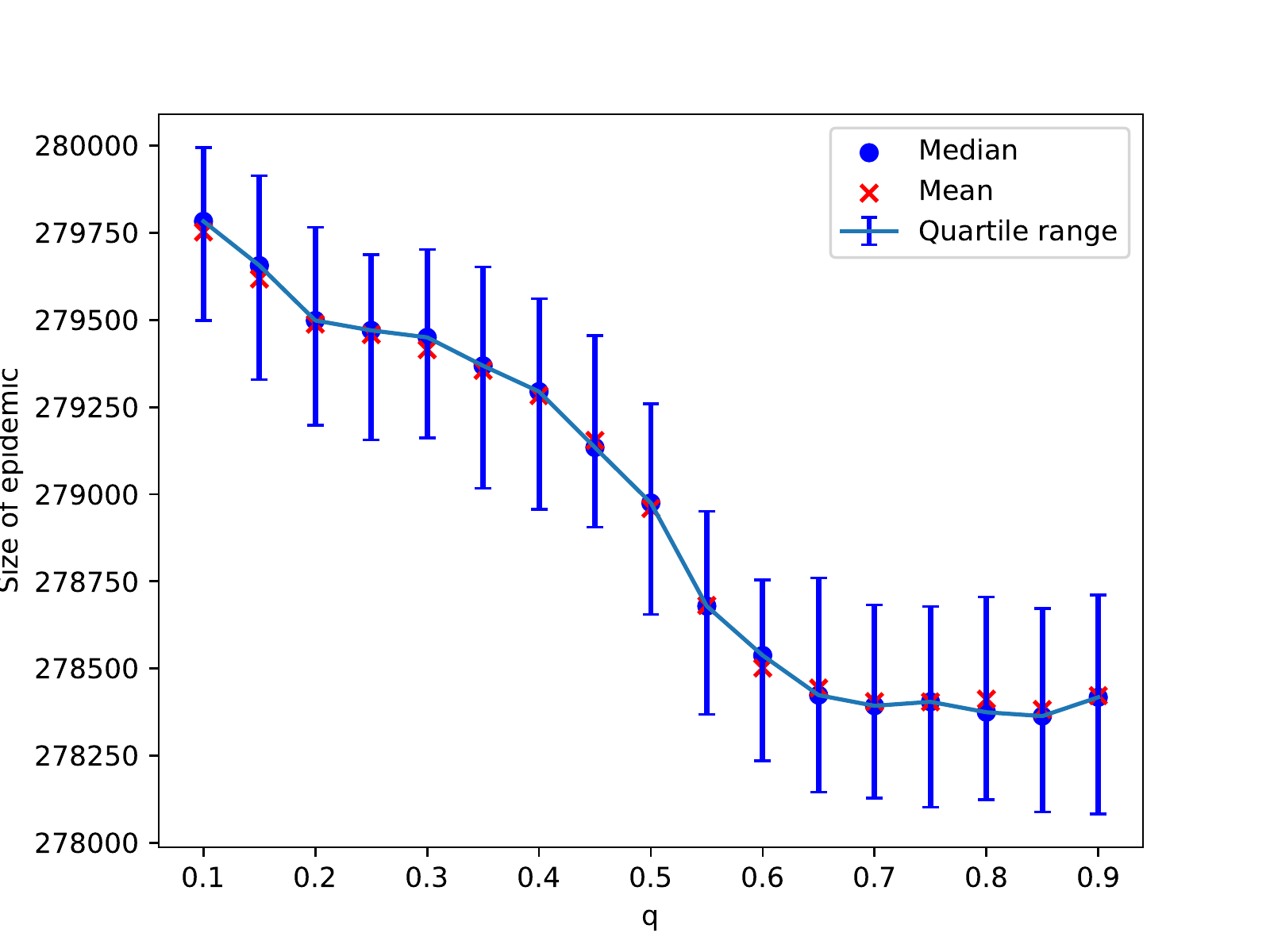}} 
\subfigure[]{\includegraphics[width=5.5cm]{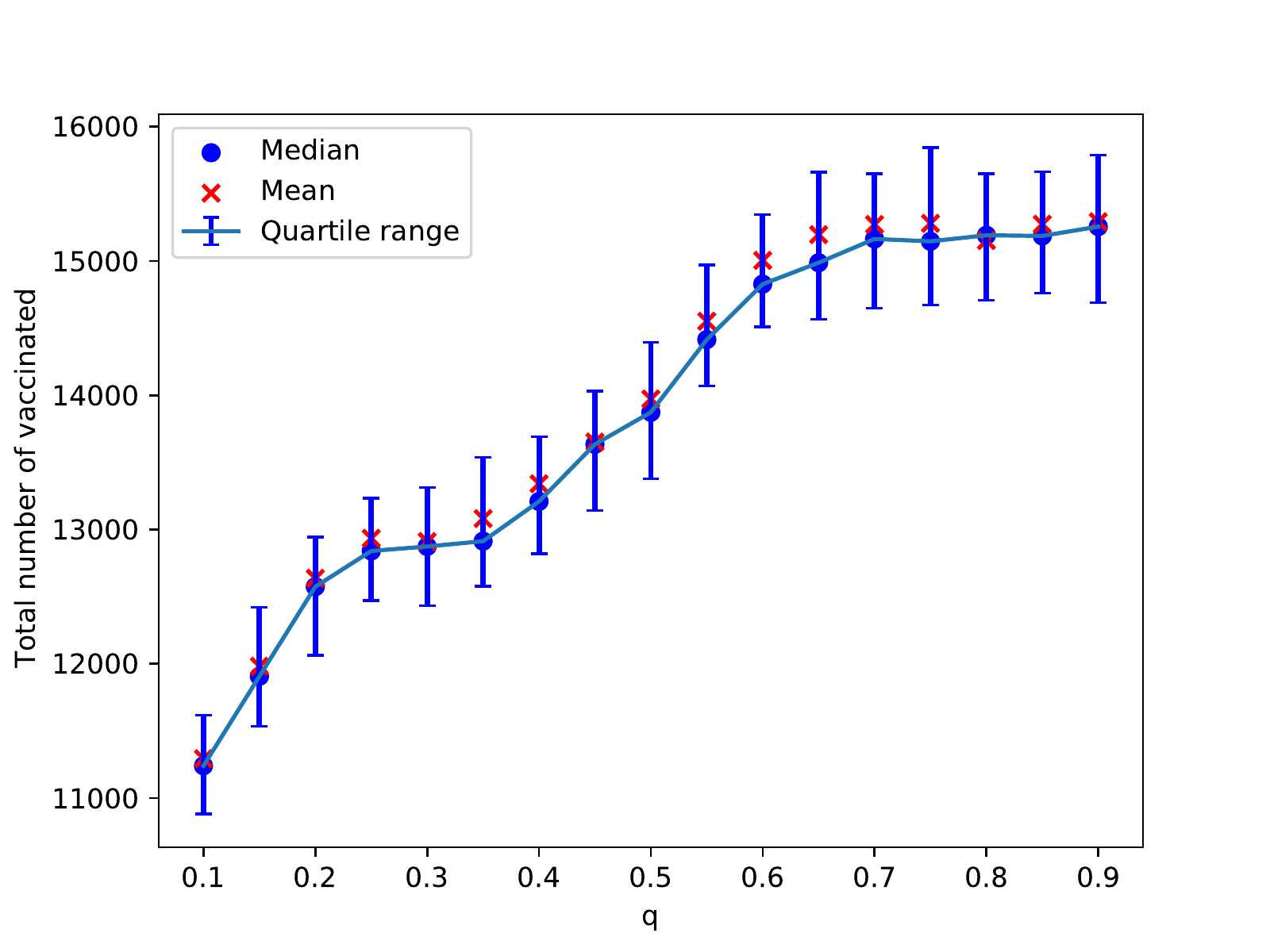}}
\subfigure[]{\includegraphics[width=5.5cm]{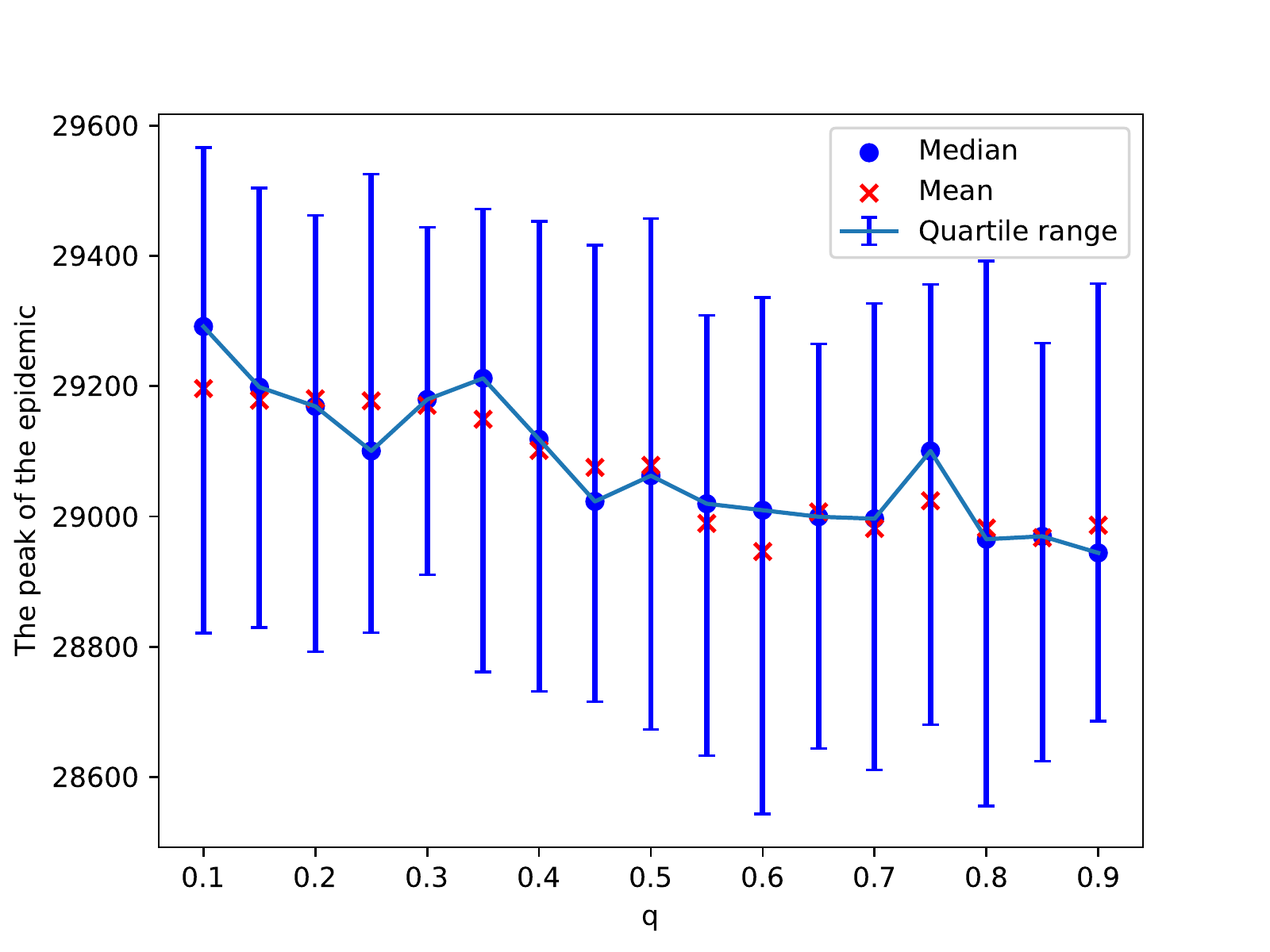}} 
\subfigure[]{\includegraphics[width=5.5cm]{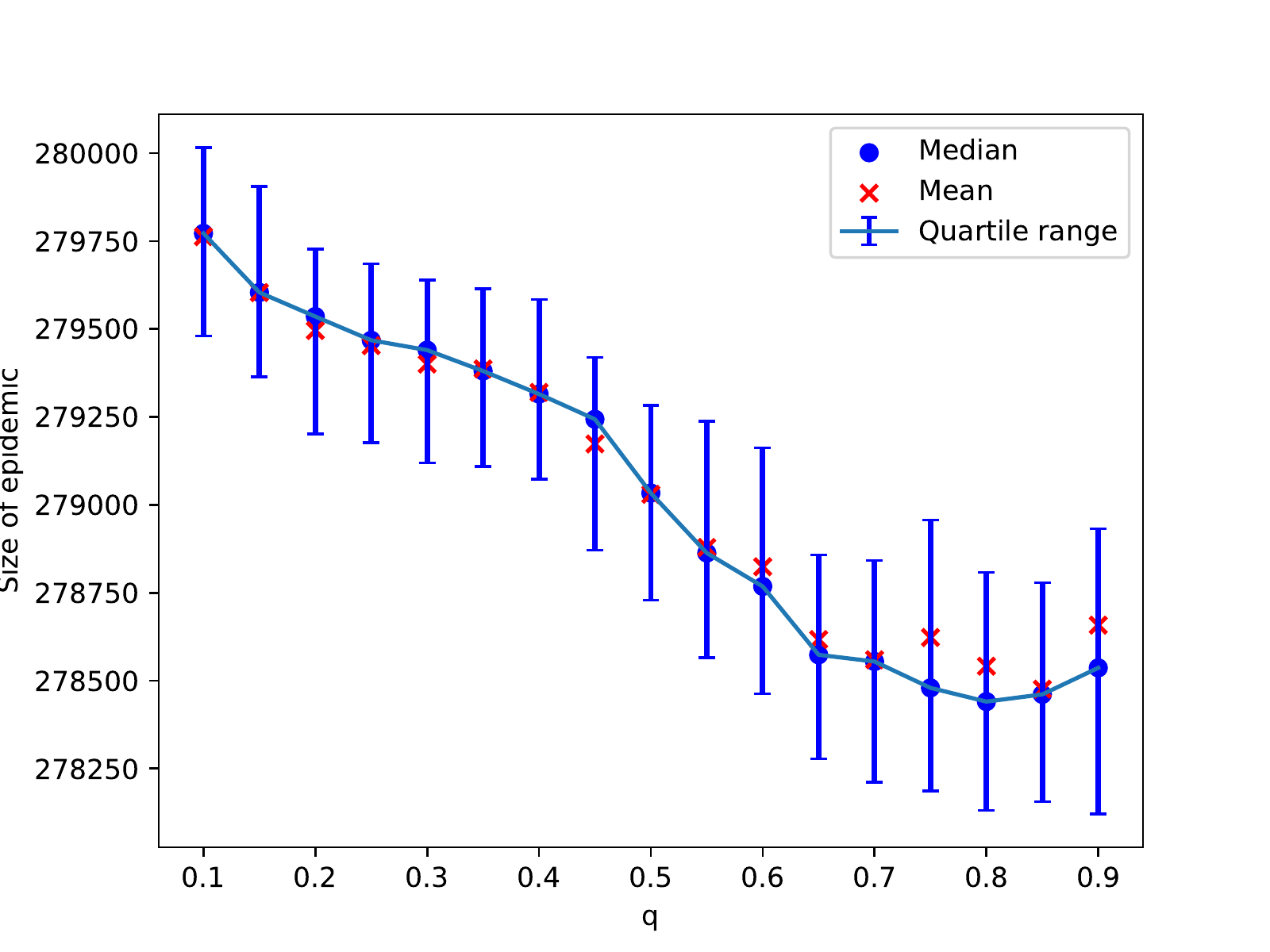}} 
\subfigure[]{\includegraphics[width=5.5cm]{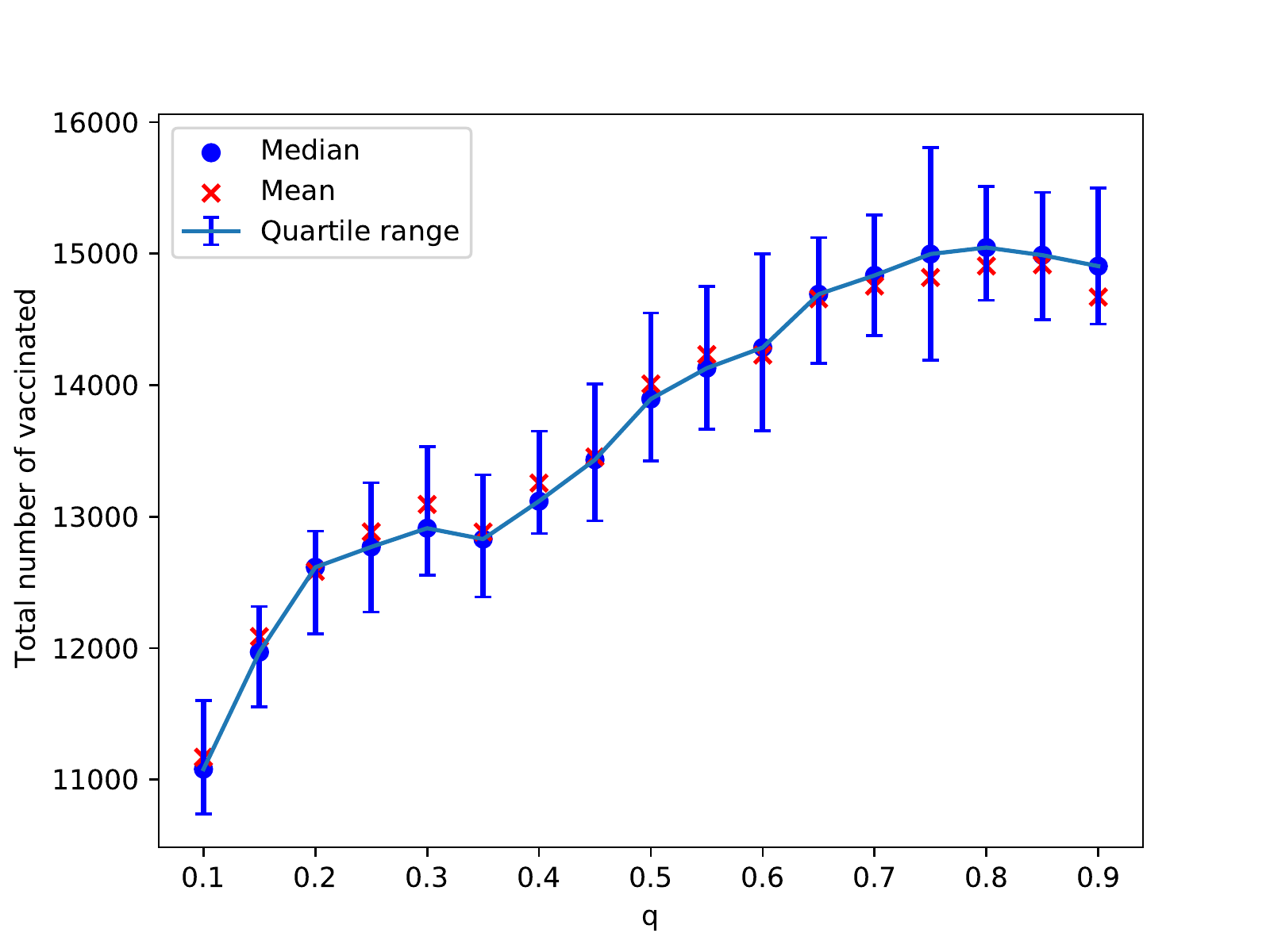}}
\subfigure[]{\includegraphics[width=5.5cm]{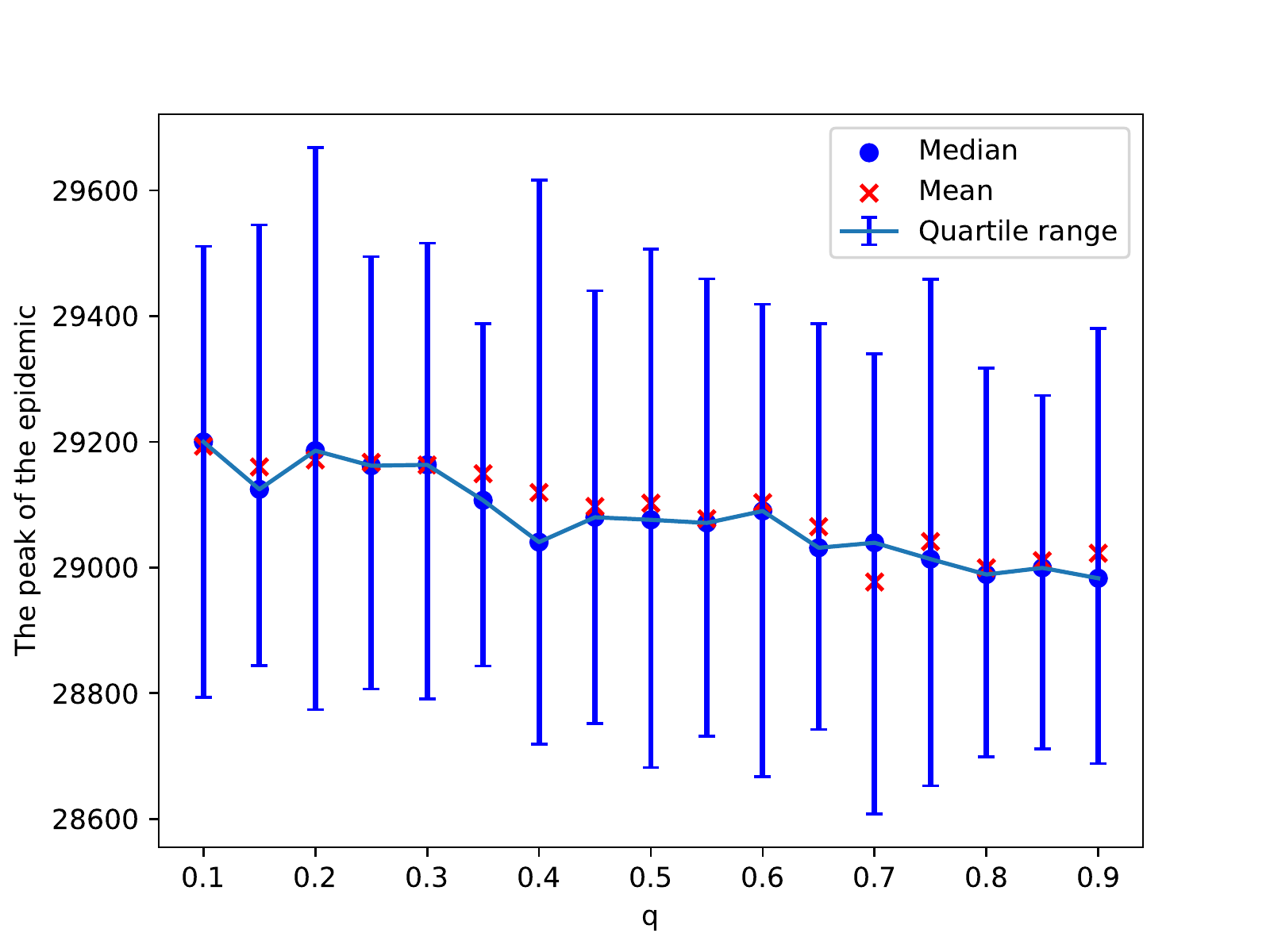}}
\subfigure[]{\includegraphics[width=5.5cm]{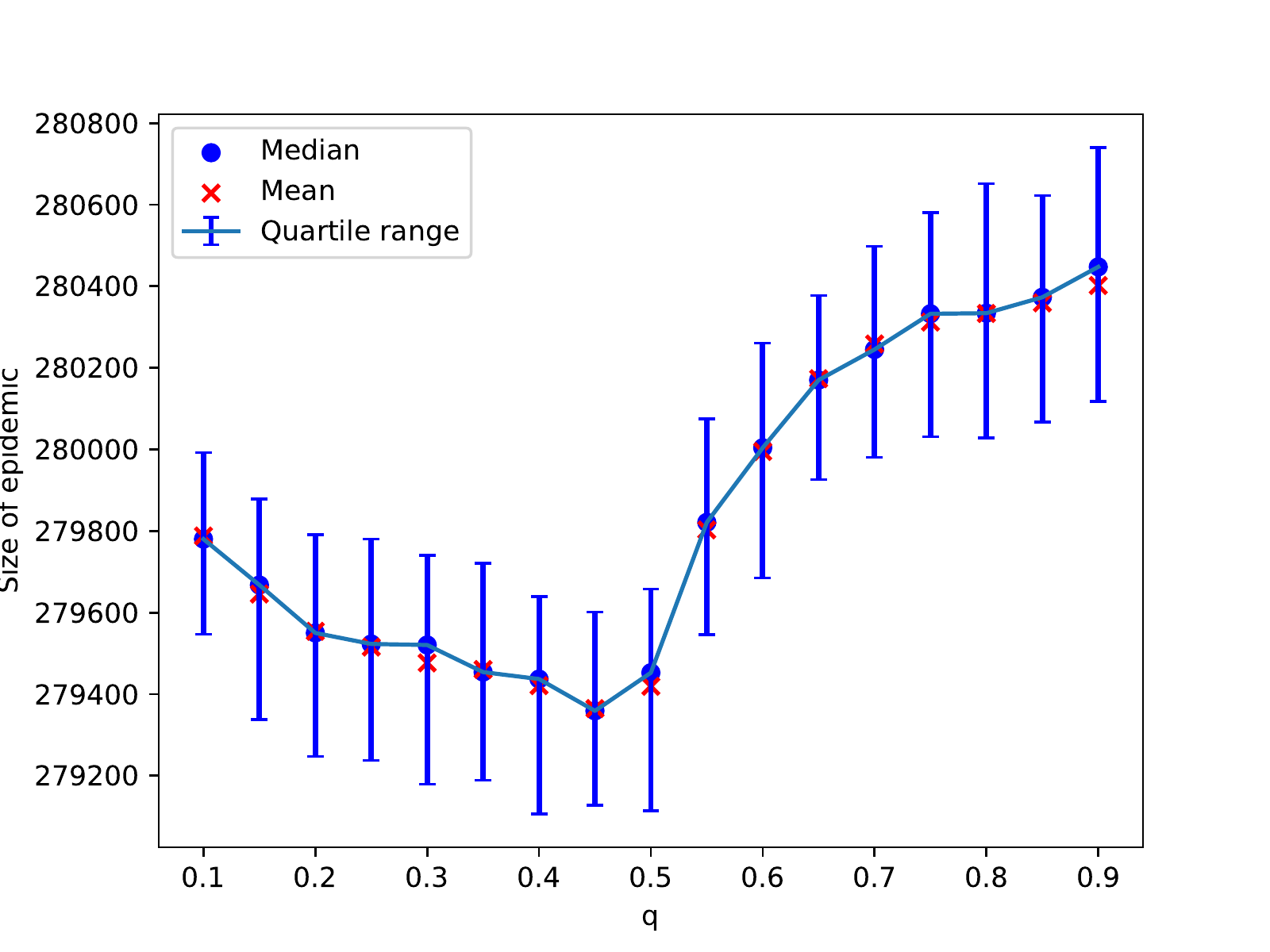}} 
\subfigure[]{\includegraphics[width=5.5cm]{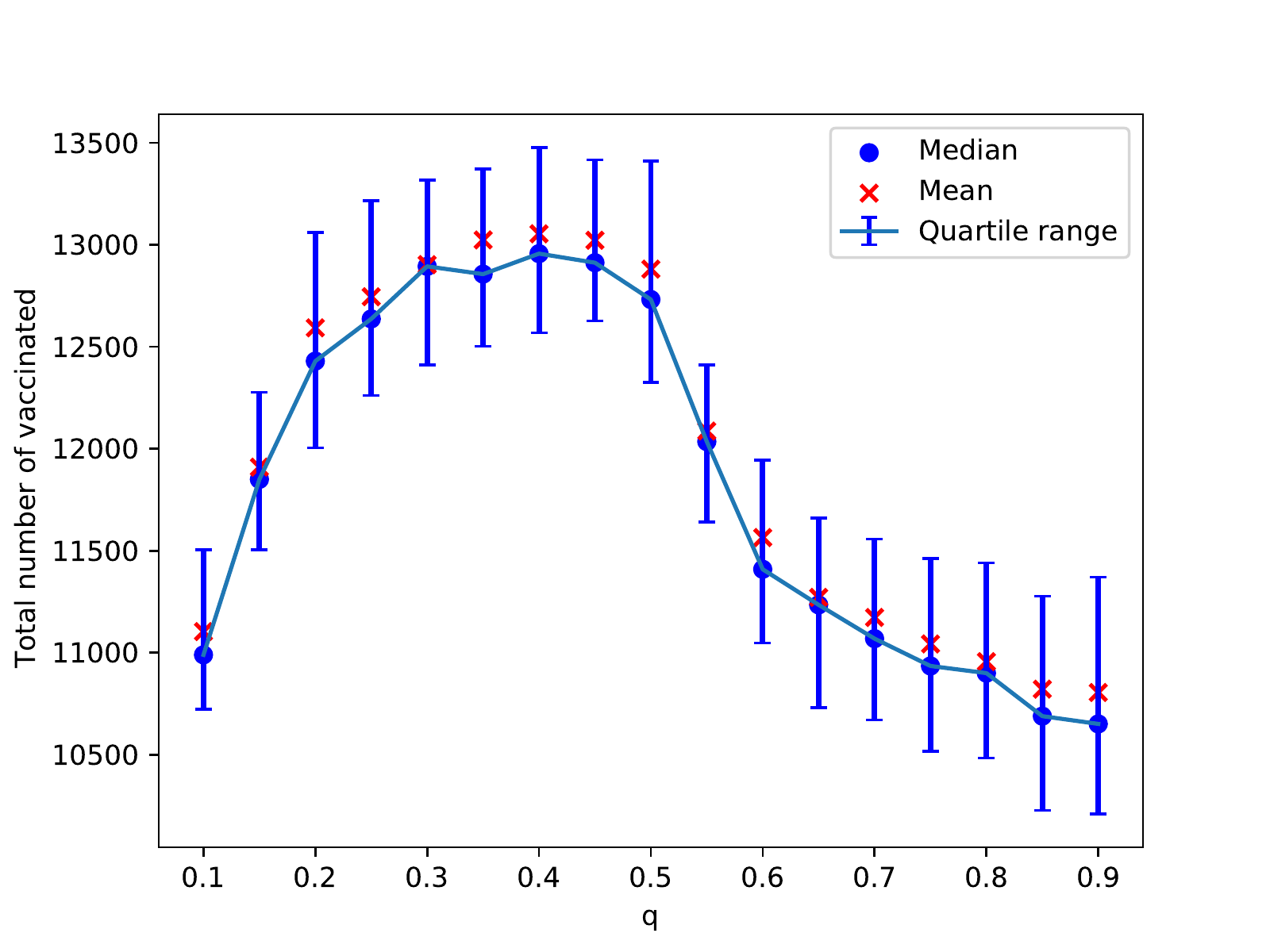}}
\subfigure[]{\includegraphics[width=5.5cm]{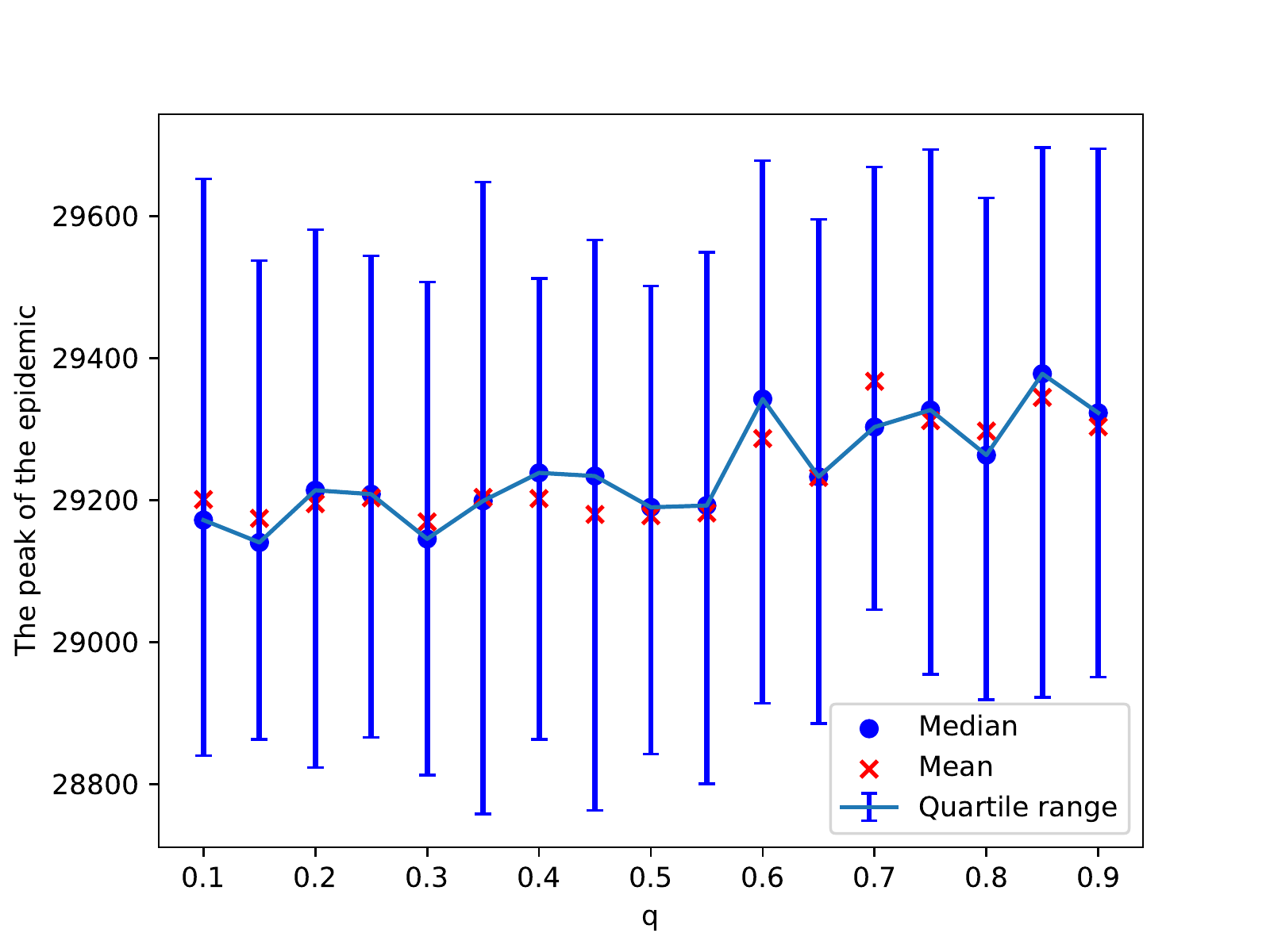}}
\caption{{\bf Fig. 3. Simulations of sizes of the epidemic, the total number of vaccinated children, and the peak of the epidemic on Erd\H{o}s-R\'{e}nyi (random) network model (ERN) for different values of $q$.} Simulations are done using $P_{adv}=.0001$ in  (a), (b), and (c), $P_{adv}=.001$ in (d), (e), and (f), and $P_{adv}=.01$ in (g), (h) and (i). In all of the simulations $\rho=.001$.}\label{fig2}
\end{figure}

In case of Barab\'{a}si-Albert network model (BAN) and using the Bayesian updating rule in equation \eqref{qeq0}, the sizes of the epidemics are smaller than those in ERNs, compare panels of Fig. 2 to Fig. 4, and panels of Fig. 3 to Fig. 5. That is against the known fact that disease spreads faster on BANs. But that might be due to the increase in vaccine uptake that we observe in BANs in contrast to ERNs. The difference between BANs and ERNs is also true for the values of $q$. When the probability of adverse event increases to $P_{adv}=.01$, vaccine uptake levels dropped from those levels when $P_{adv}=.0001$ and $P_{adv}=.001$, compare panel (h) of Fig. 4 with panels (b) and (e). However, in that case, the learning probability causes a pressure to increase vaccine uptake even if it causes more adverse events; compare panel Fig. 4 (h) to panel Fig. 2 (h).  

\begin{figure}[H]
\centering
\subfigure[]{\includegraphics[width=5.5cm]{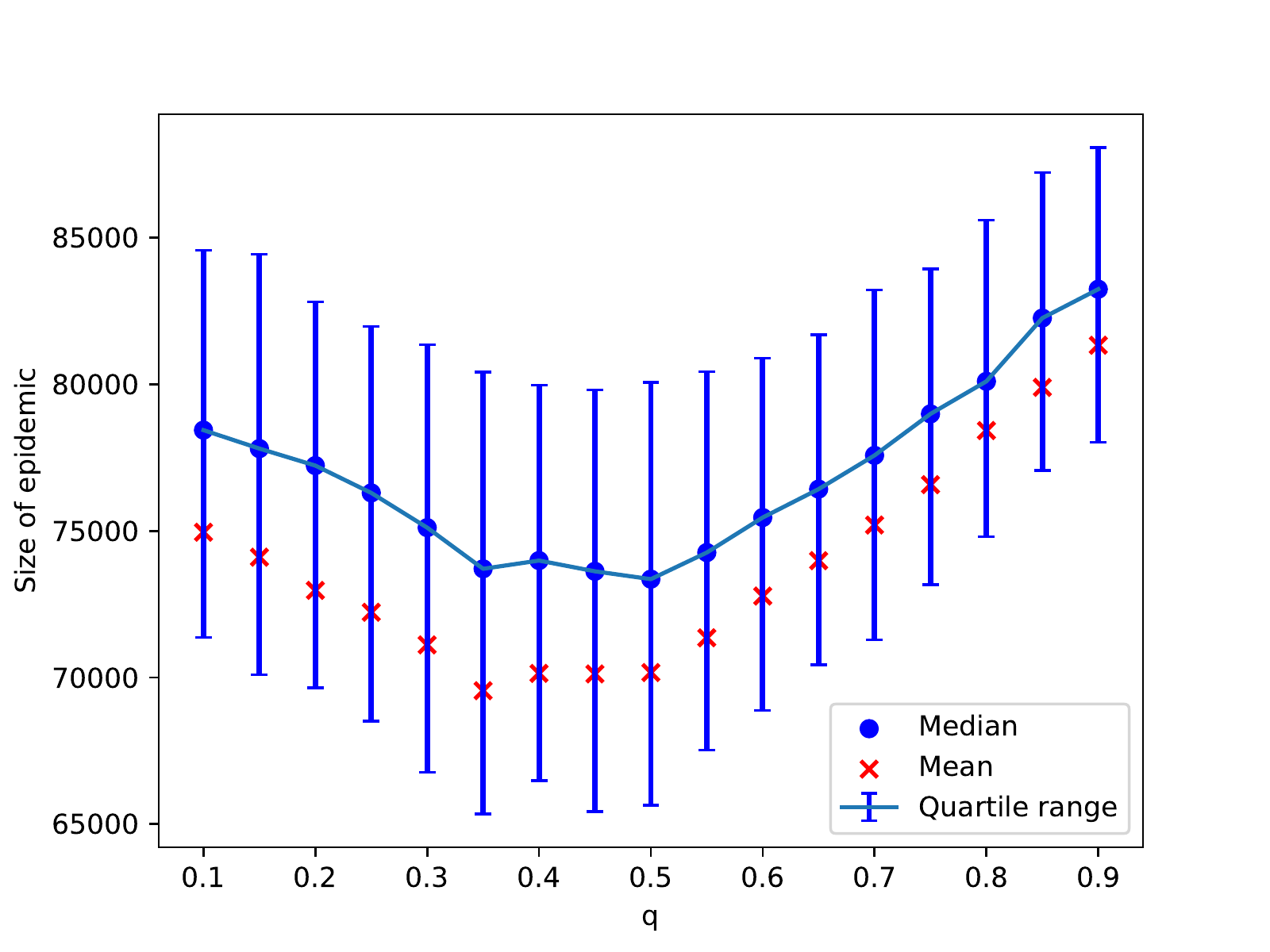}} 
\subfigure[]{\includegraphics[width=5.5cm]{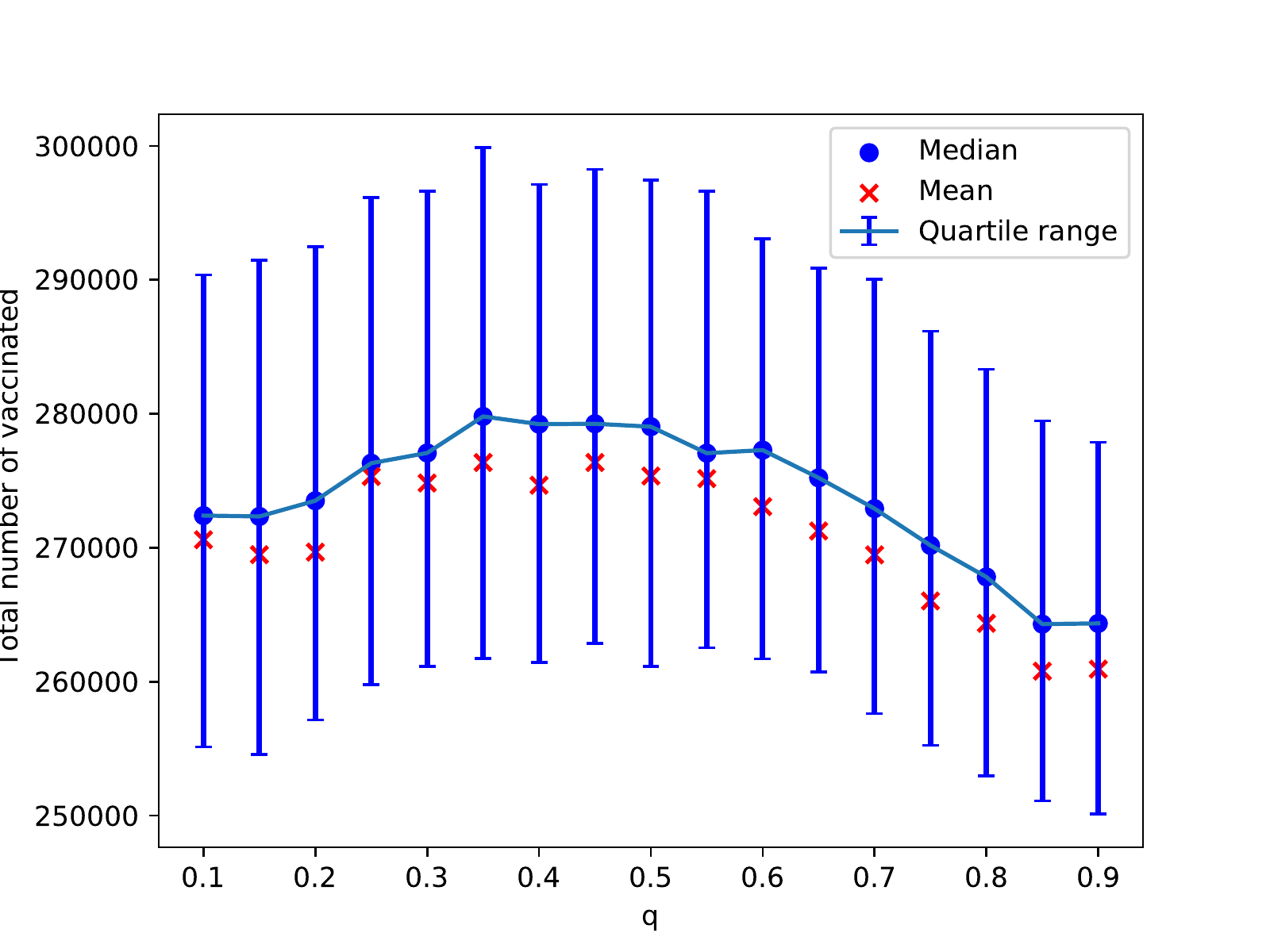}}
\subfigure[]{\includegraphics[width=5.5cm]{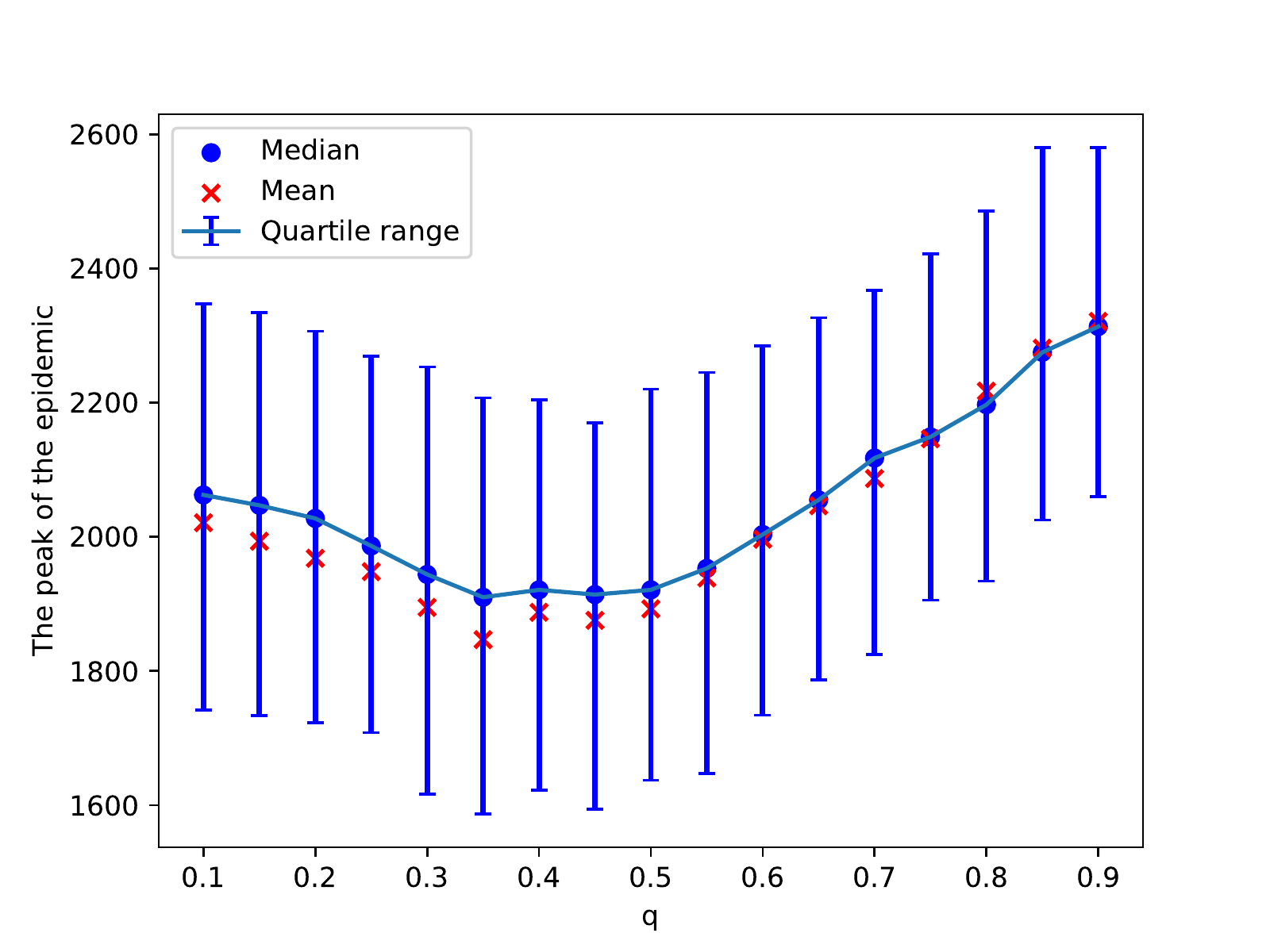}}
\subfigure[]{\includegraphics[width=5.5cm]{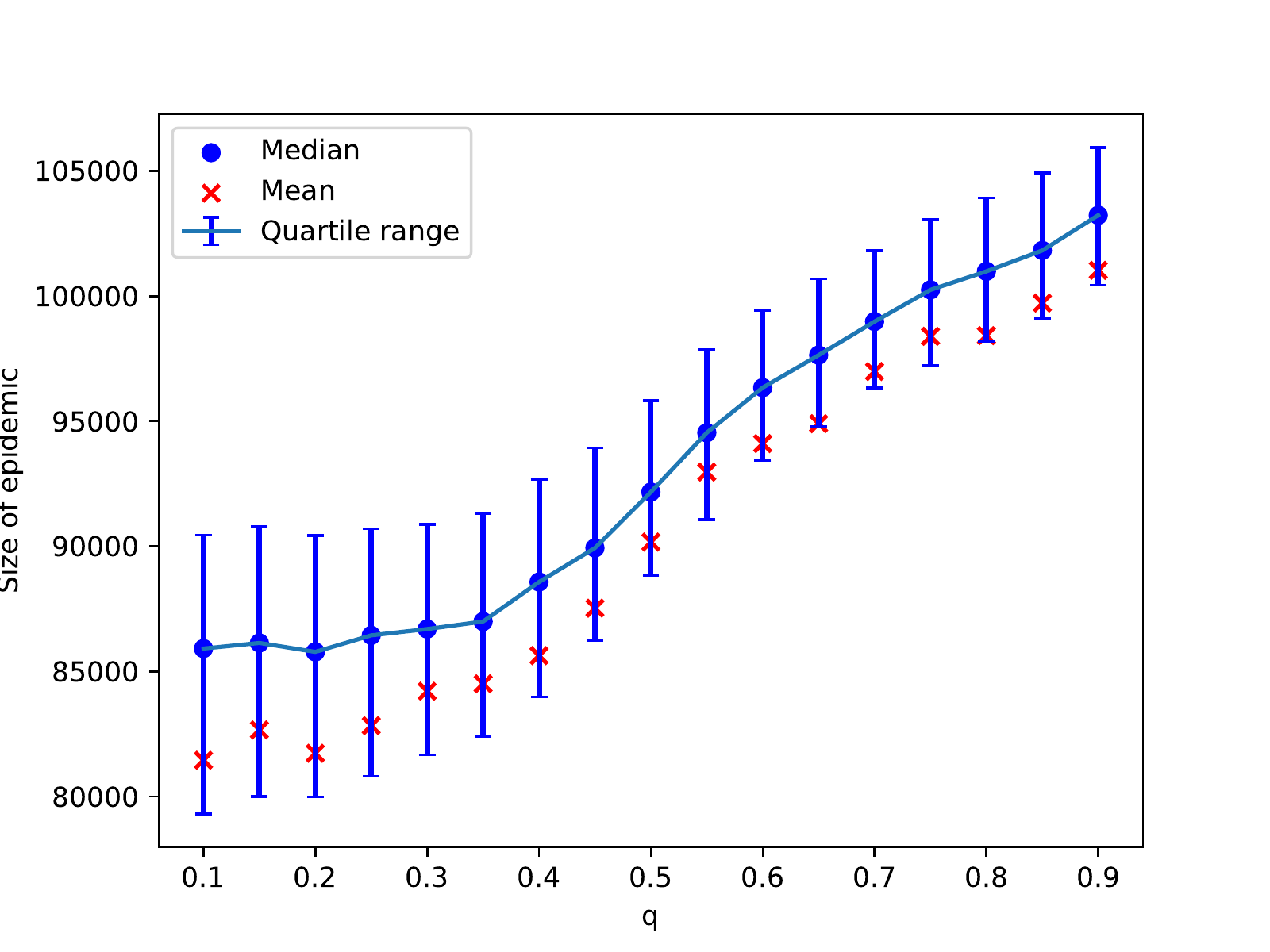}} 
\subfigure[]{\includegraphics[width=5.5cm]{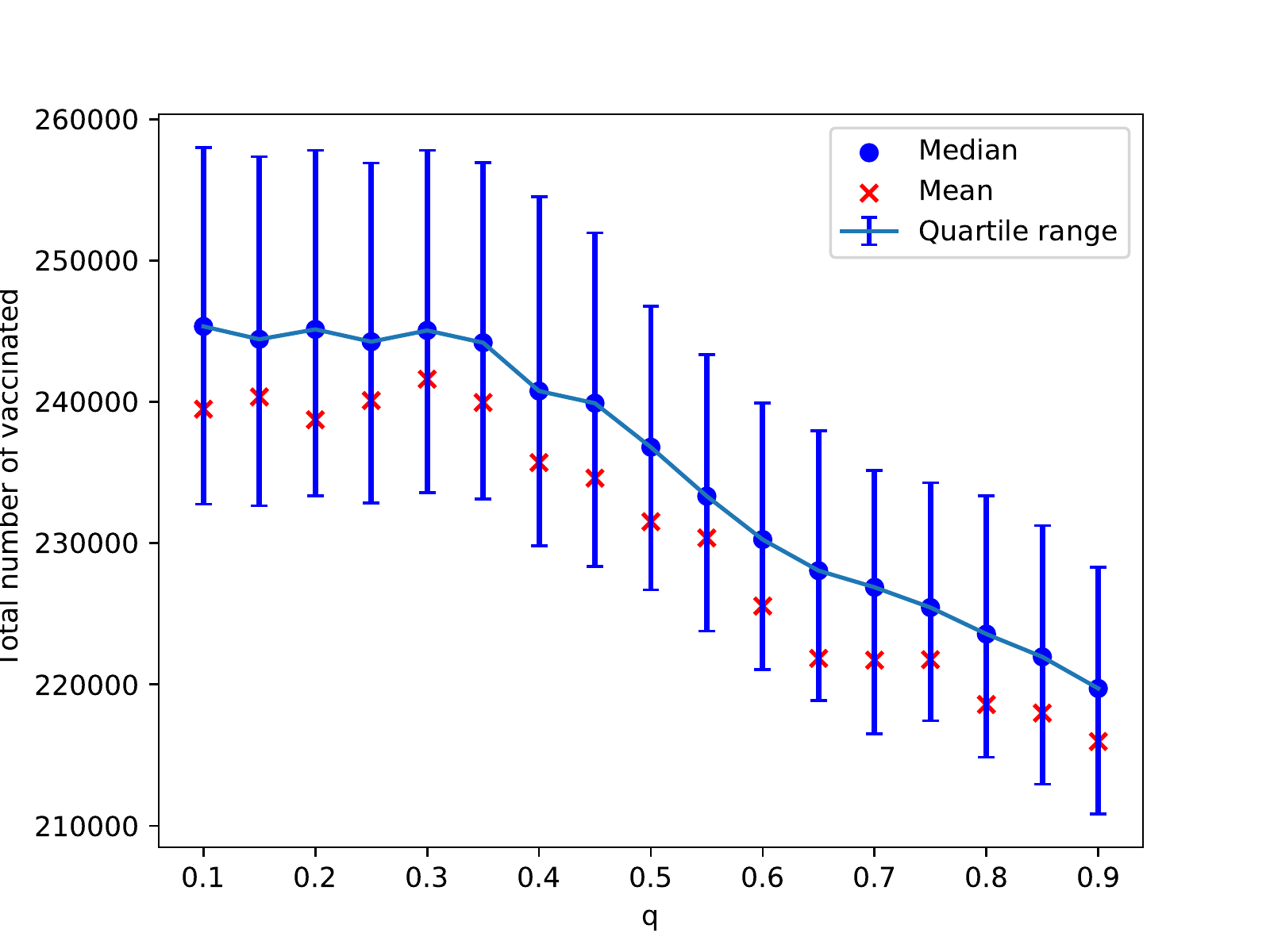}}
\subfigure[]{\includegraphics[width=5.5cm]{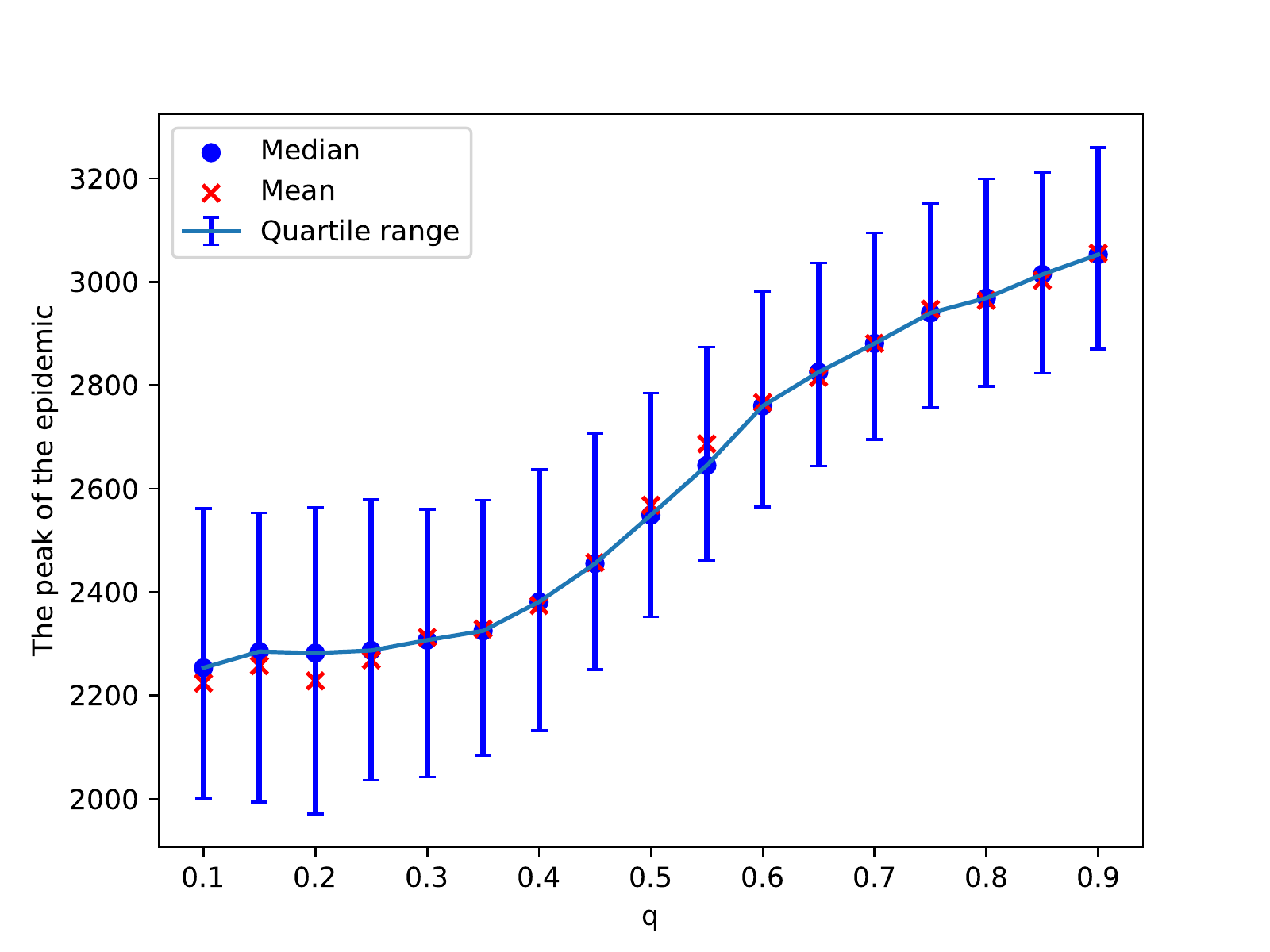}}
\subfigure[]{\includegraphics[width=5.5cm]{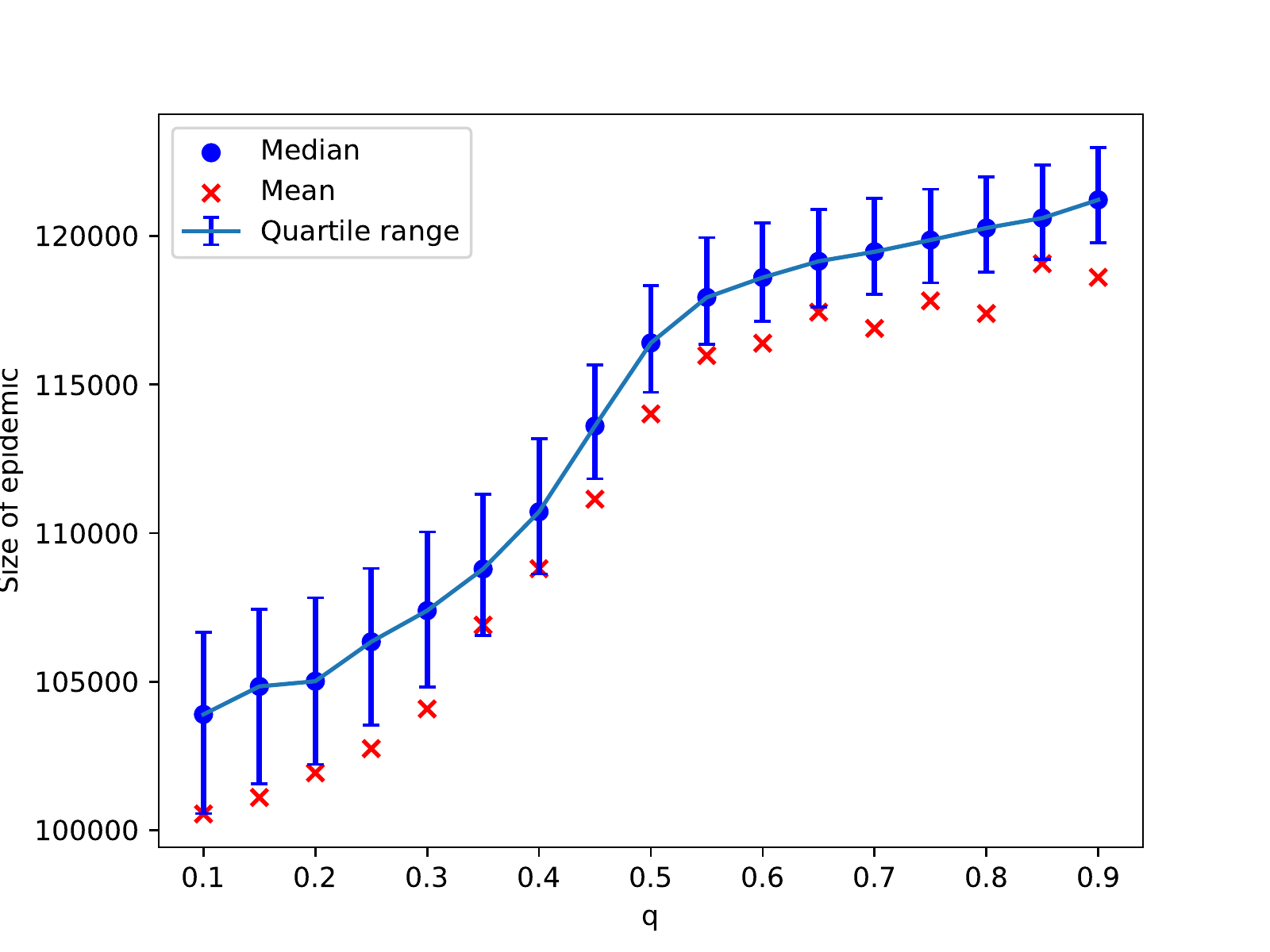}} 
\subfigure[]{\includegraphics[width=5.5cm]{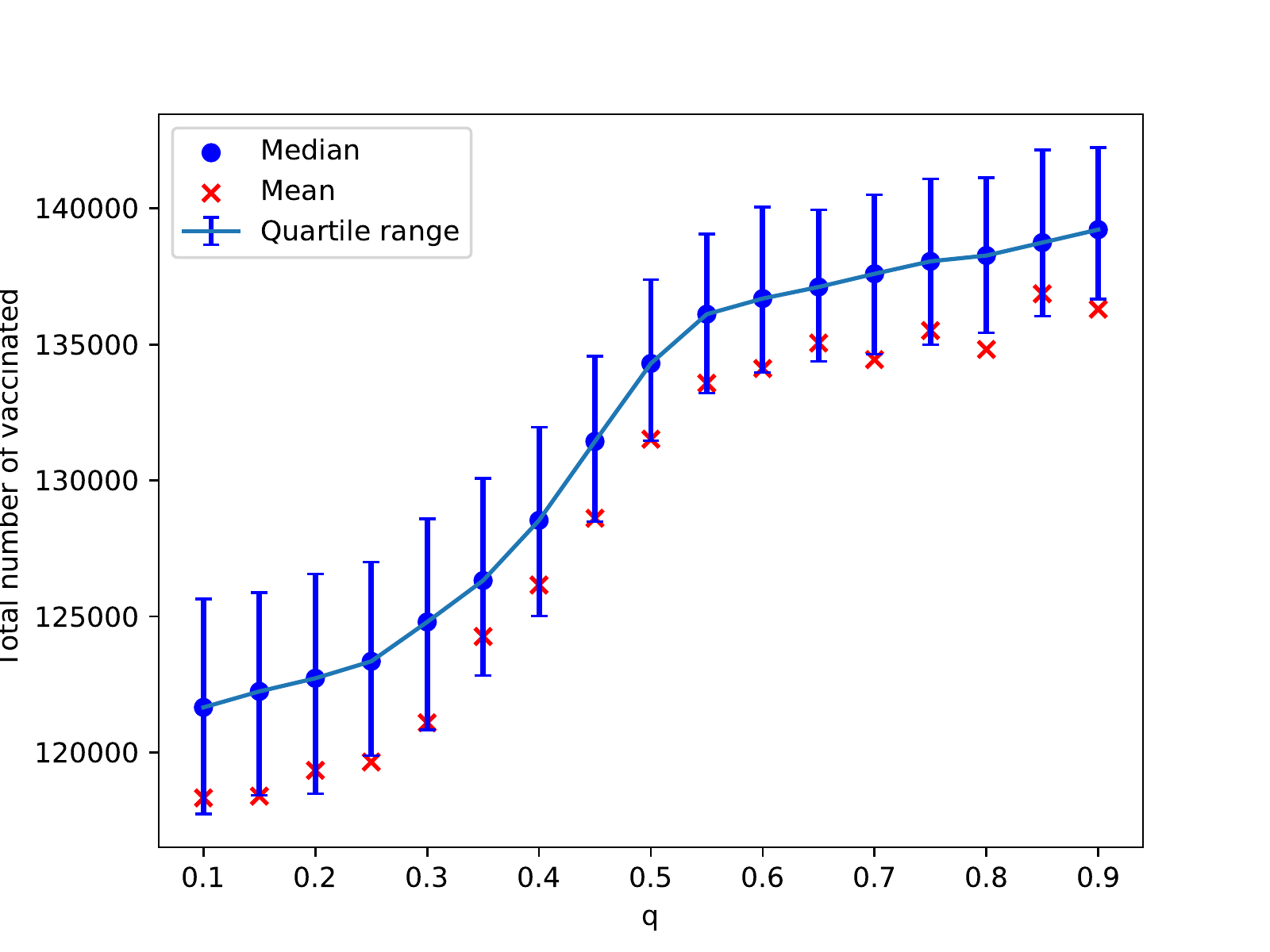}}
\subfigure[]{\includegraphics[width=5.5cm]{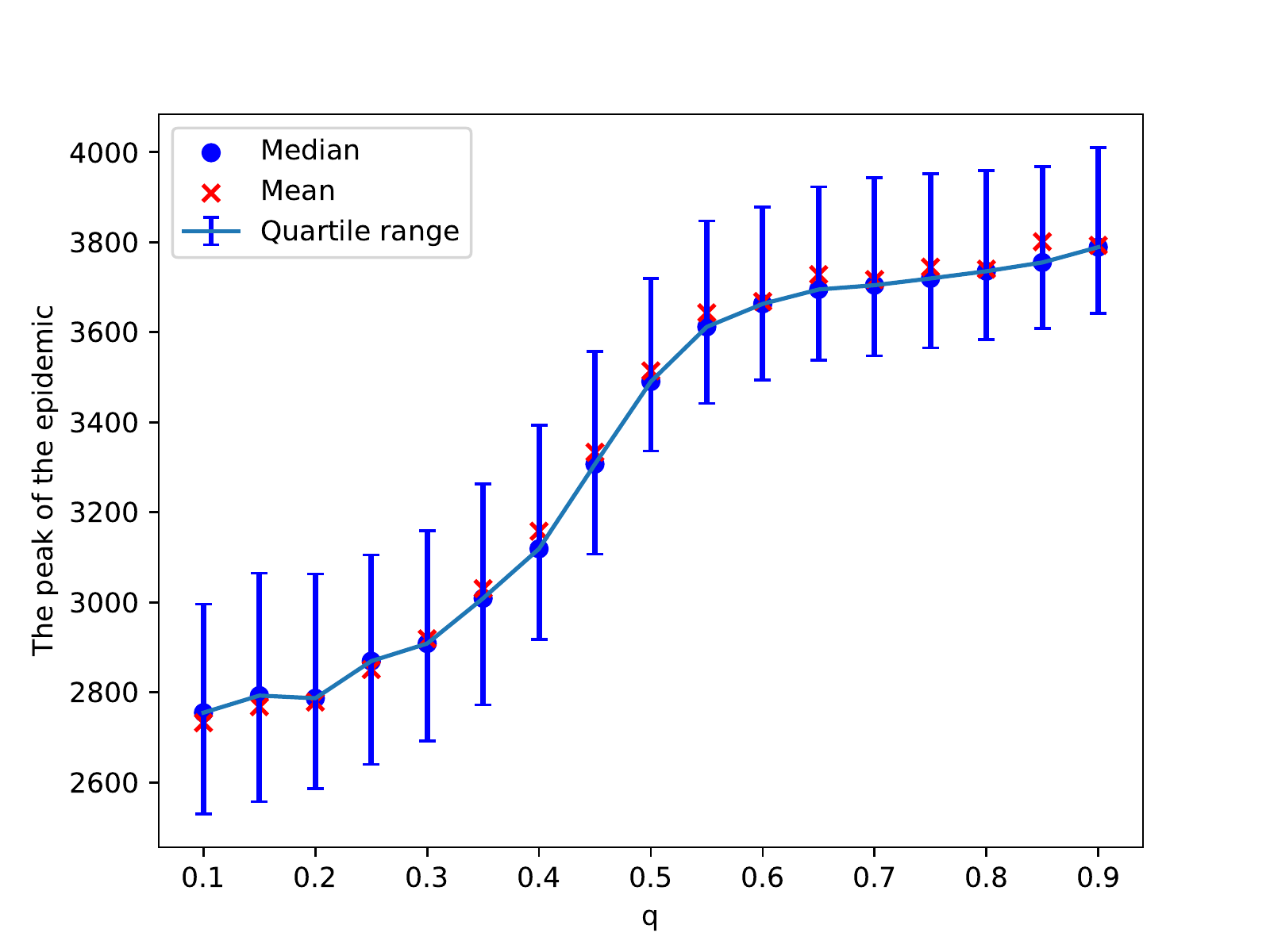}} 
\caption{{\bf Fig. 4. Simulations of sizes of the epidemic, the total number of vaccinated children, and the peak of the epidemic on Barab\'{a}si-Albert network model (BAN) for different values of $q$.} Simulations are done using $P_{adv}=.0001$ in  (a), (b), and (c), $P_{adv}=.001$ in (d), (e), and (f), and $P_{adv}=.01$ in (g), (h) and (i). In all of the simulations $\rho=.01$. }\label{fig3}
\end{figure}

Again, those patterns disappear when we assume that vaccines are scarce, and they are only available for one child in every 1000 children every day. Like in the ERNs, the outcome uncertainty increases, compare panels of Fig. 4 to Fig. 5. But in contrast to the ERNs, the learning probability does not have a significant influence over the vaccine uptake nor the size and peak of epidemics; compare panels of Fig. 3 to Fig. 5.

\begin{figure}[H]
\centering
\subfigure[]{\includegraphics[width=5.5cm]{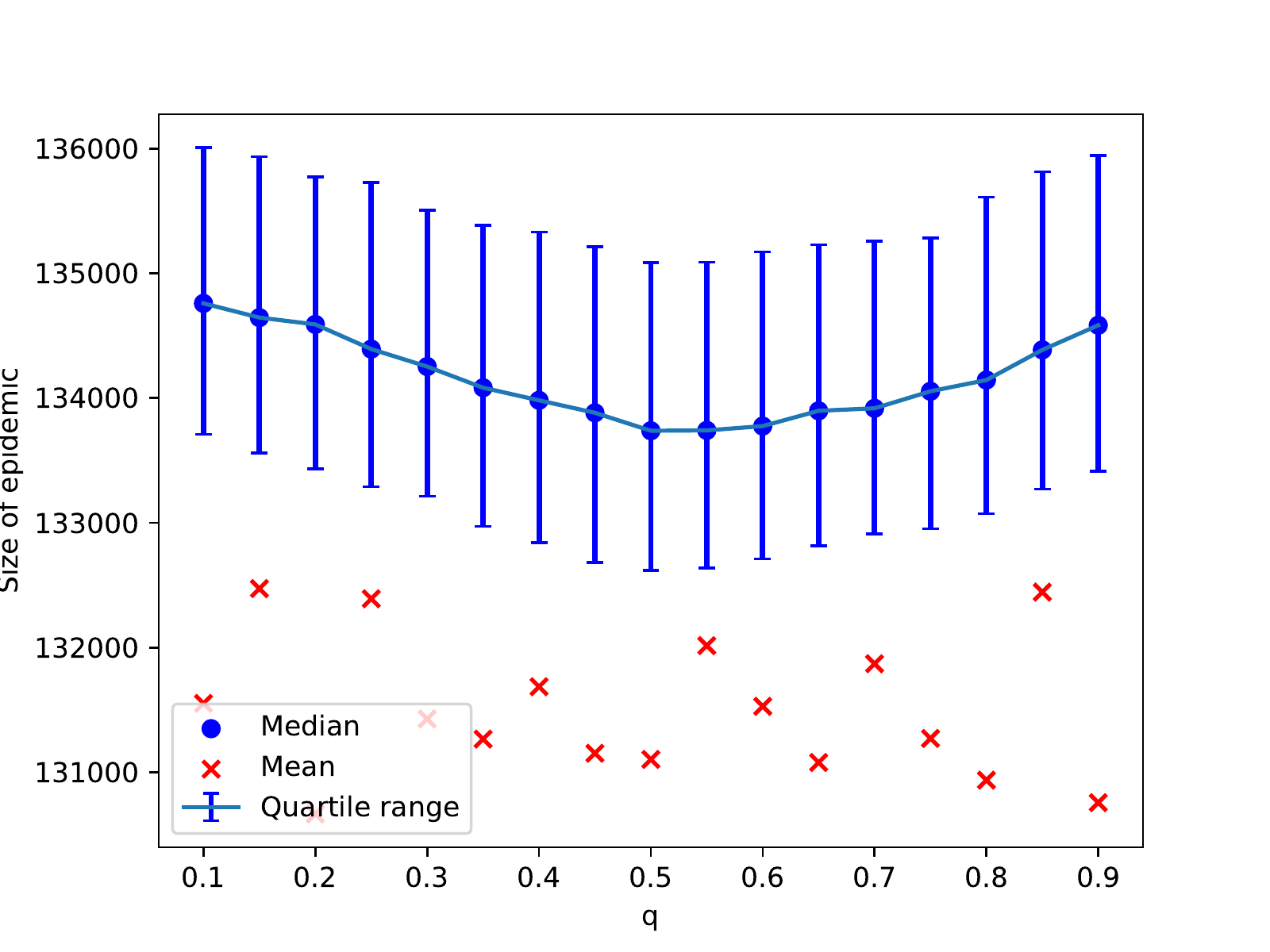}} 
\subfigure[]{\includegraphics[width=5.5cm]{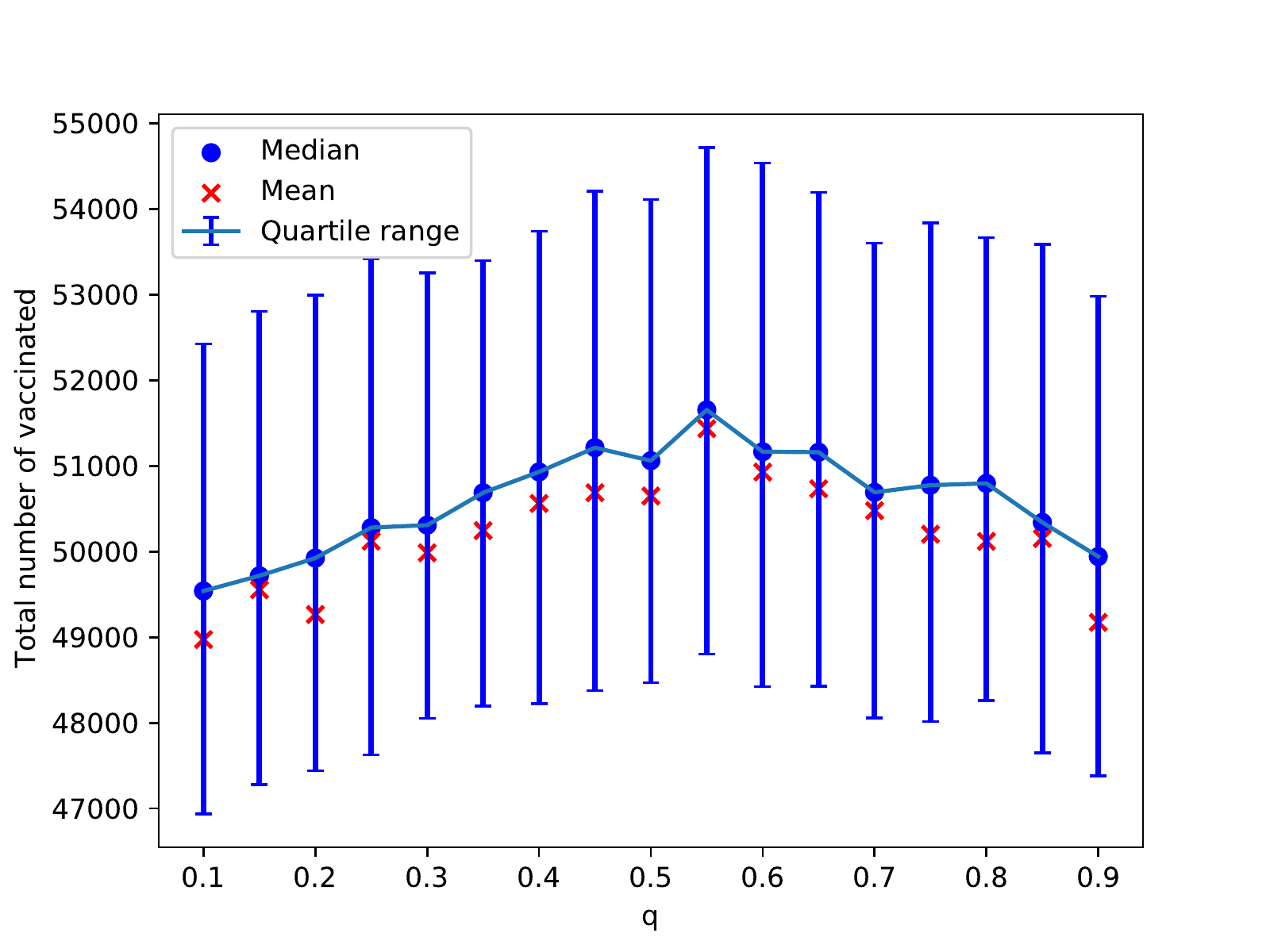}}
\subfigure[]{\includegraphics[width=5.5cm]{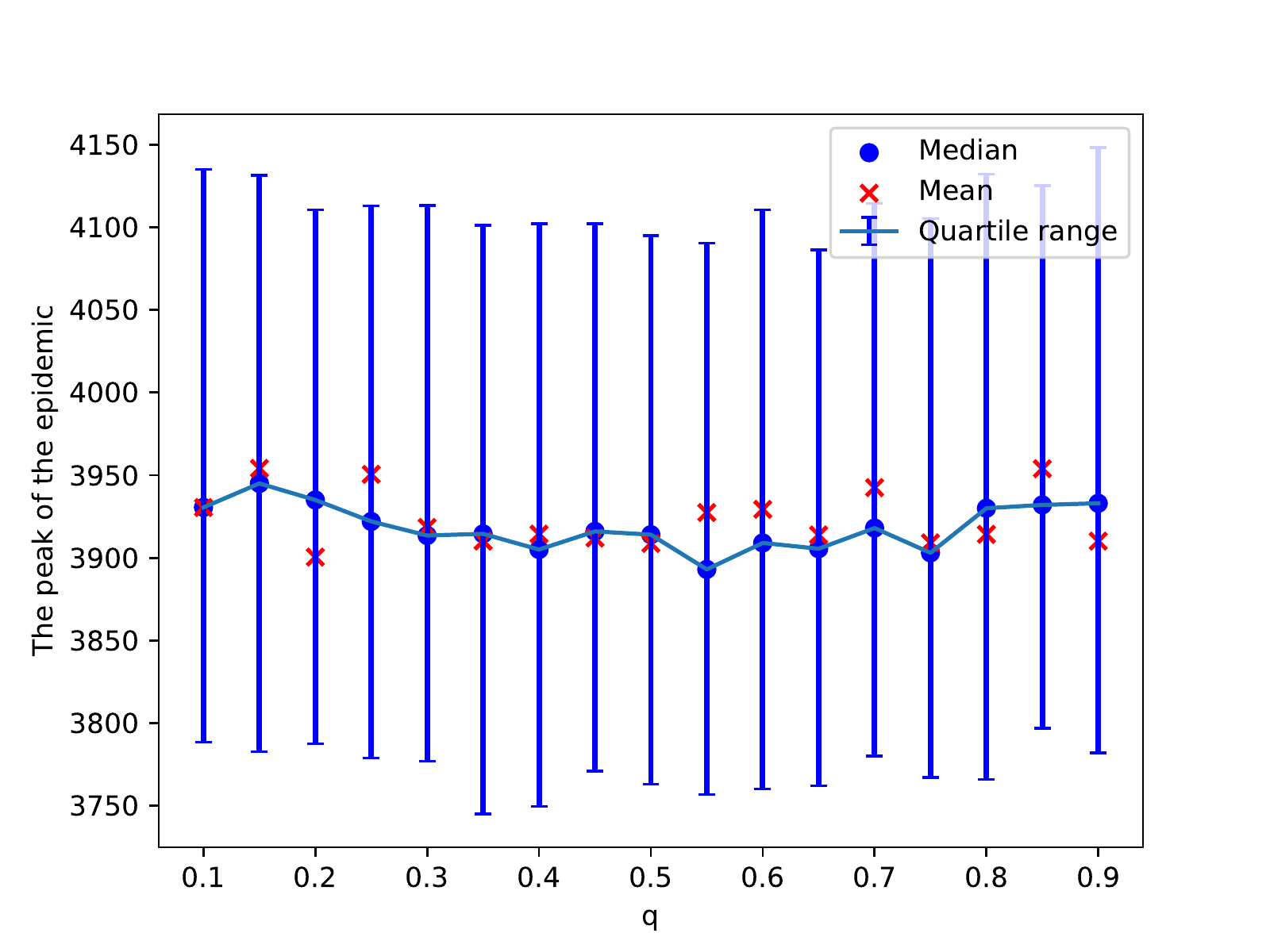}}
\subfigure[]{\includegraphics[width=5.5cm]{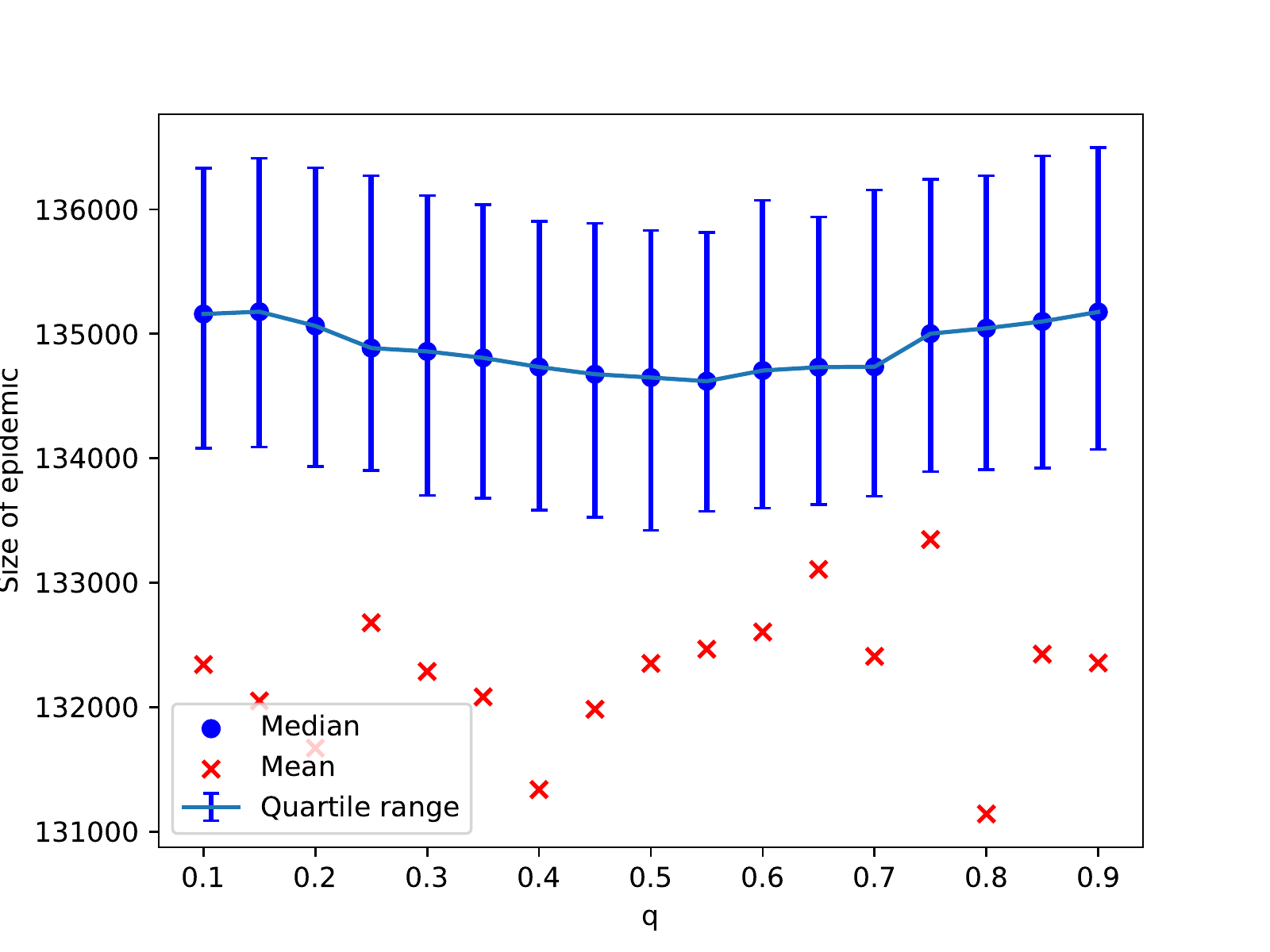}} 
\subfigure[]{\includegraphics[width=5.5cm]{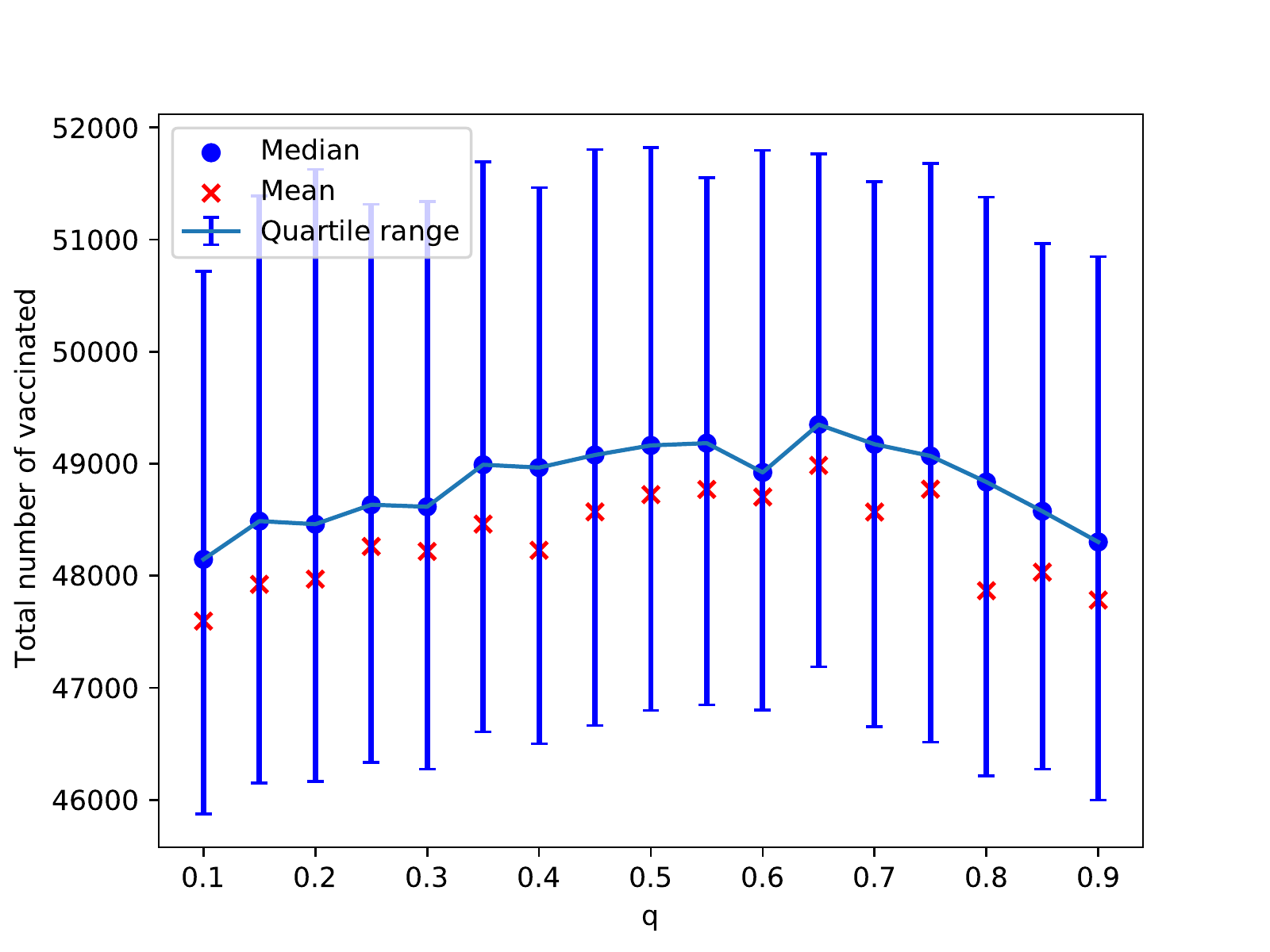}}
\subfigure[]{\includegraphics[width=5.5cm]{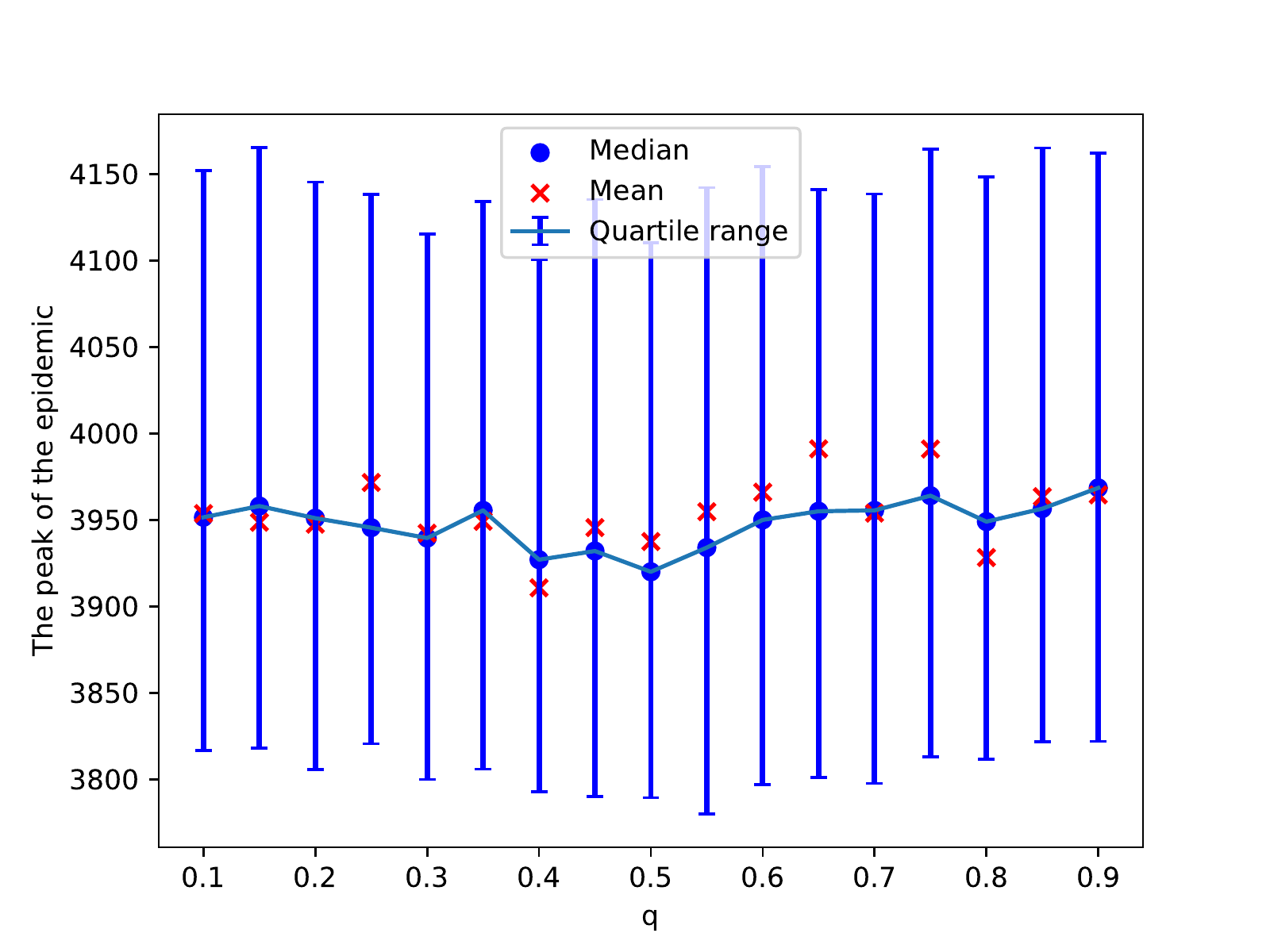}}
\subfigure[]{\includegraphics[width=5.5cm]{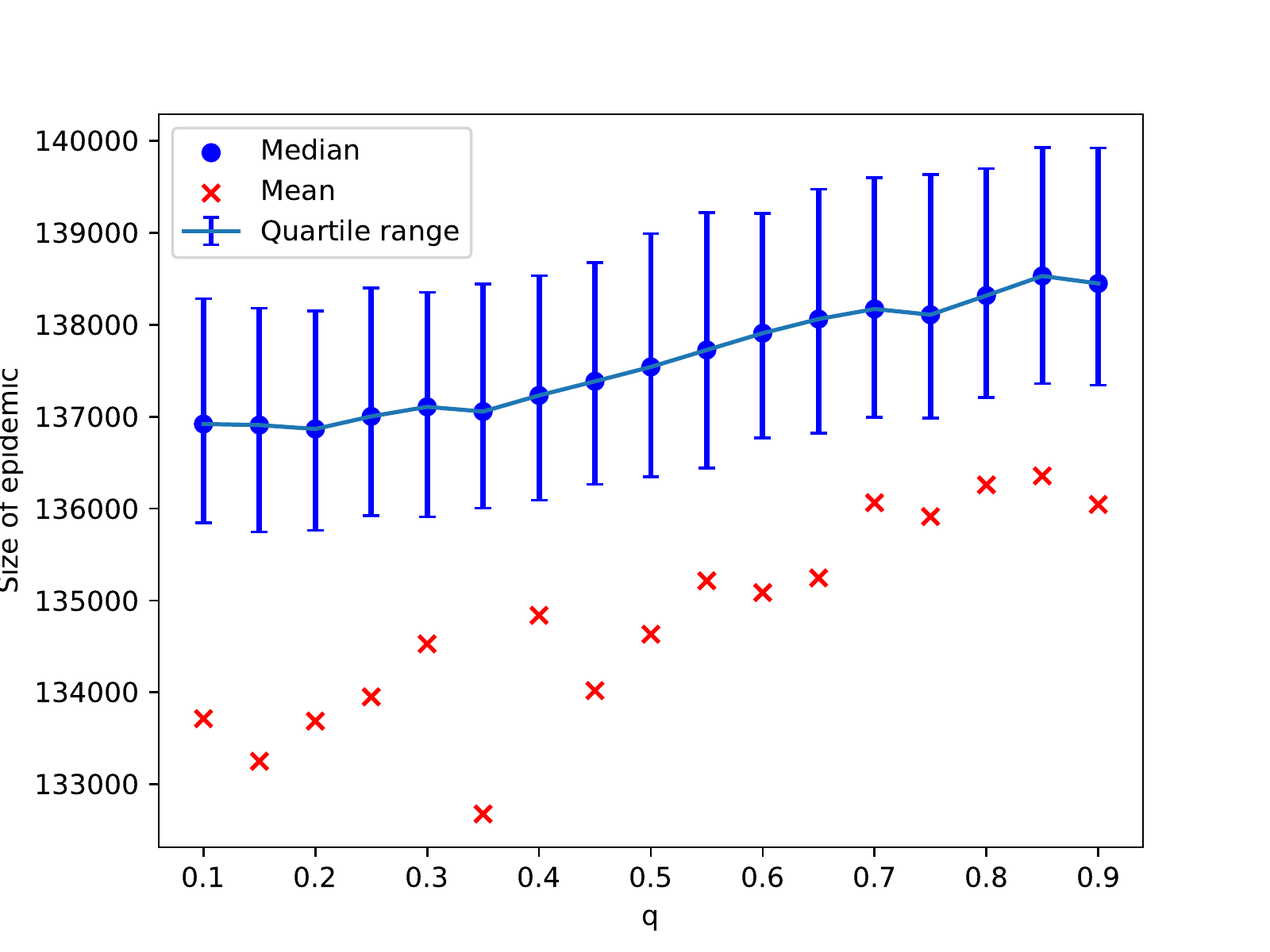}} 
\subfigure[]{\includegraphics[width=5.5cm]{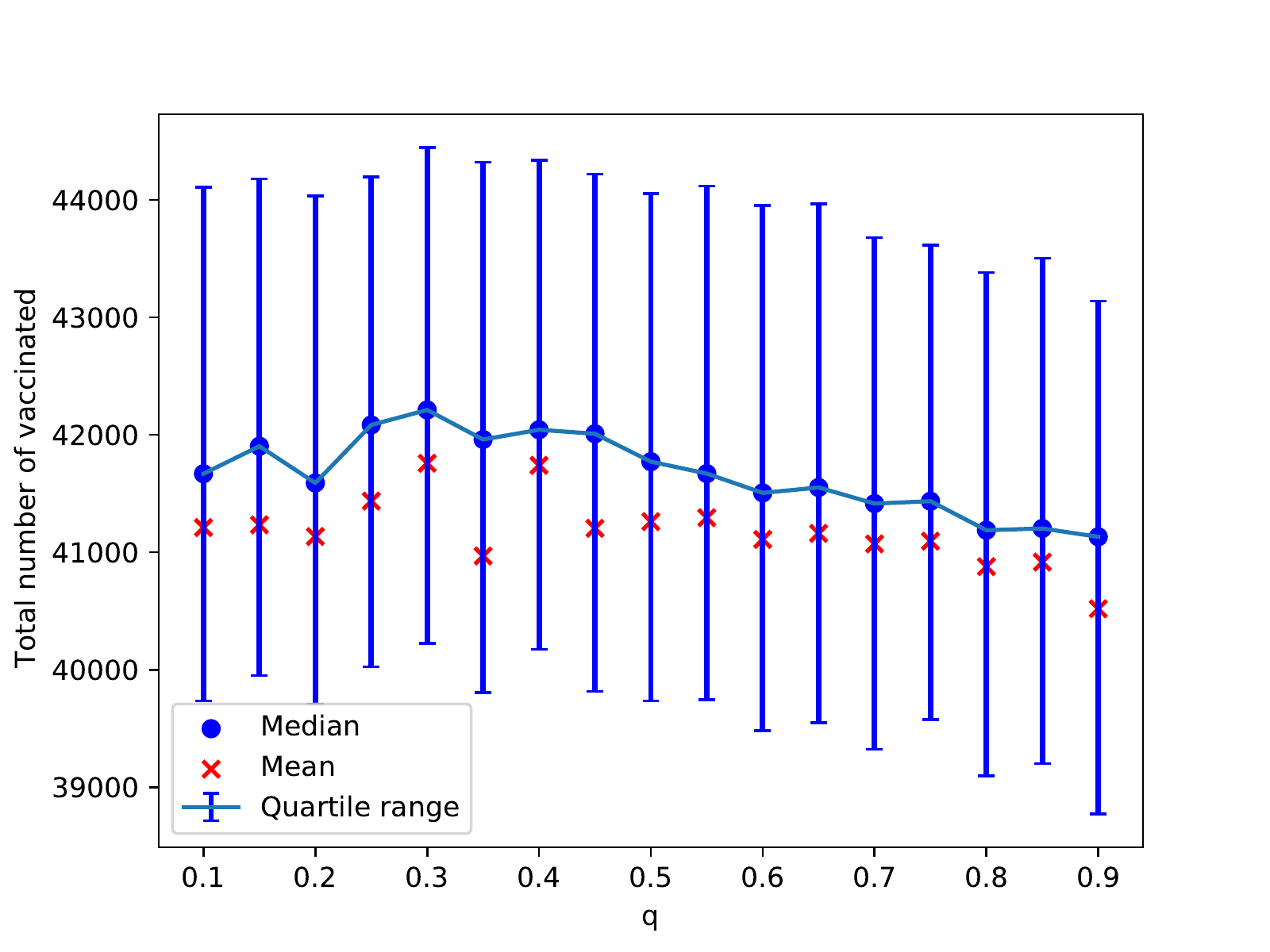}}
\subfigure[]{\includegraphics[width=5.5cm]{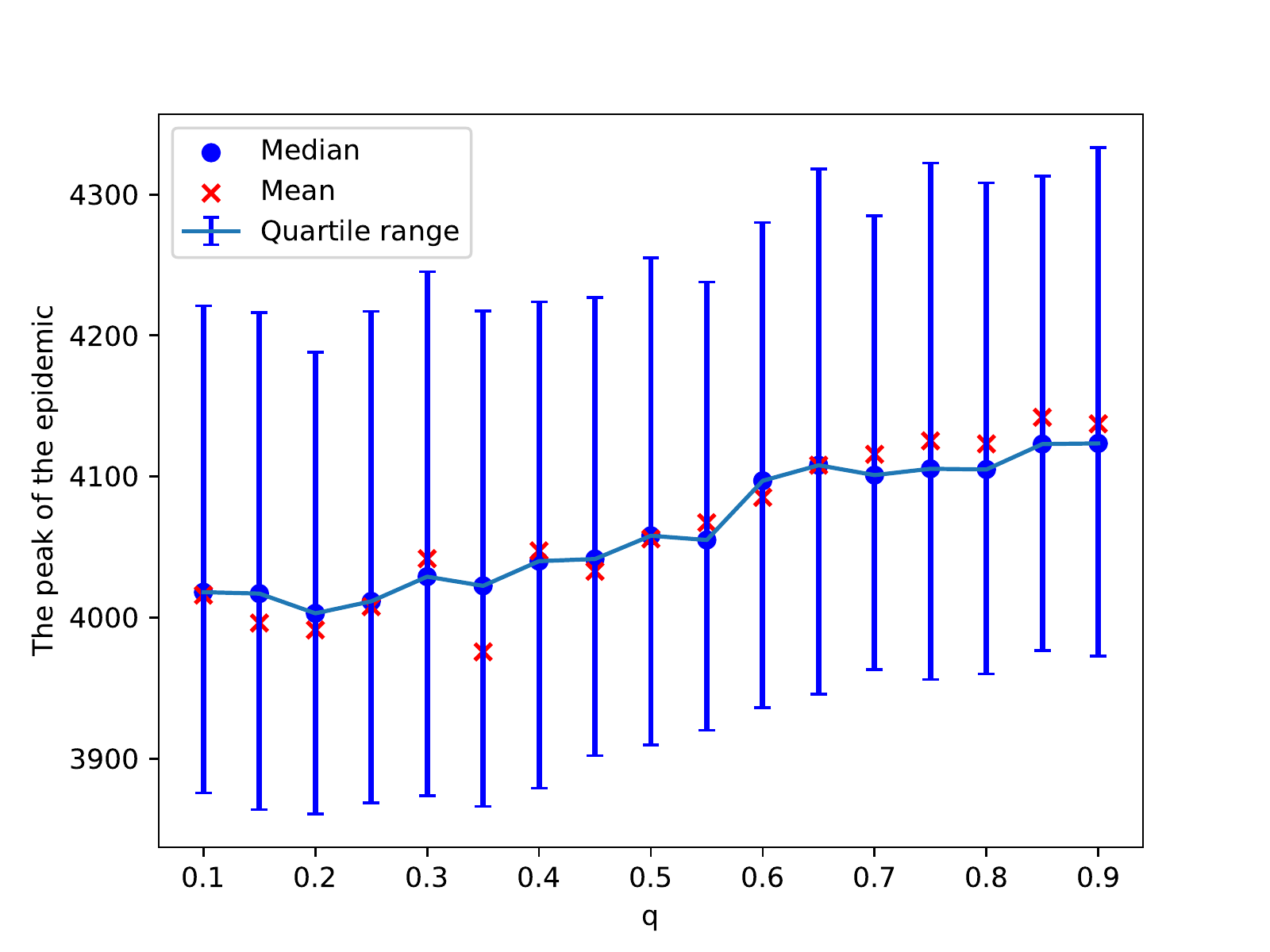}} 
\caption{{\bf Fig. 5. Simulations of sizes of the epidemic, the total number of vaccinated children, and the peak of the epidemic on Barab\'{a}si-Albert network model (BAN) for different values of $q$.} Simulations are done using $P_{adv}=.0001$ in  (a), (b), and (c), $P_{adv}=.001$ in (d), (e), and (f), and $P_{adv}=.01$ in (g), (h) and (i). In all of the simulations $\rho=.001$.}\label{fig4}
\end{figure}

The parameter planes in Fig. 7 for the relevance of the disease $\alpha$ and the relevance of the adverse event of the vaccine $\gamma$ to the rational choice component and for different values of $q$ show consistent patterns with the simulations in Fig. 2. That means that for diseases and opinions spreading on ERN, the learning effect effectively suppresses the rational perception of the parent about the reward of vaccination. Parameter planes in Figs. S2 and S3 show consistent patterns as well.

\begin{figure}[H]
\subfigure[]{\includegraphics[width=5.5cm]{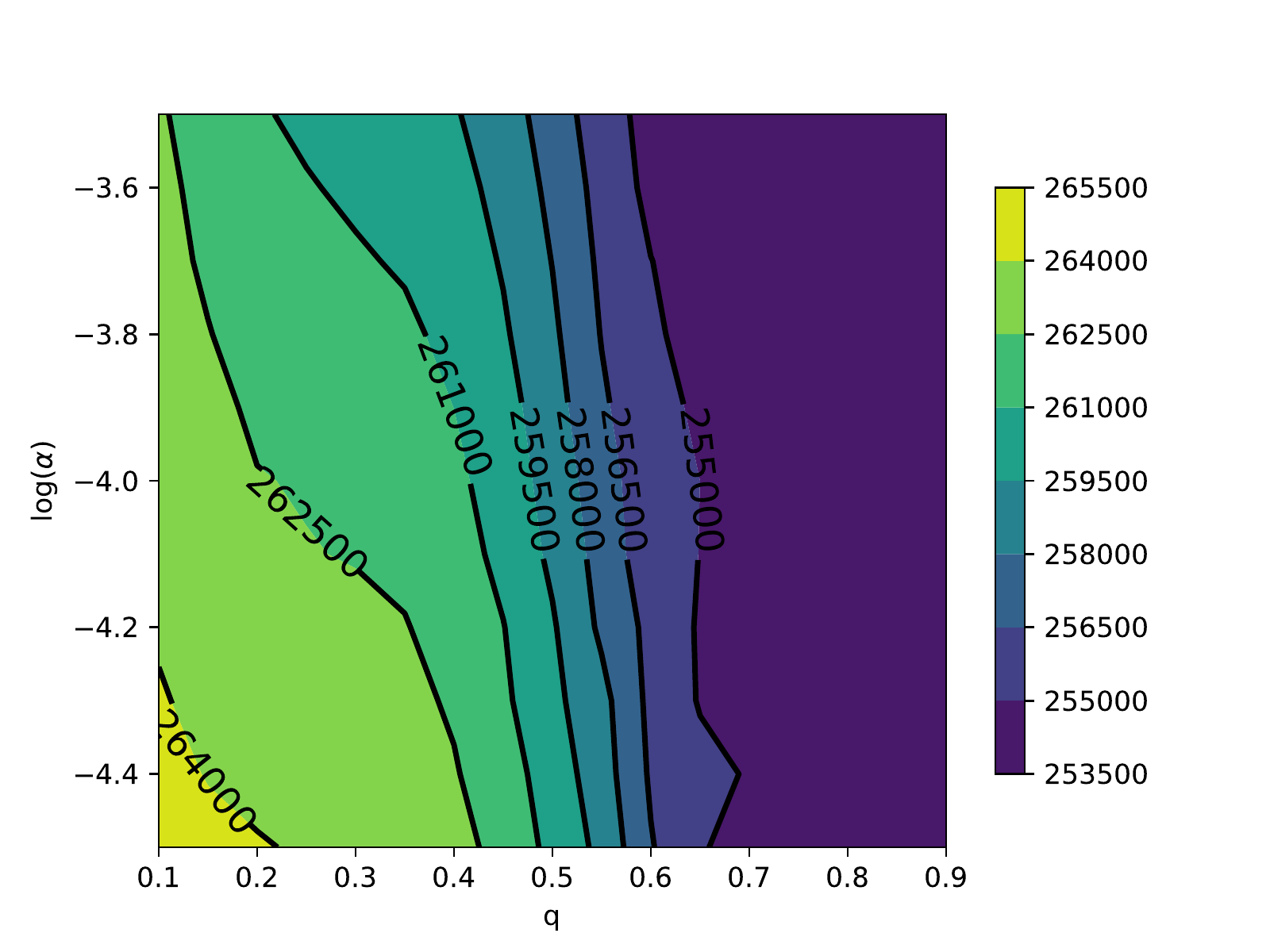}} 
\subfigure[]{\includegraphics[width=5.5cm]{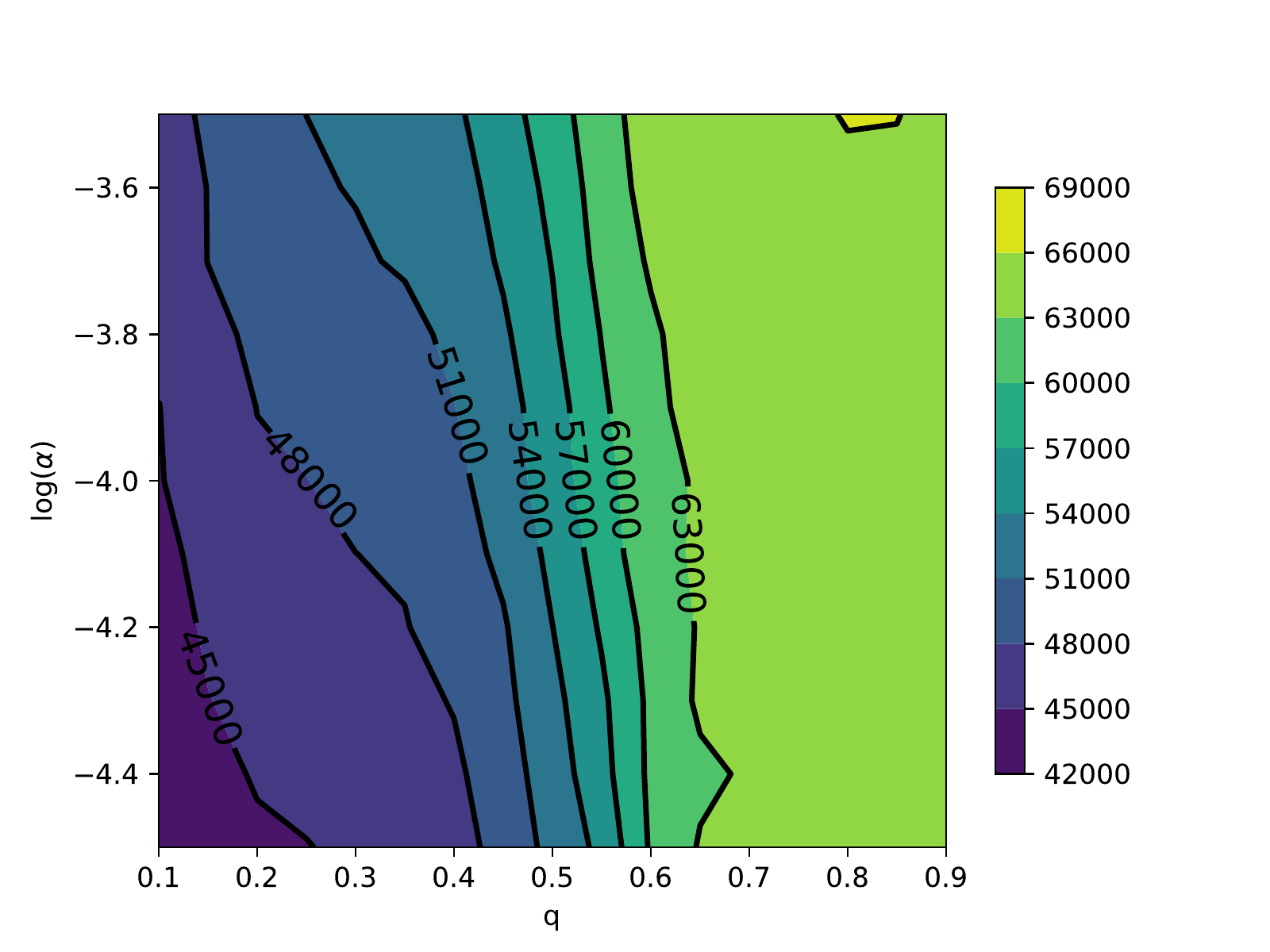}}
\subfigure[]{\includegraphics[width=5.5cm]{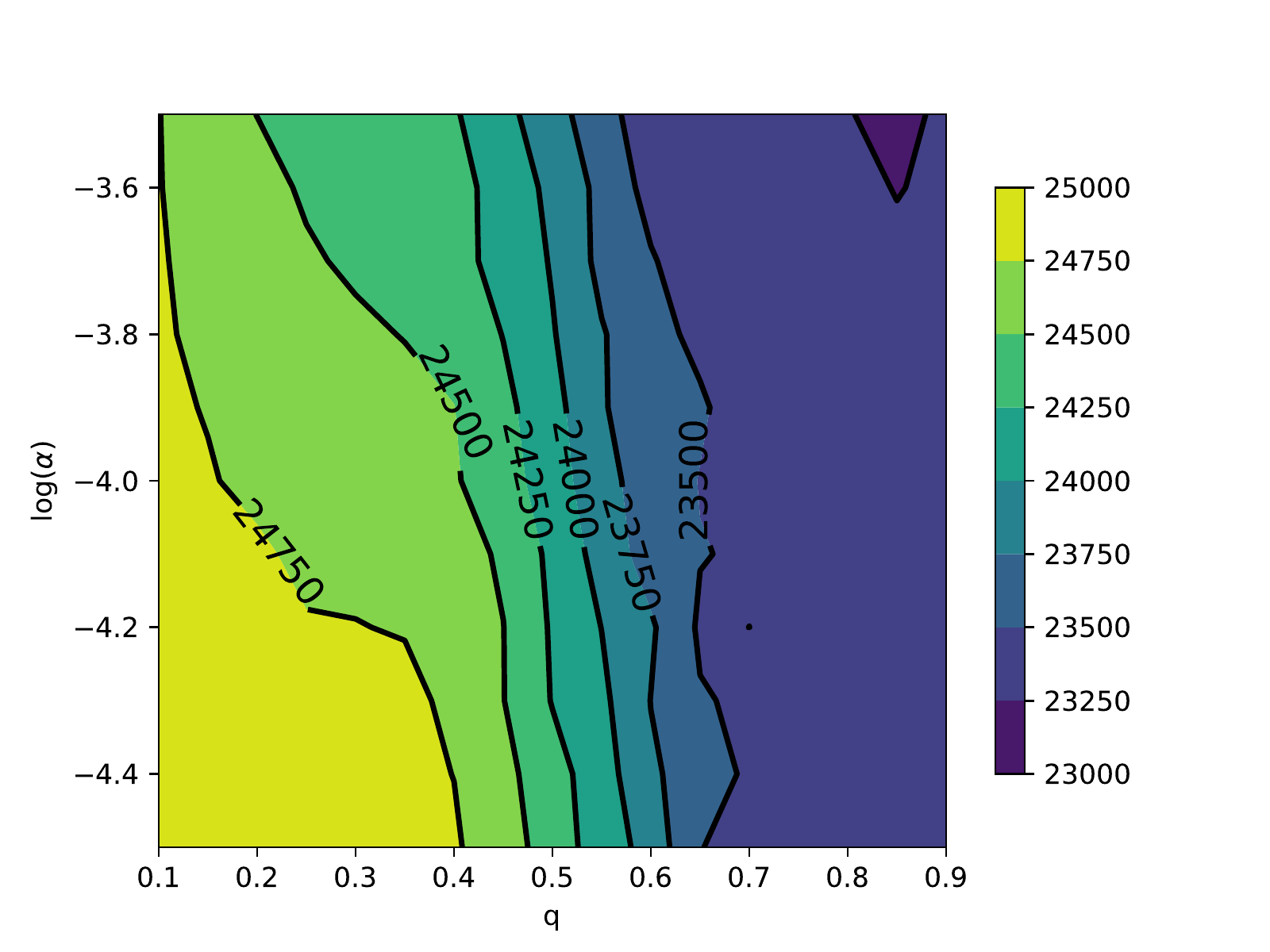}}
\subfigure[]{\includegraphics[width=5.5cm]{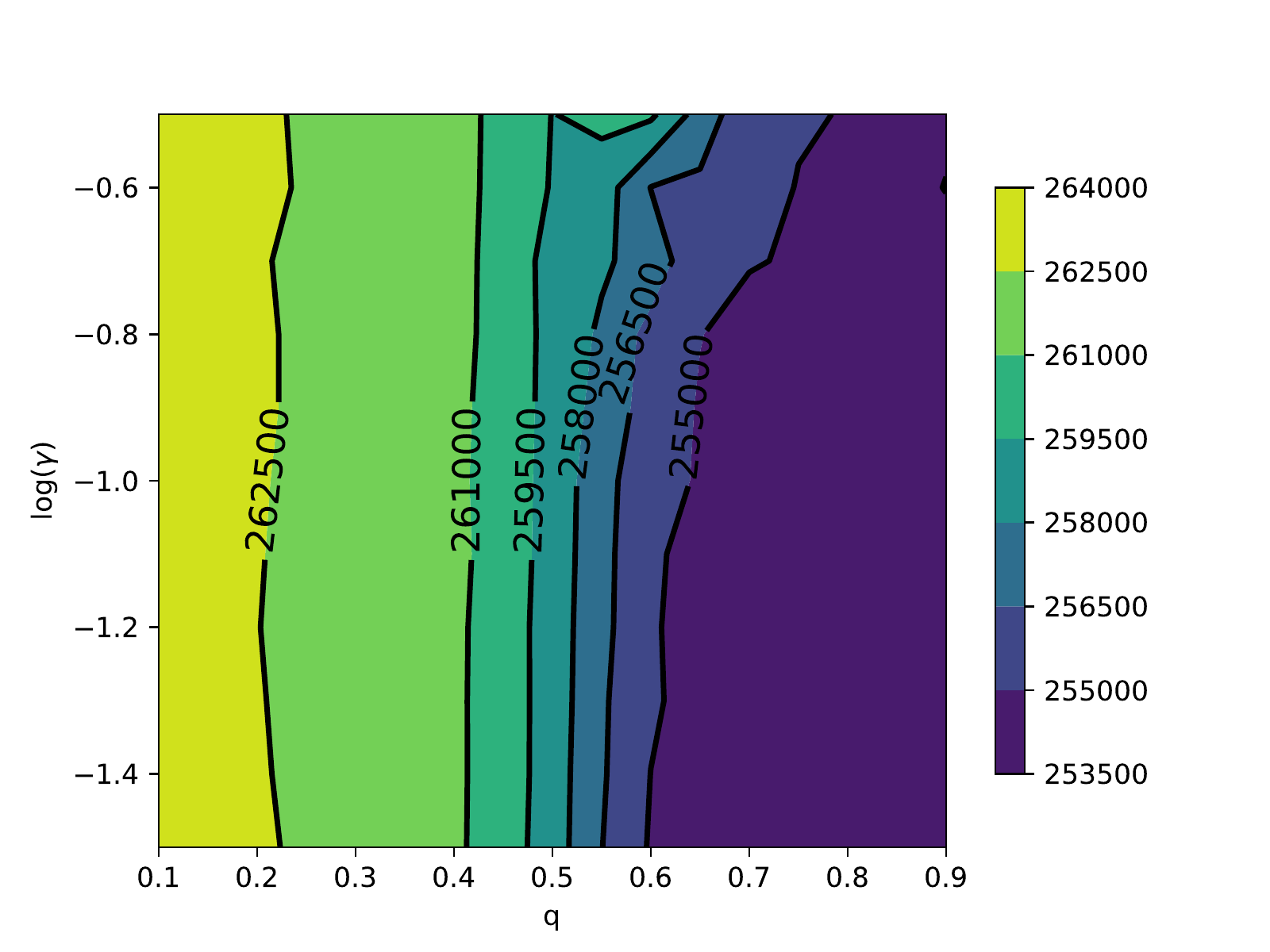}}
\subfigure[]{\includegraphics[width=5.5cm]{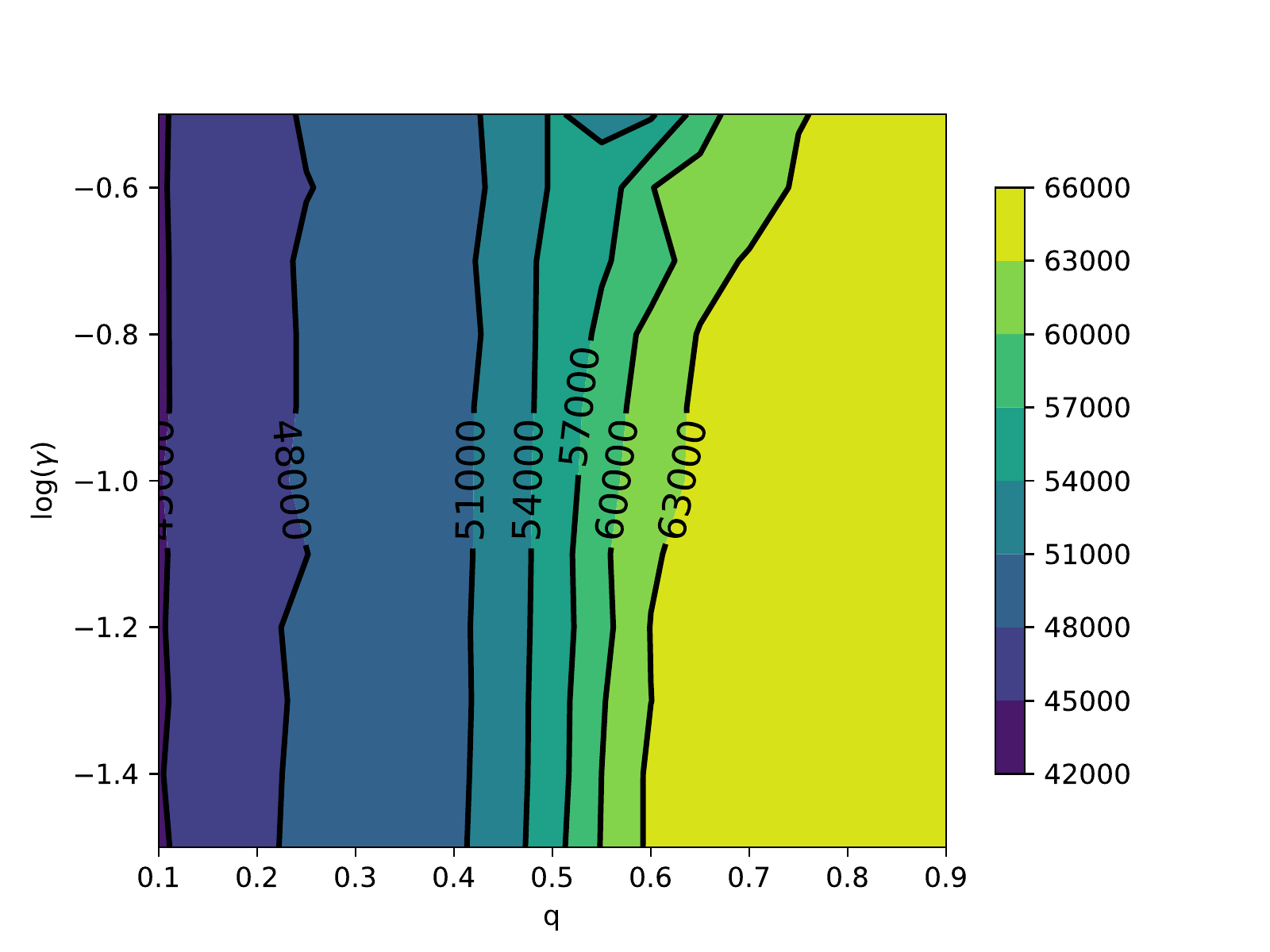}} \subfigure[]{\includegraphics[width=5.5cm]{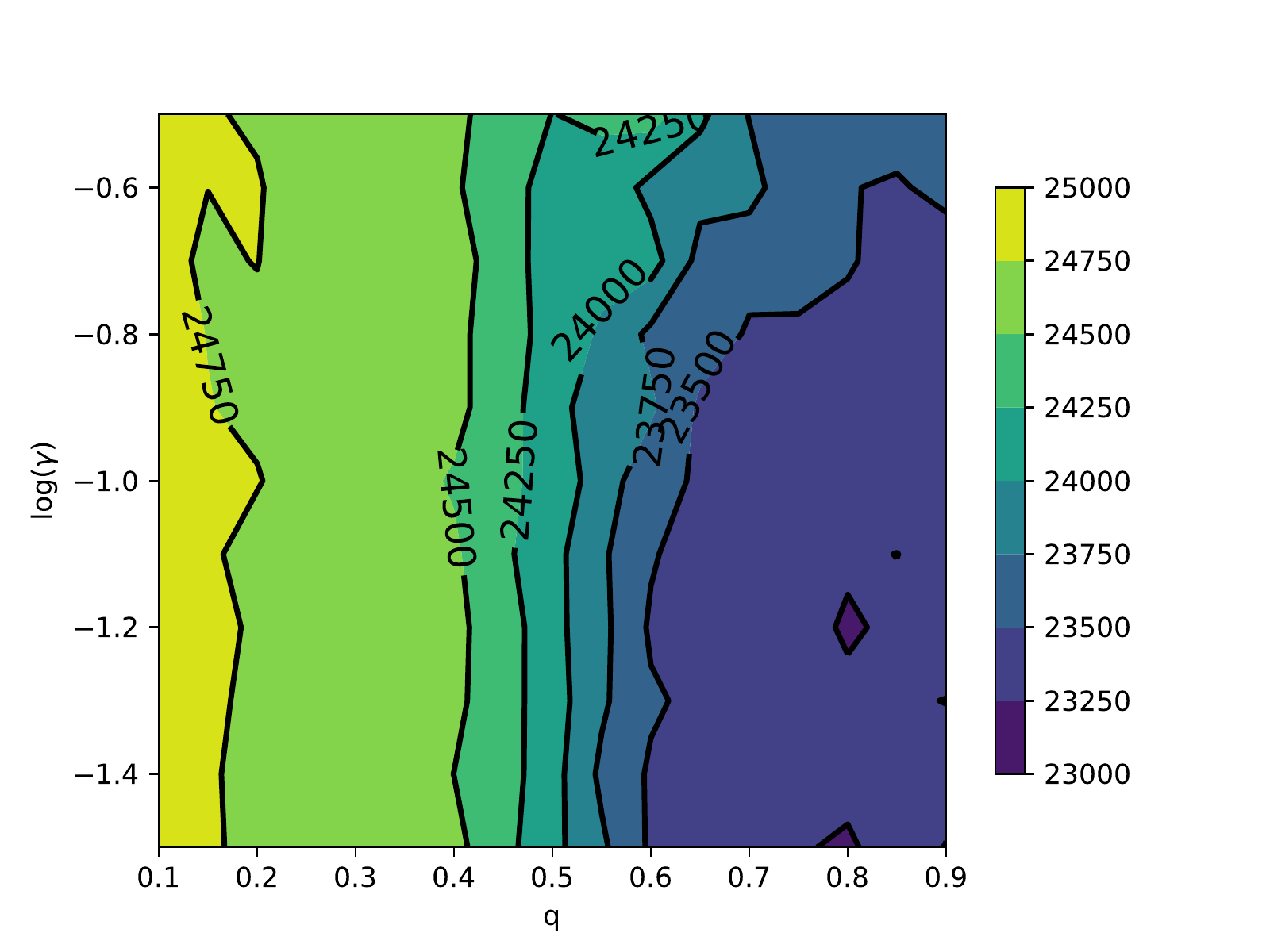}}

\caption{{\bf Fig. 6. Parameter planes of $\alpha$ against $q$ and $\gamma$ against $q$ for epidemic sizes, the total number of vaccinated children and the peak of the epidemic on Erd\H{o}s-R\'{e}nyi (random) network model (ERN).} In all simulations, the median value of the simulations is used to plot the parameter planes that are performed at $P_{adv}=.0001$ and $\rho=.01$.}\label{fig6}
\end{figure}

Similarly, the parameter planes in Fig. 8 for the relevance of the disease $\alpha$ and the relevance of the vaccine's adverse event $\gamma$ to the rational choice component and for different values of $q$ show consistent patterns with the simulations in Fig. 4. Qualitatively, for diseases and opinion spreading on BAN, vaccine uptake and epidemics can have a symmetric reaction to high and low learning probability. Quantitatively, the parents' rational perception of the payoff of vaccination show influence over vaccine uptake and epidemics. As perceived disease risk increases, vaccine uptake increases, and epidemic sizes and peaks decrease, Fig. 8 (a), (b), and (c). And as the perceived risk of adverse events from vaccine increases, vaccine uptake decreases, and epidemic sizes and peaks increase, Fig. 8 (d), (e), and (f). The parameter planes in Figs. S4 and S5 also show consistent patterns.

\begin{figure}[H]
\subfigure[]{\includegraphics[width=5.5cm]{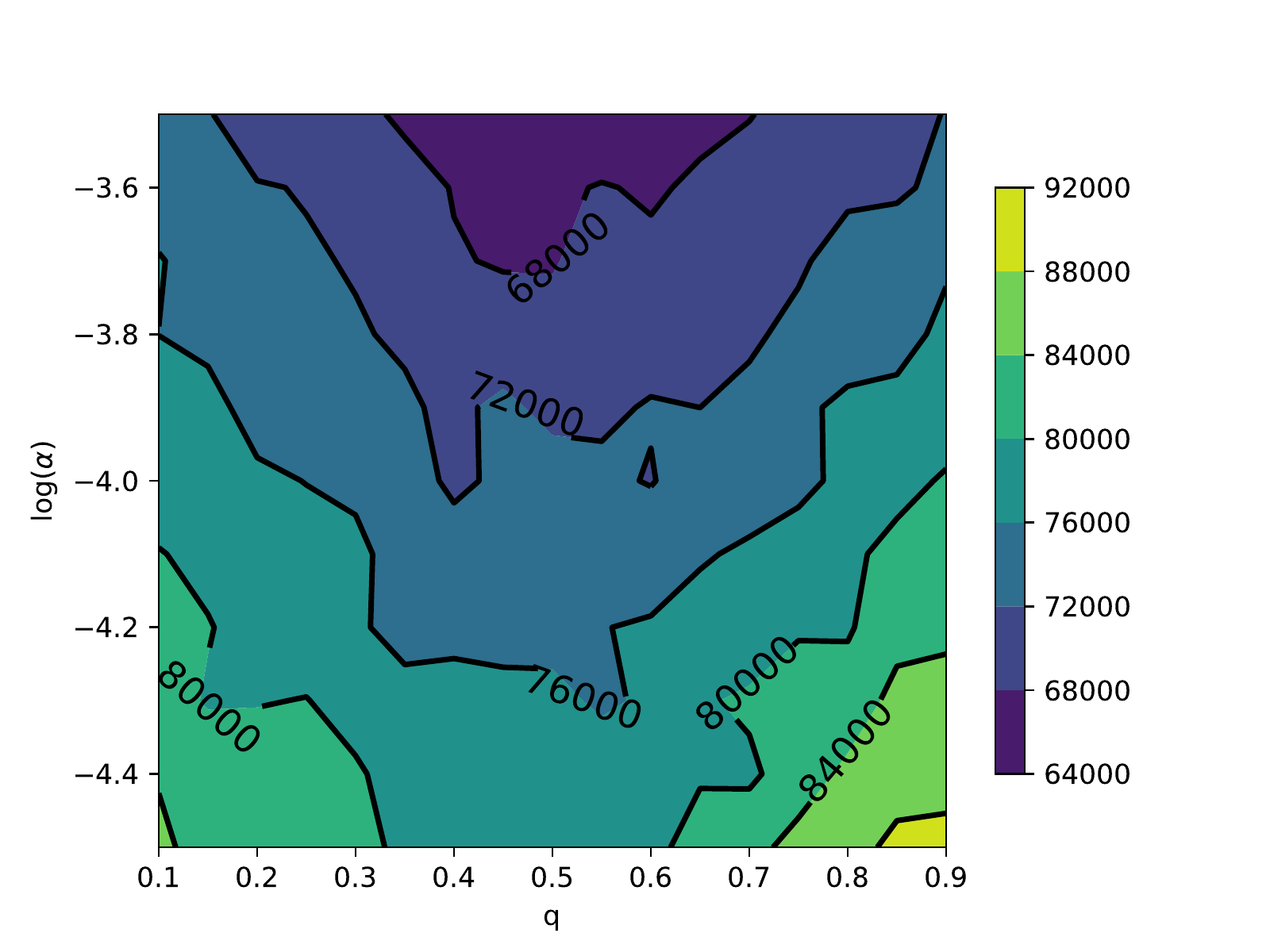}} \subfigure[]{\includegraphics[width=5.5cm]{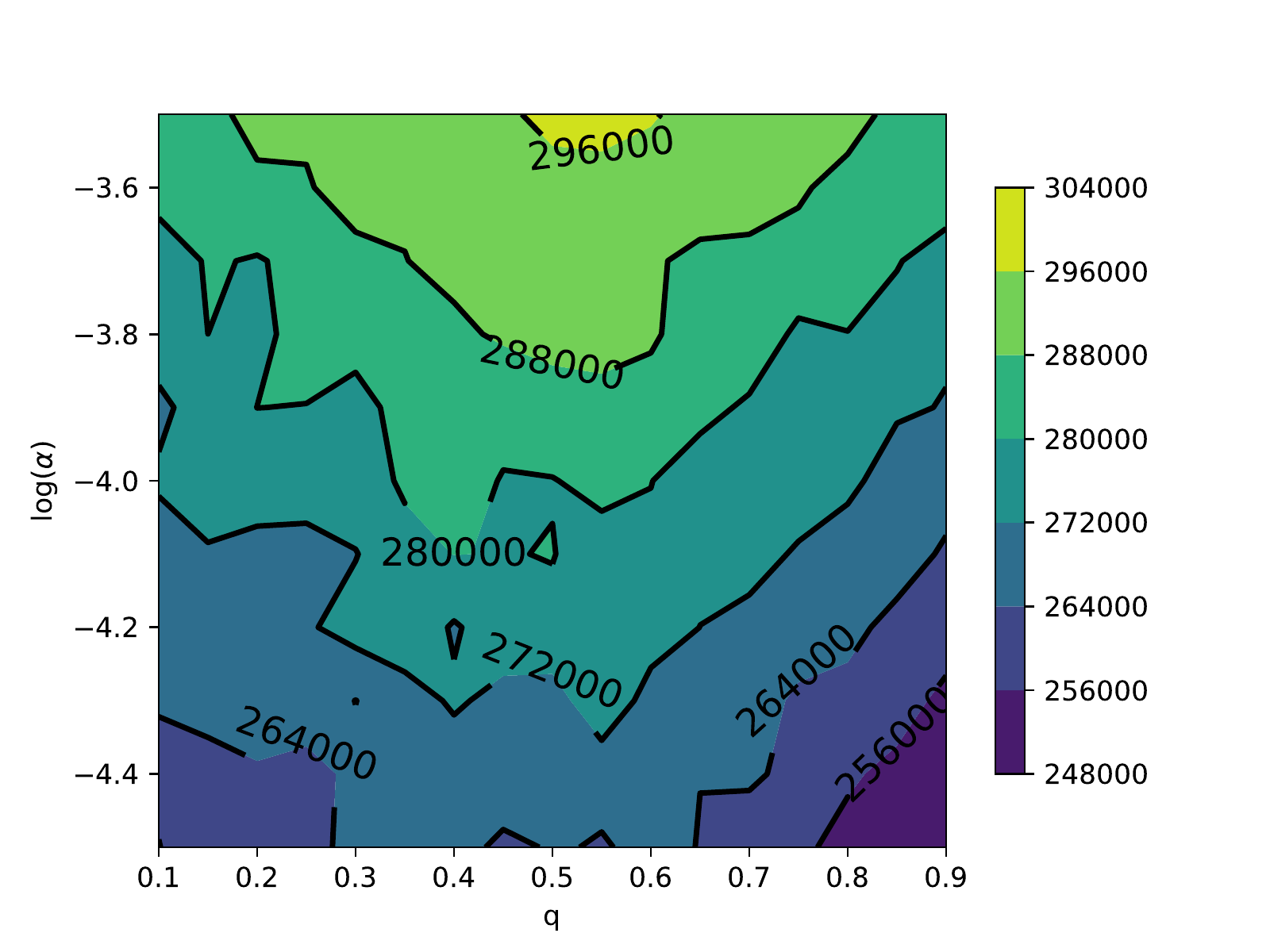}}
\subfigure[]{\includegraphics[width=5.5cm]{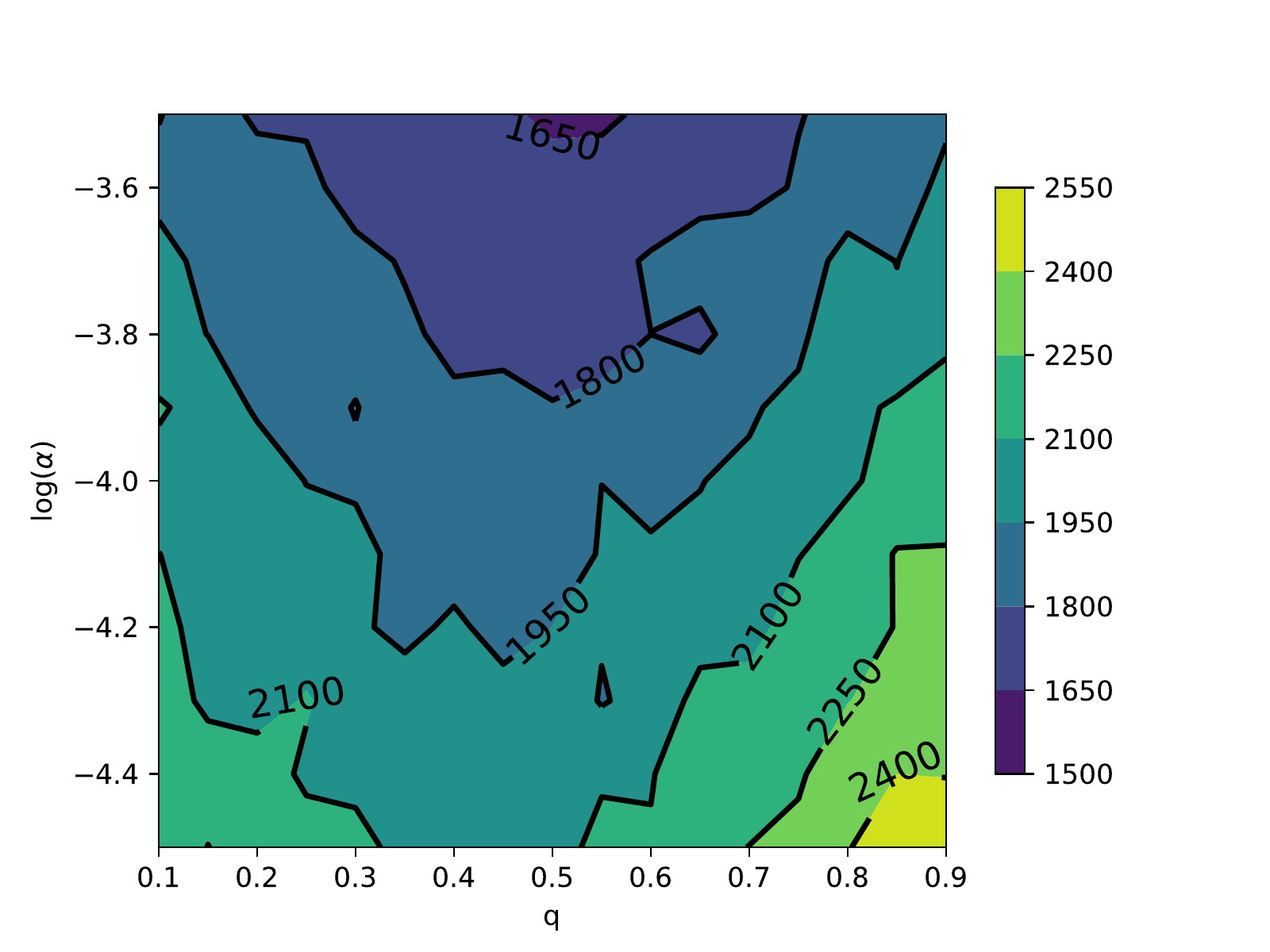}}
\subfigure[]{\includegraphics[width=5.5cm]{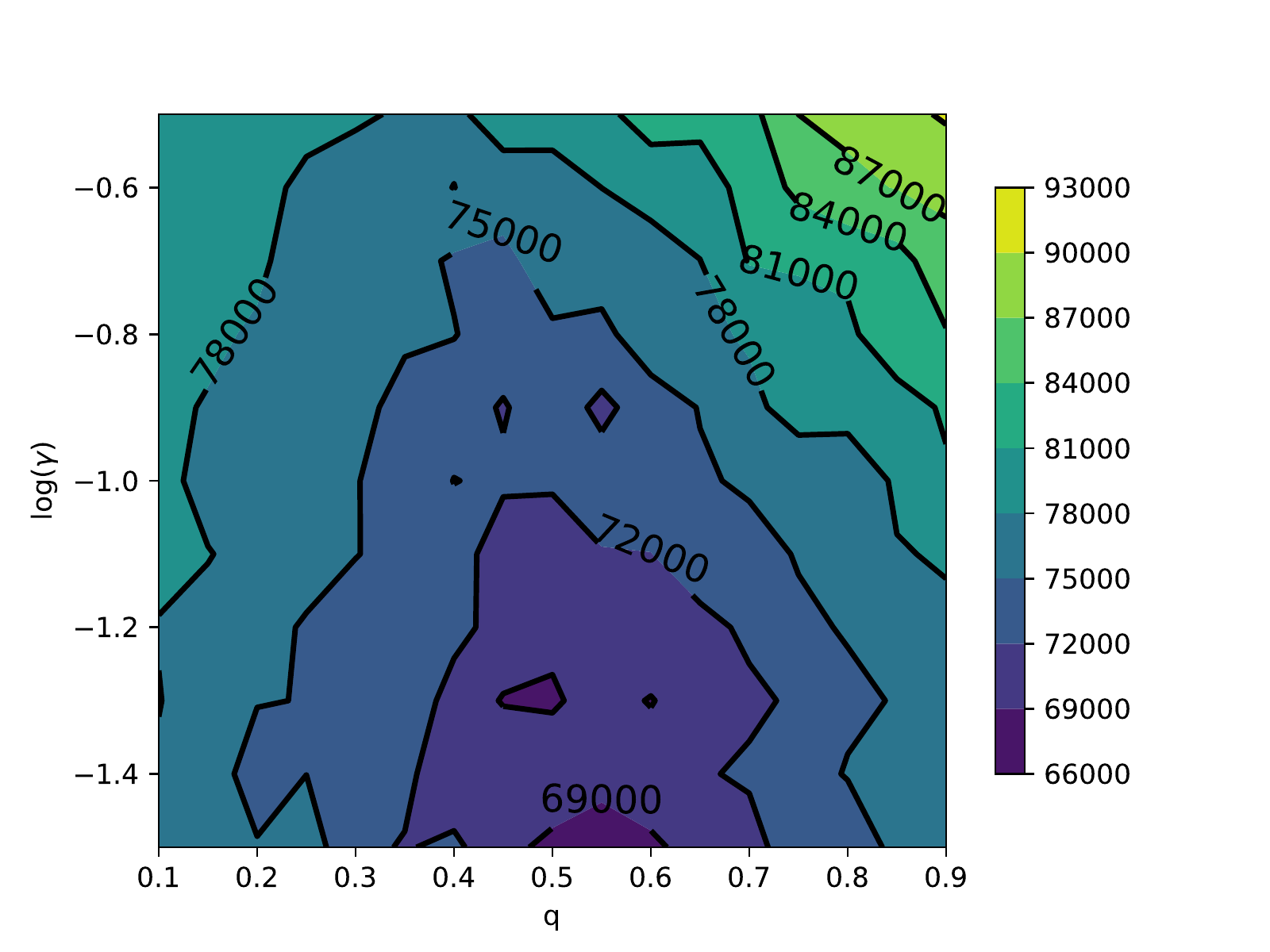}}
\subfigure[]{\includegraphics[width=5.5cm]{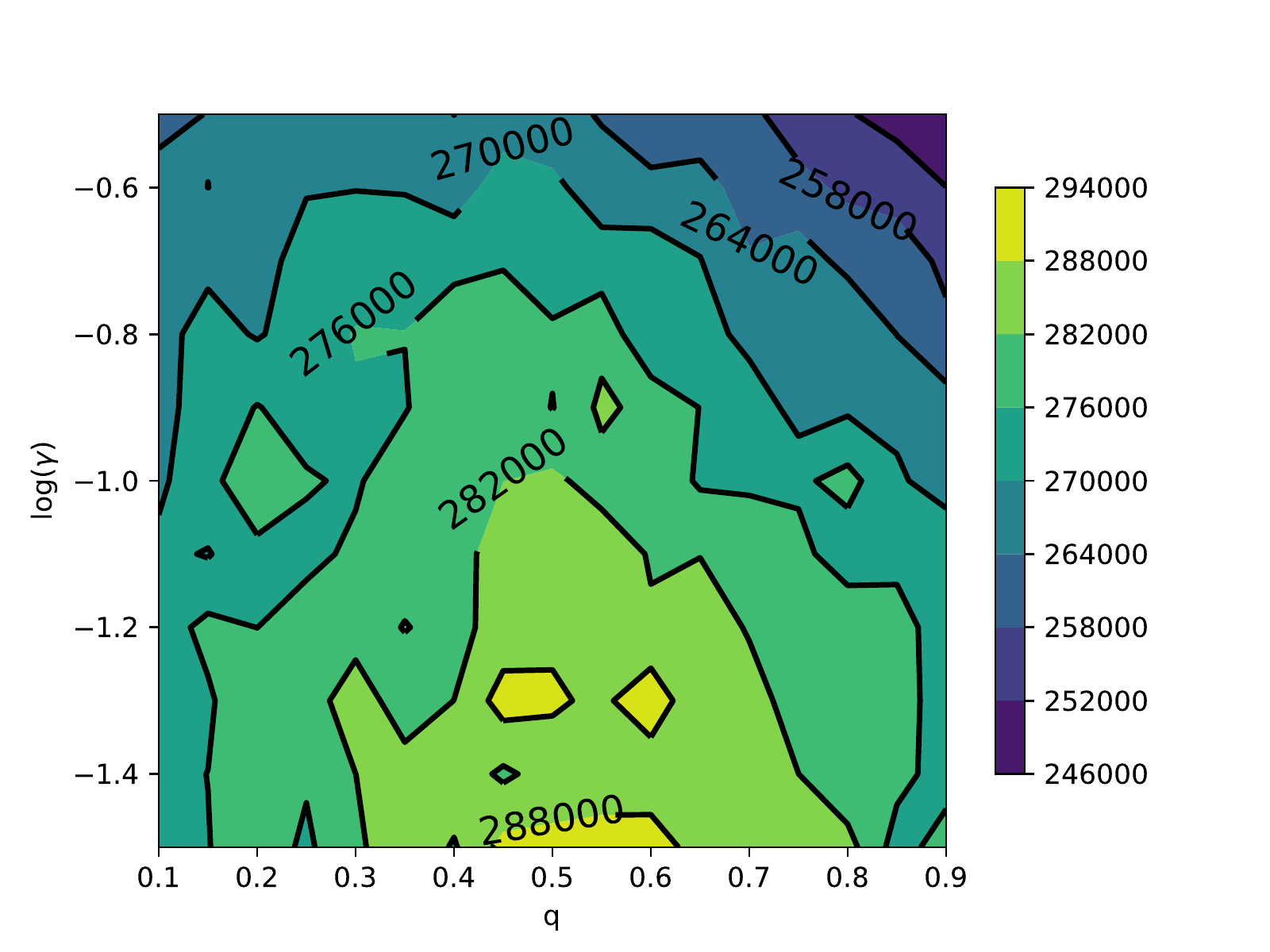}} \subfigure[]{\includegraphics[width=5.5cm]{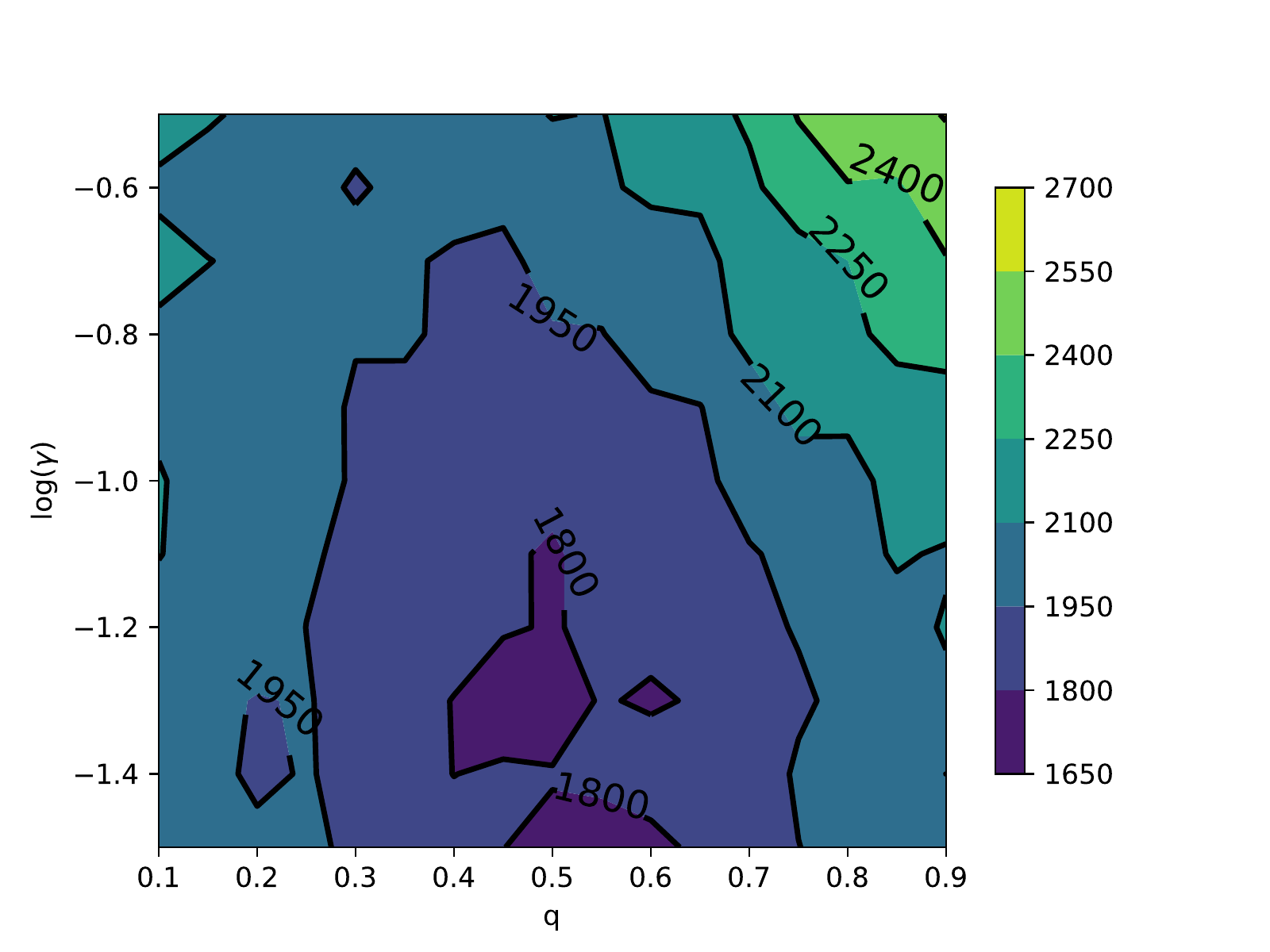}}

\caption{{\bf Fig. 7. Parameter planes of $\alpha$ against $q$ and $\gamma$ against $q$ for epidemic sizes, total number of vaccinated children and the peak of the epidemic on Barab\'{a}si-Albert network model (BAN).} In all simulations, the median value of the simulations is used to plot the parameter planes that are performed in $P_{adv}=.0001$ and $\rho=.01$.}\label{fig7}
\end{figure}

The probability of learning of parents can lead to a vaccination consensus in the case of Erd\H{o}s-R\'{e}nyi (random) network model (ERN) and Barab\'{a}si-Albert network model (BAN) shown in Figs. 8 (a) and (b) as the number of vaccinators on the last day of the epidemic. However, a moderate to high learning probability is required in the case of ERN. The pattern stays exactly the same as in Fig. 8 (a) for ERN when $P_{adv}=.01$, data are not shown to avoid redundancy. It changes drastically, however, for BAN when $P_{adv}=.01$, Fig. 8 (c). In the latter, fewer parents end up accepting to vaccinate their children with a decline over the increase in the learning probability.

\begin{figure}[H]
\subfigure[]{\includegraphics[width=8cm]{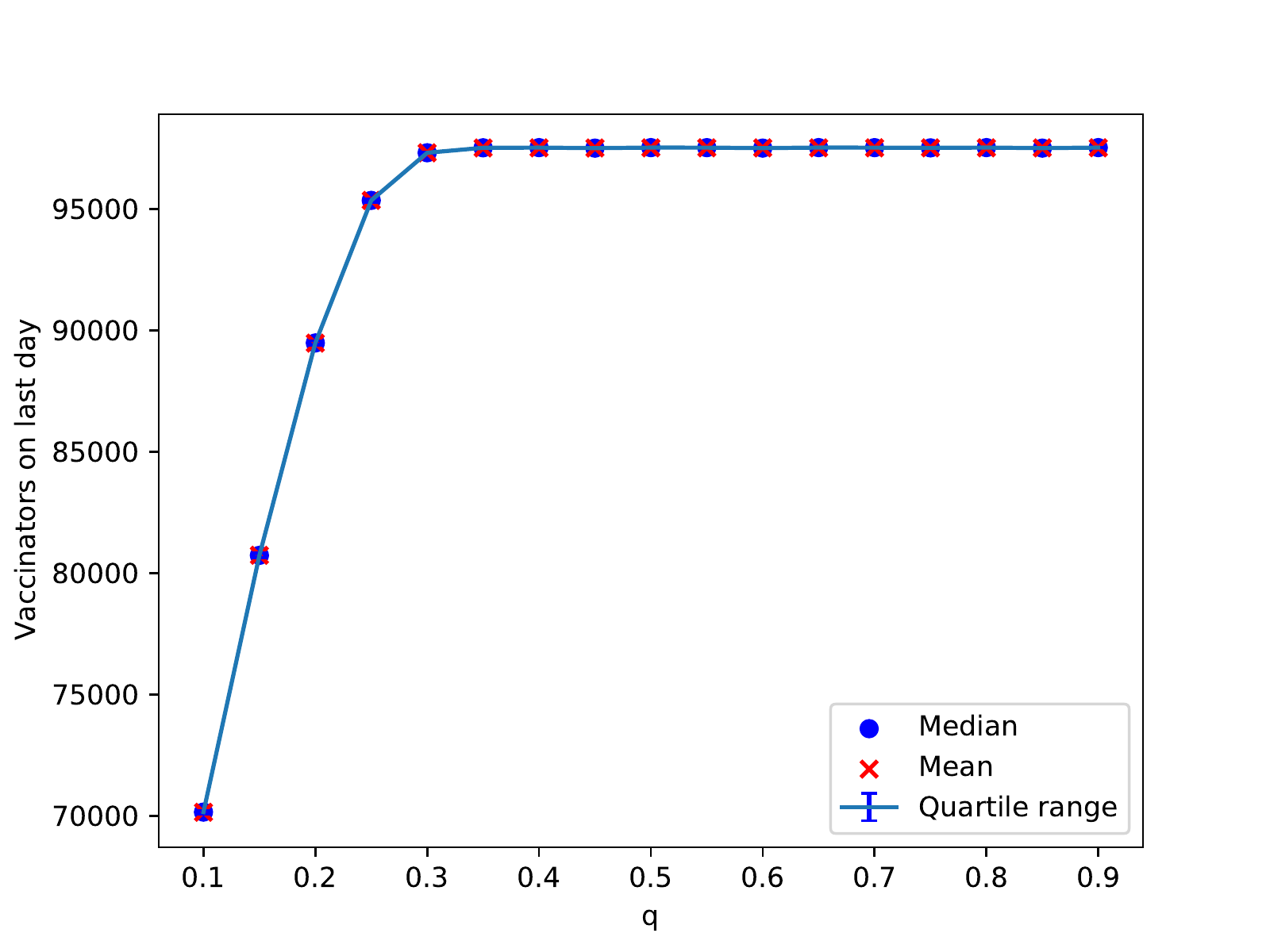}}
\subfigure[]{\includegraphics[width=8cm]{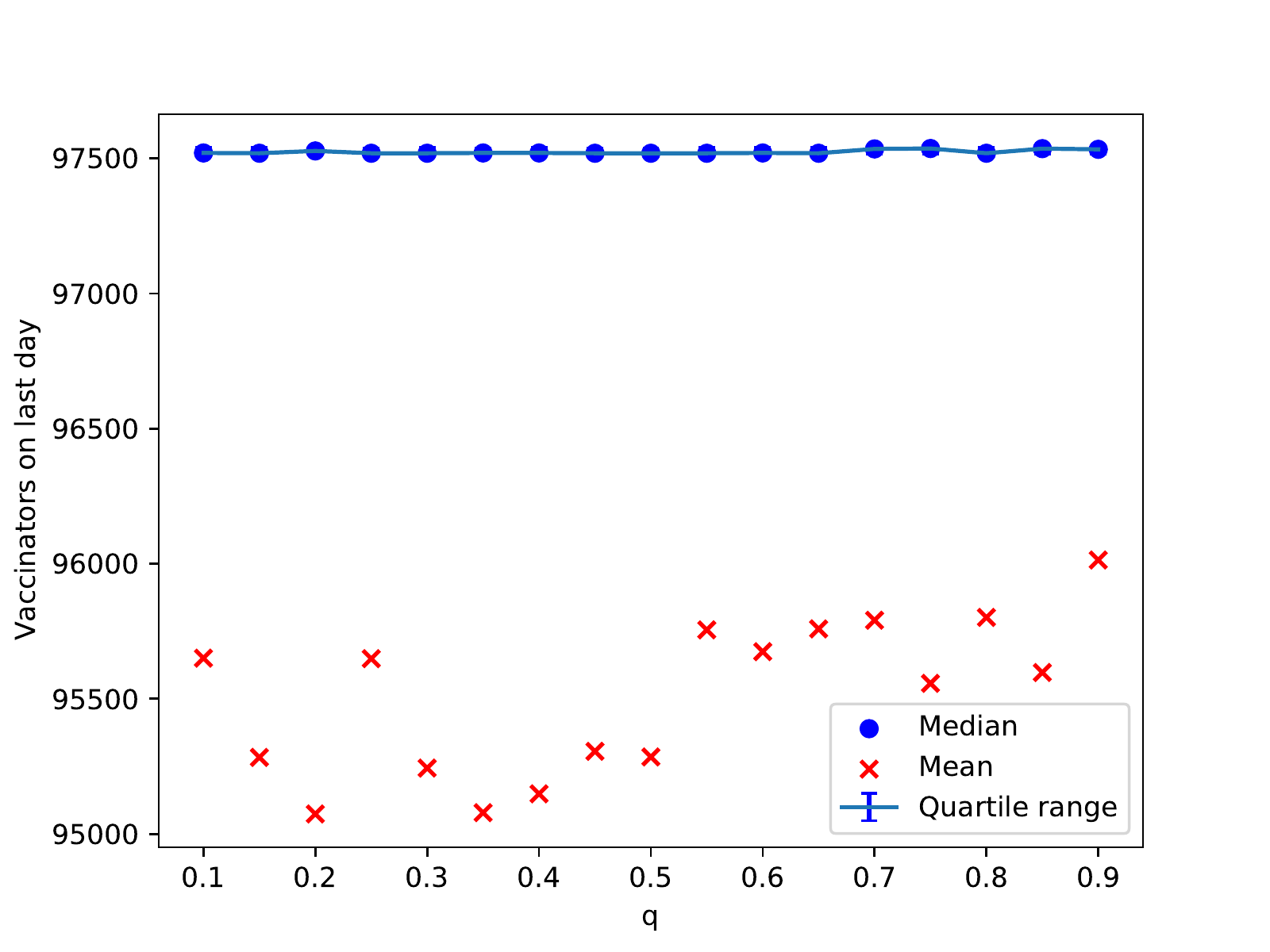}}
\subfigure[]{\includegraphics[width=8cm]{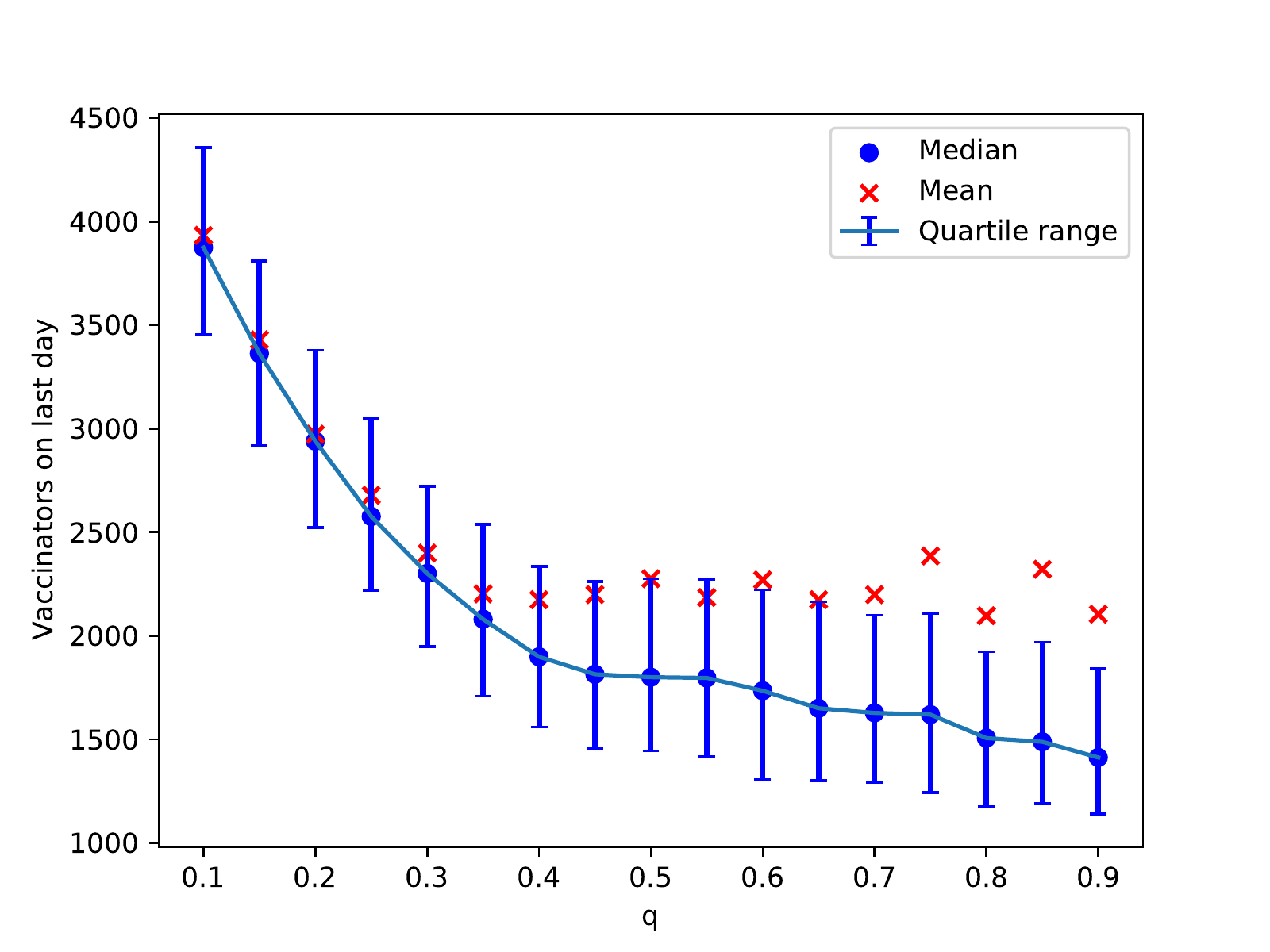}}
\caption{{\bf Fig. 8. Simulations of the final total number of vaccinators on Erd\H{o}s-R\'{e}nyi (random) network model (ERN) and Barab\'{a}si-Albert network model (BAN) for different values of $q$.} Simulations are done on Erd\H{o}s-R\'{e}nyi (random) network model (ERN) using $P_{adv}=.0001$ in (a). It was done on Barab\'{a}si-Albert network model (BAN) when using $P_{adv}=.0001$  in (b) and using $P_{adv}=.01$ in (c). In all of the simulations $\rho=.01$.}\label{fig5}
\end{figure}

Up to this point, the simulations of the model assumed that the population had a homogeneous culture. That is, parents show their actual preference or strategy with probabilities $q_{i,j} \in (q- .05,q+ .05)$ for one fixed value of $q$. In the next part, we examine the effect of the population having a cultural attribute. That attribute gives rise to a heterogeneous population, in which the population consists of two groups that use probabilities in two different regions: $.1\pm .05$ and $.9\pm .05$. We choose $.1$ and $.9$ as extreme case. We call the group of the latter type, subpopulation with attribute 1. The proportion of those with attribute 1 can affect the fate of vaccination uptake and the epidemic.

In the case of an epidemic spreading on an ERN, having a small proportion of parents with attribute 1, results in a large epidemic size and its peak, as well as less vaccine uptake. This is apparent in the simulations shown in Fig. 9. In the case of an epidemic spreading on a BAN, heterogeneity in the population's culture does not show a significant effect on the epidemics or vaccine uptake; see simulations in Fig. 10.

\begin{figure}[H]
\centering
\subfigure[]{\includegraphics[width=5.5cm]{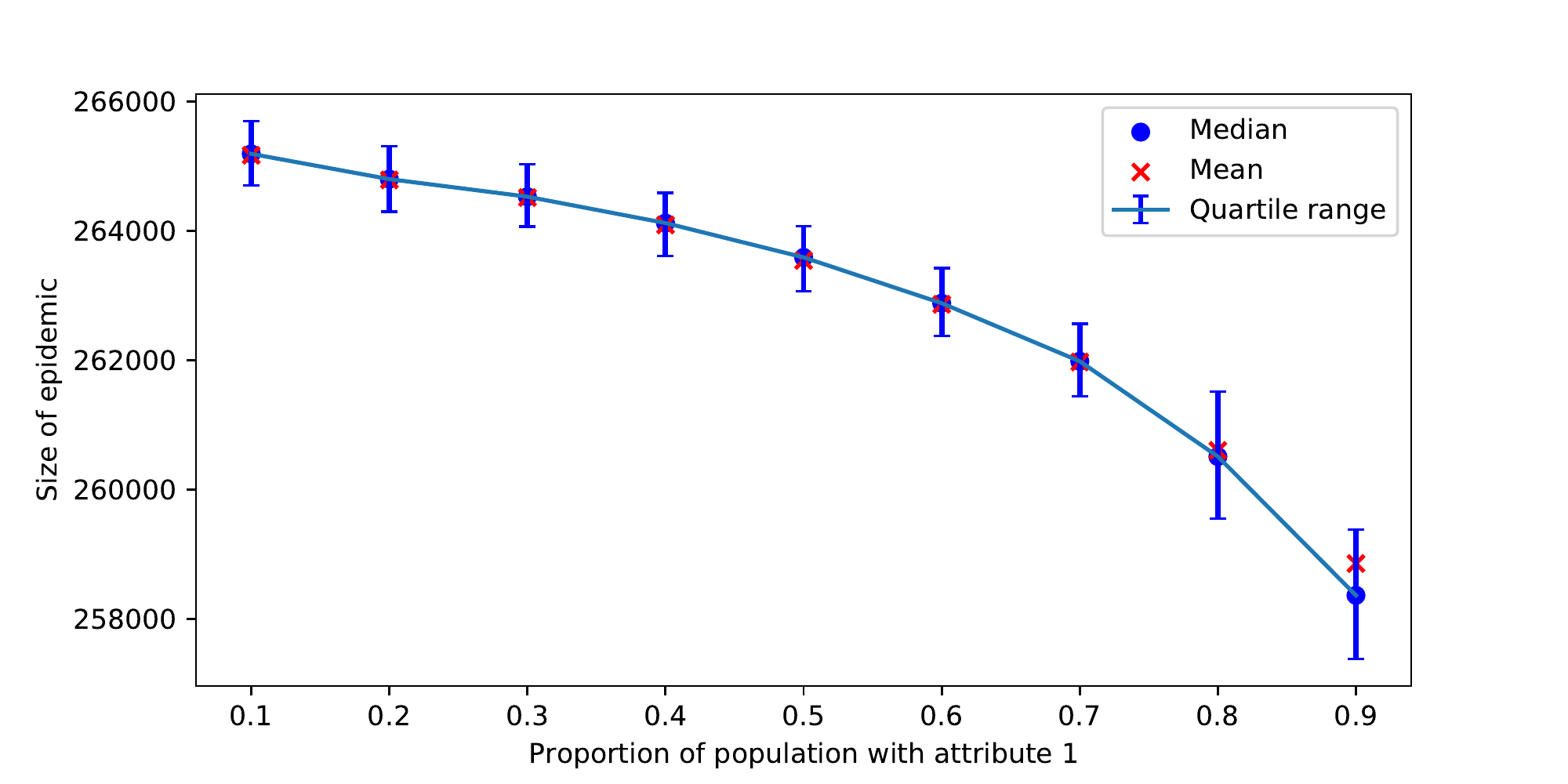}} \subfigure[]{\includegraphics[width=5.5cm]{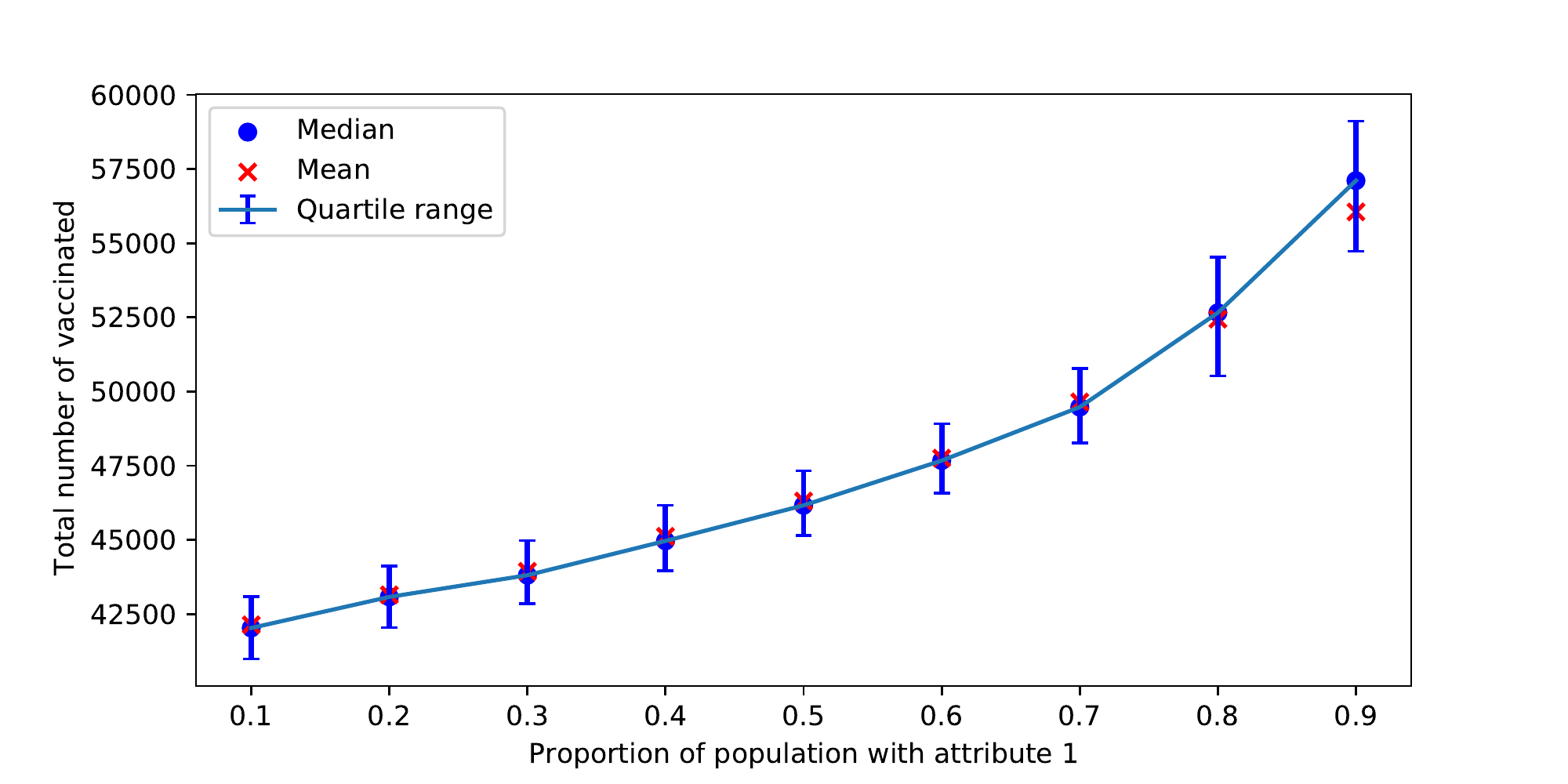}}
\subfigure[]{\includegraphics[width=5.5cm]{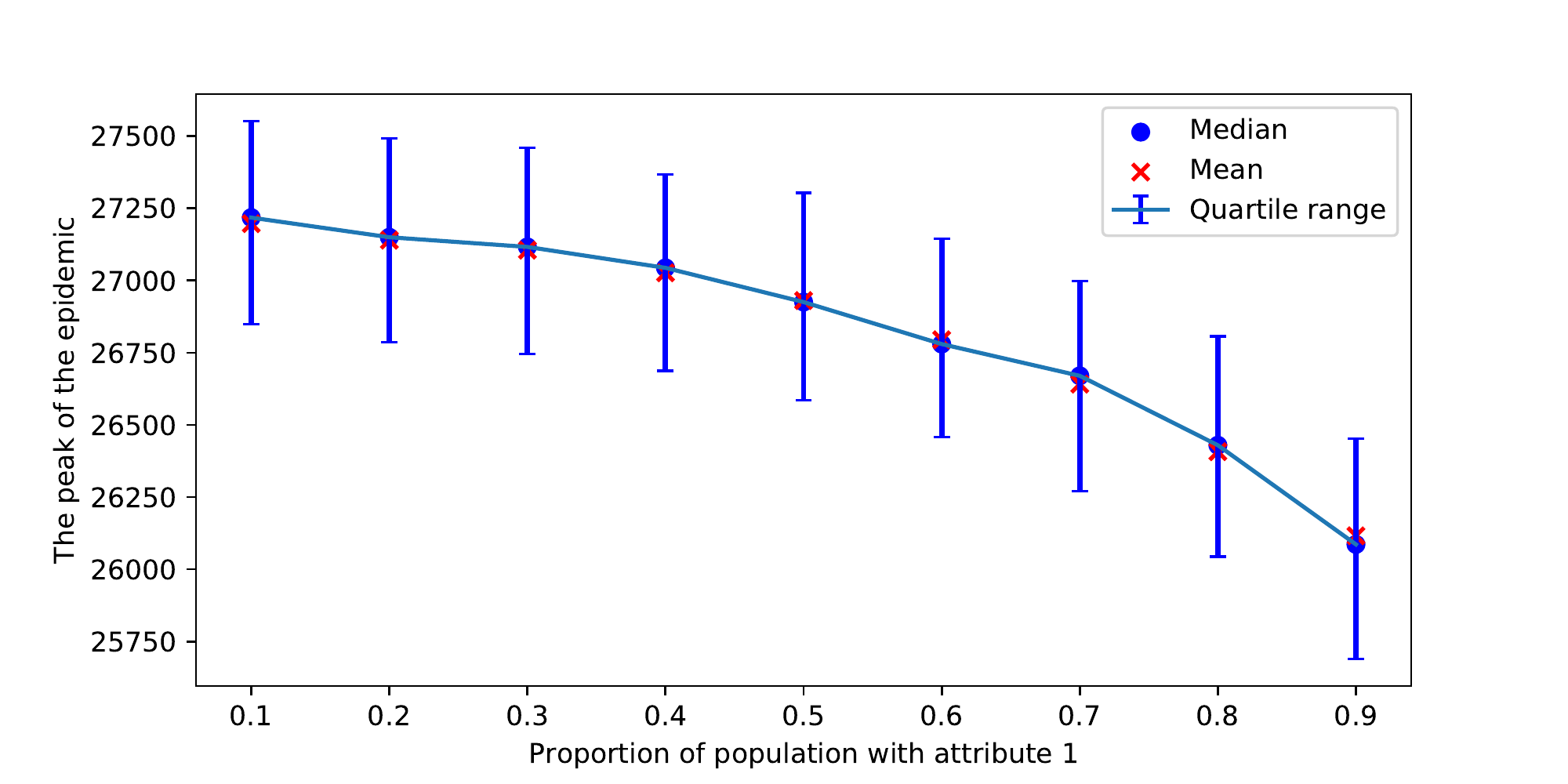}} \subfigure[]{\includegraphics[width=5.5cm]{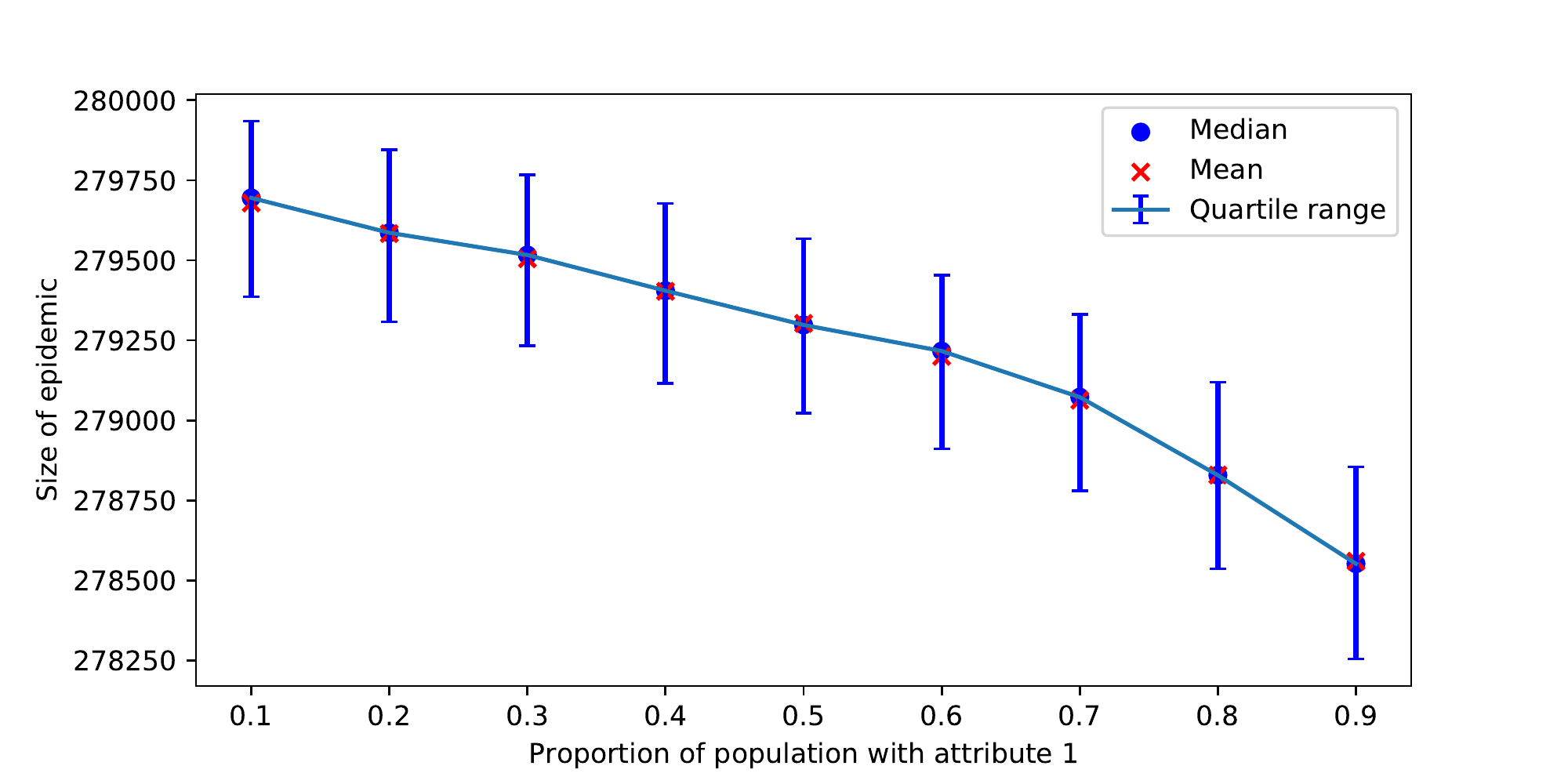}} \subfigure[]{\includegraphics[width=5.5cm]{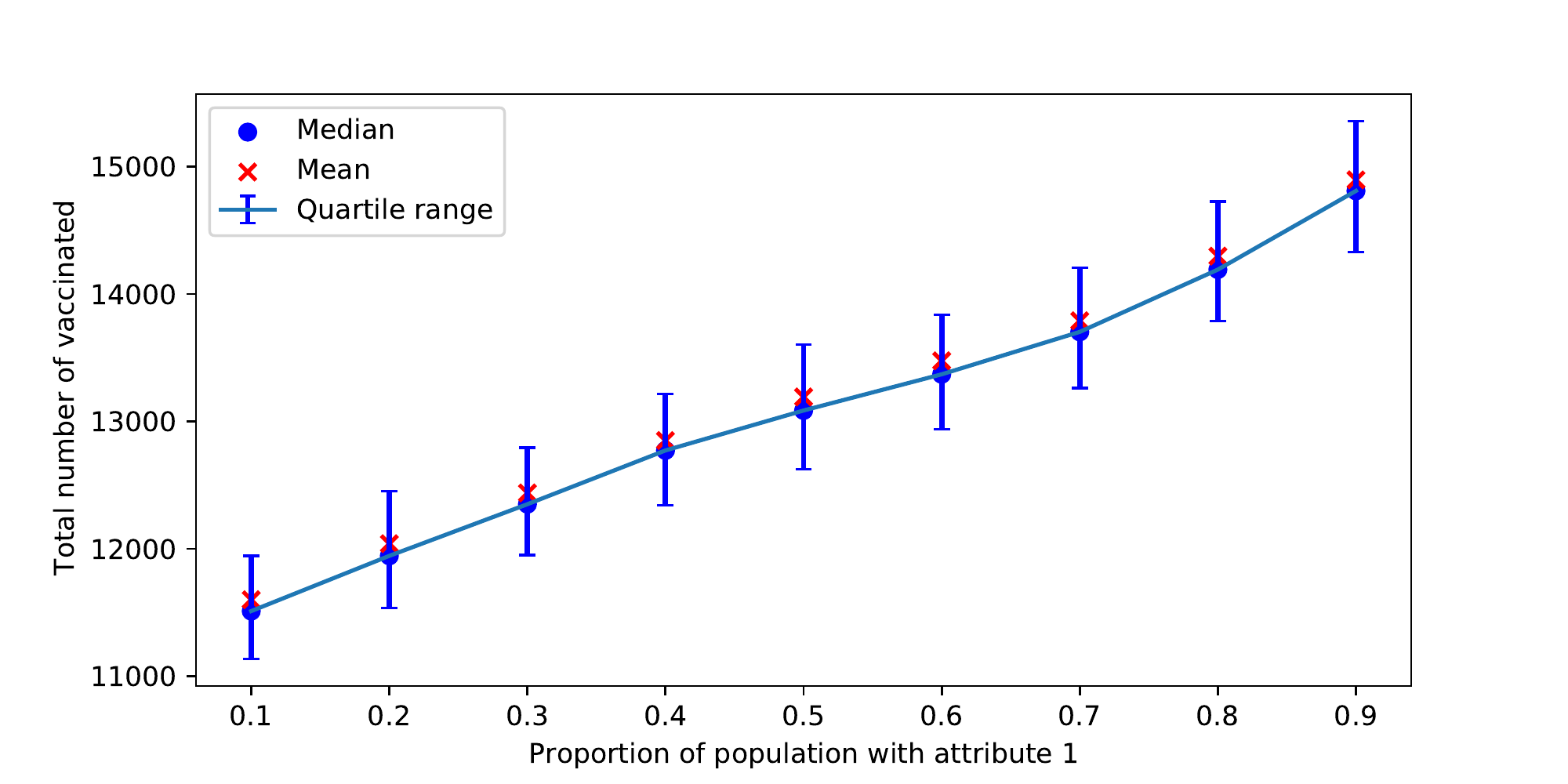}}
\subfigure[]{\includegraphics[width=5.5cm]{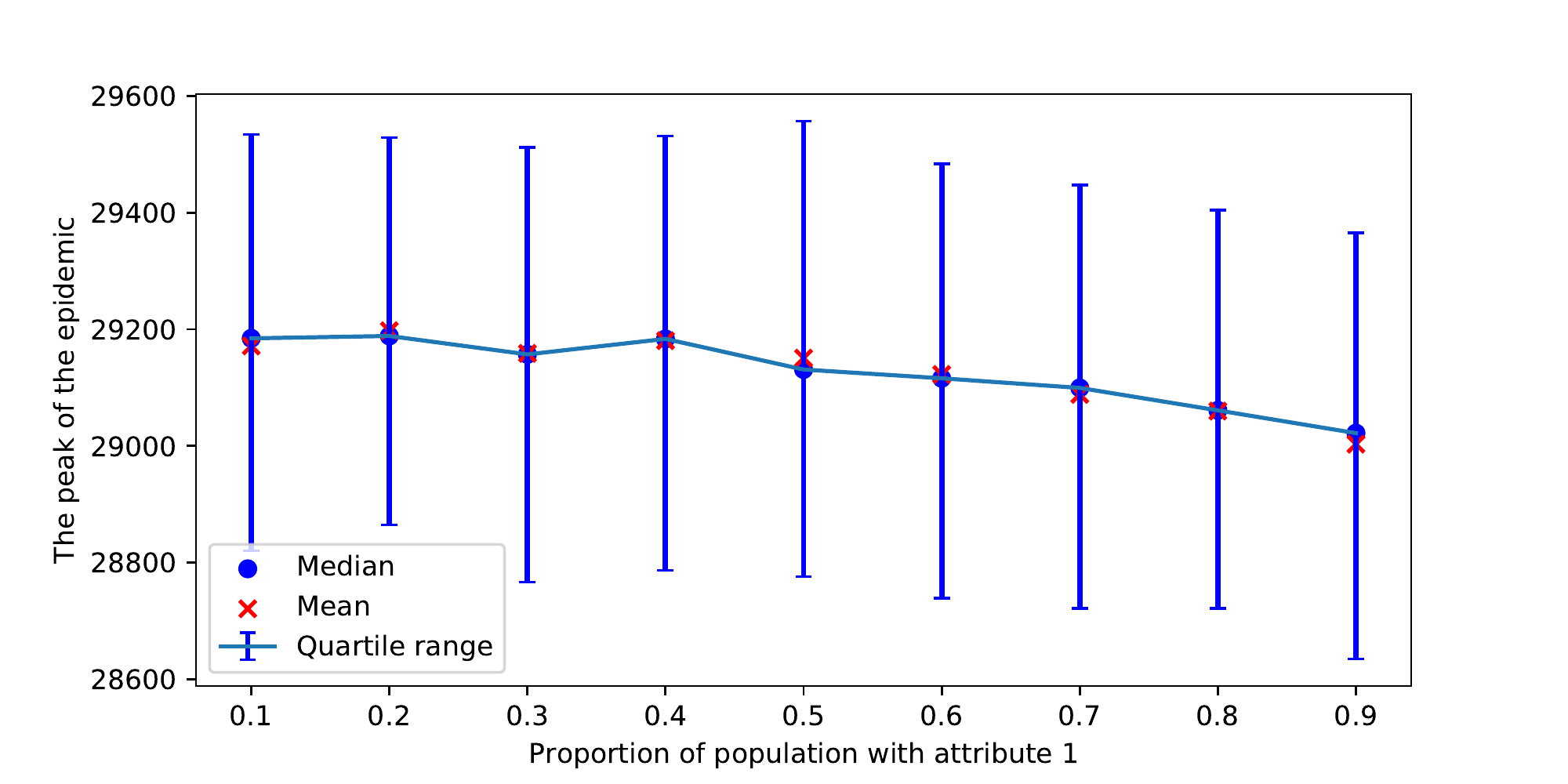}}
\caption{{\bf Fig. 9. Simulations of sizes of epidemic, total number of vaccinated children, and the peak of the epidemic on Erd\H{o}s-R\'{e}nyi (random) network model (ERN) for different proportions of attribute 1.} Simulations are done using $\rho=.01$ in  (a), (b) and (c), and $\rho=.001$ in (d), (e) and (f). In all of the simulations $P_{adv}=.0001$.}\label{fig8}
\end{figure}

\begin{figure}[H]
\centering
\subfigure[]{\includegraphics[width=5.5cm]{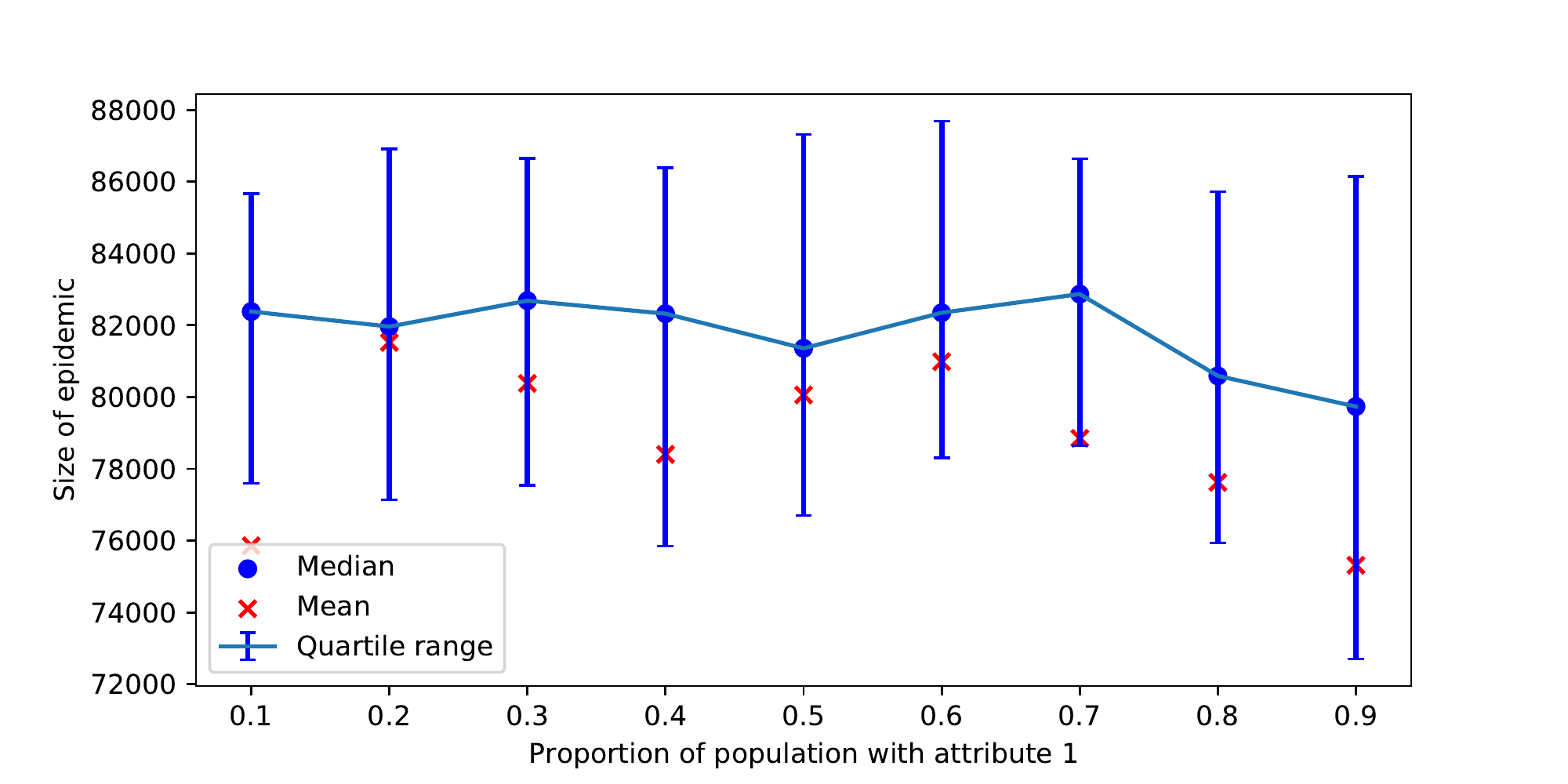}} \subfigure[]{\includegraphics[width=5.5cm]{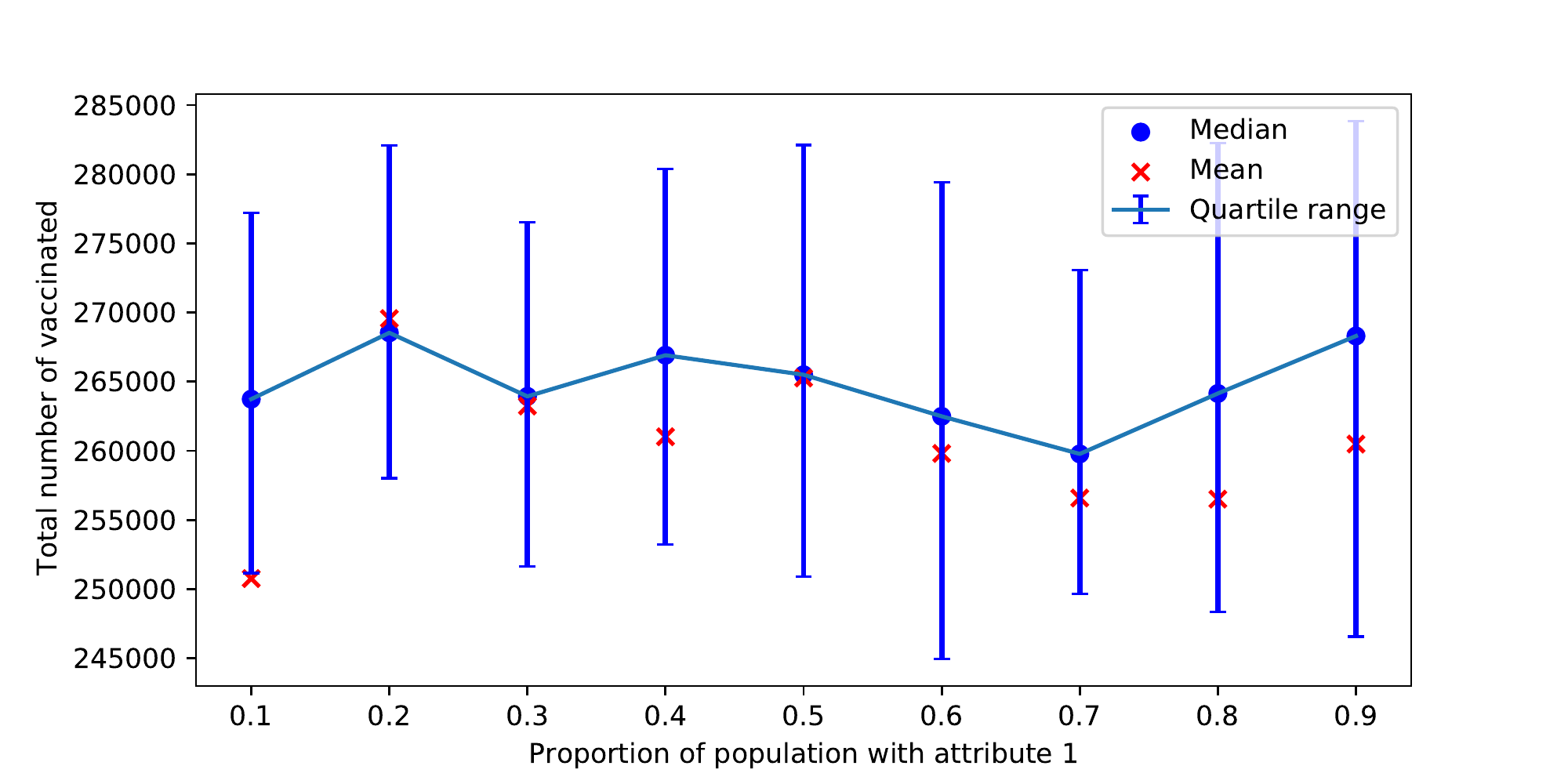}}
\subfigure[]{\includegraphics[width=5.5cm]{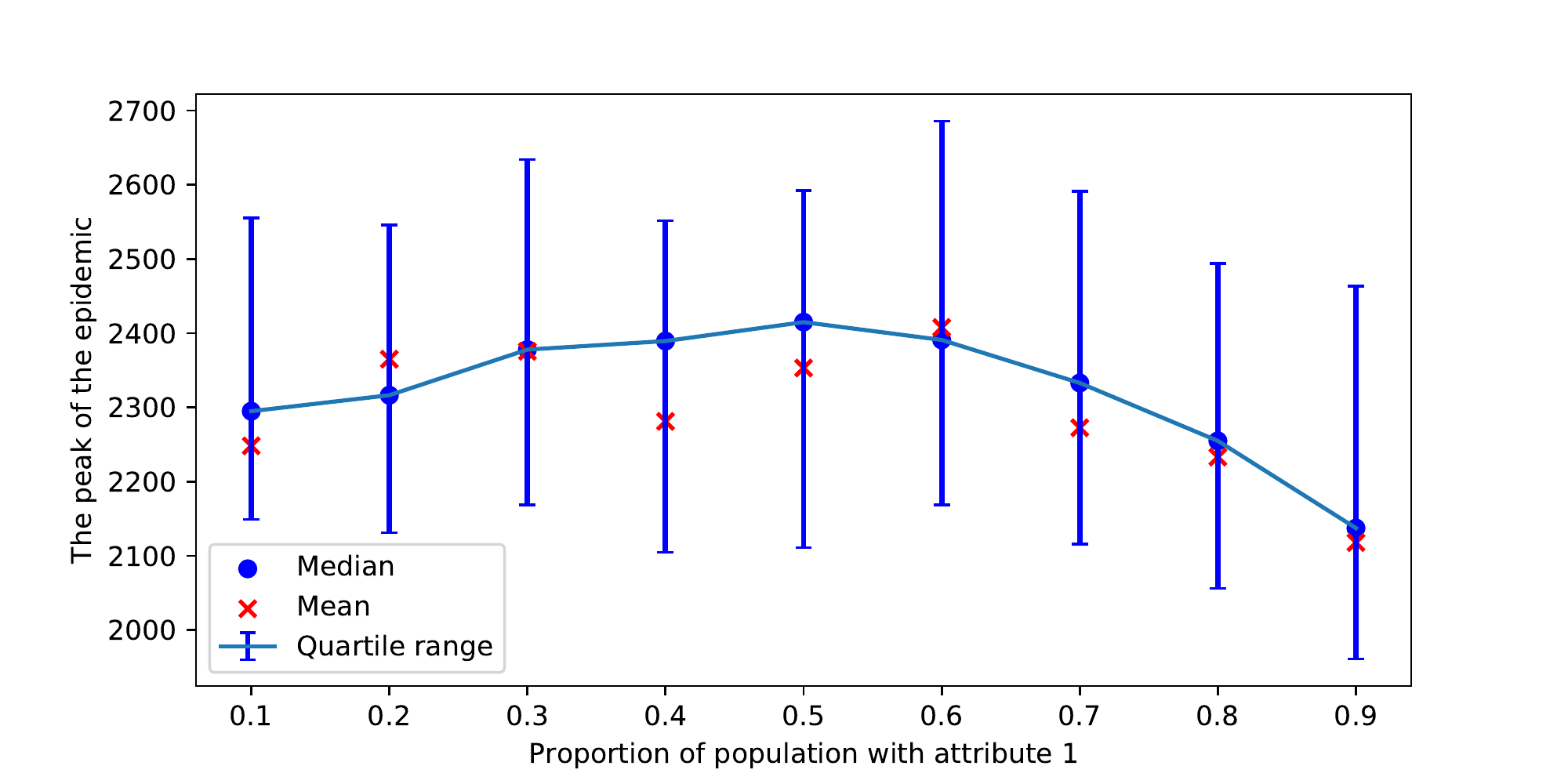}}
\subfigure[]{\includegraphics[width=5.5cm]{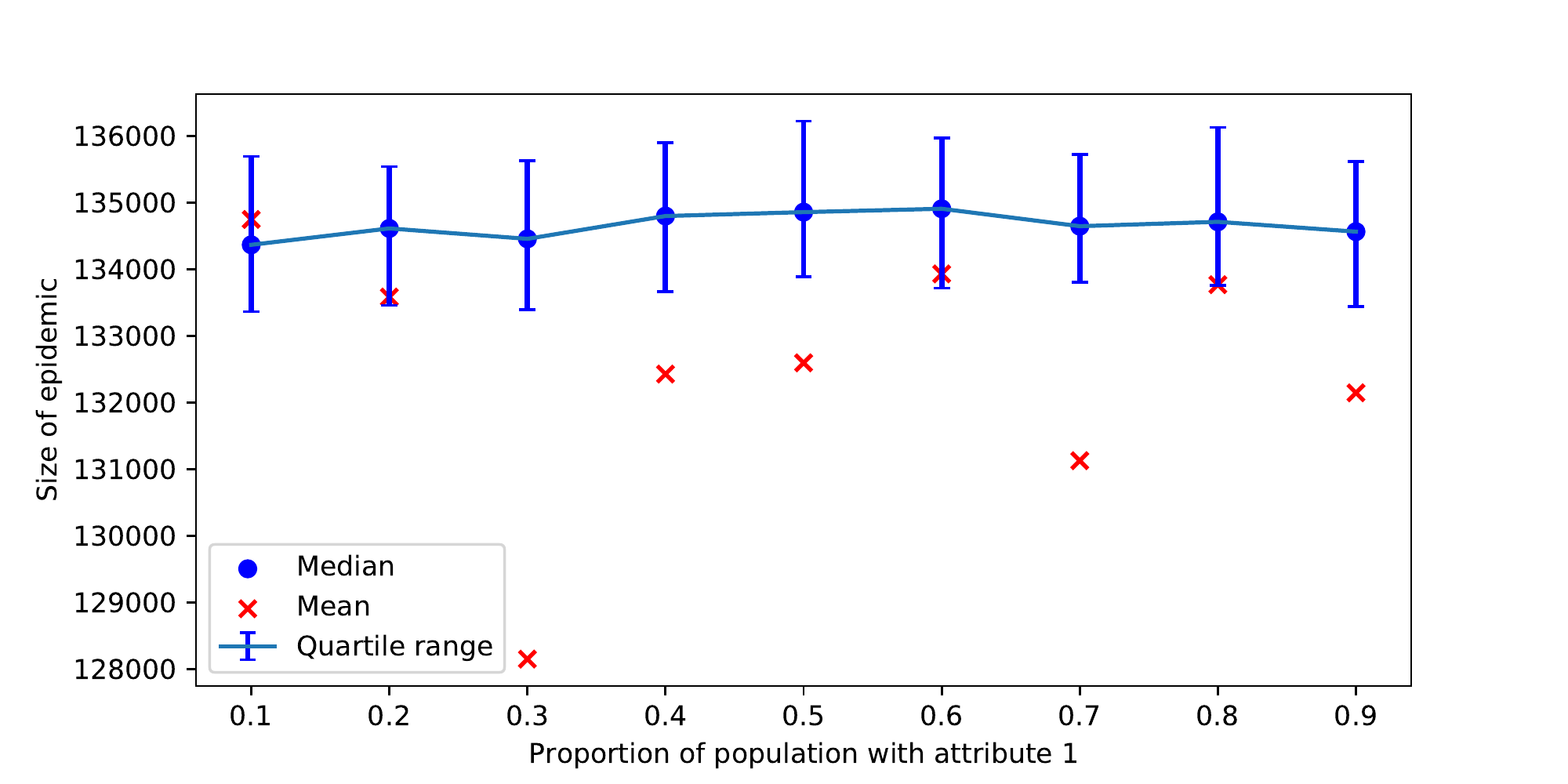}} \subfigure[]{\includegraphics[width=5.5cm]{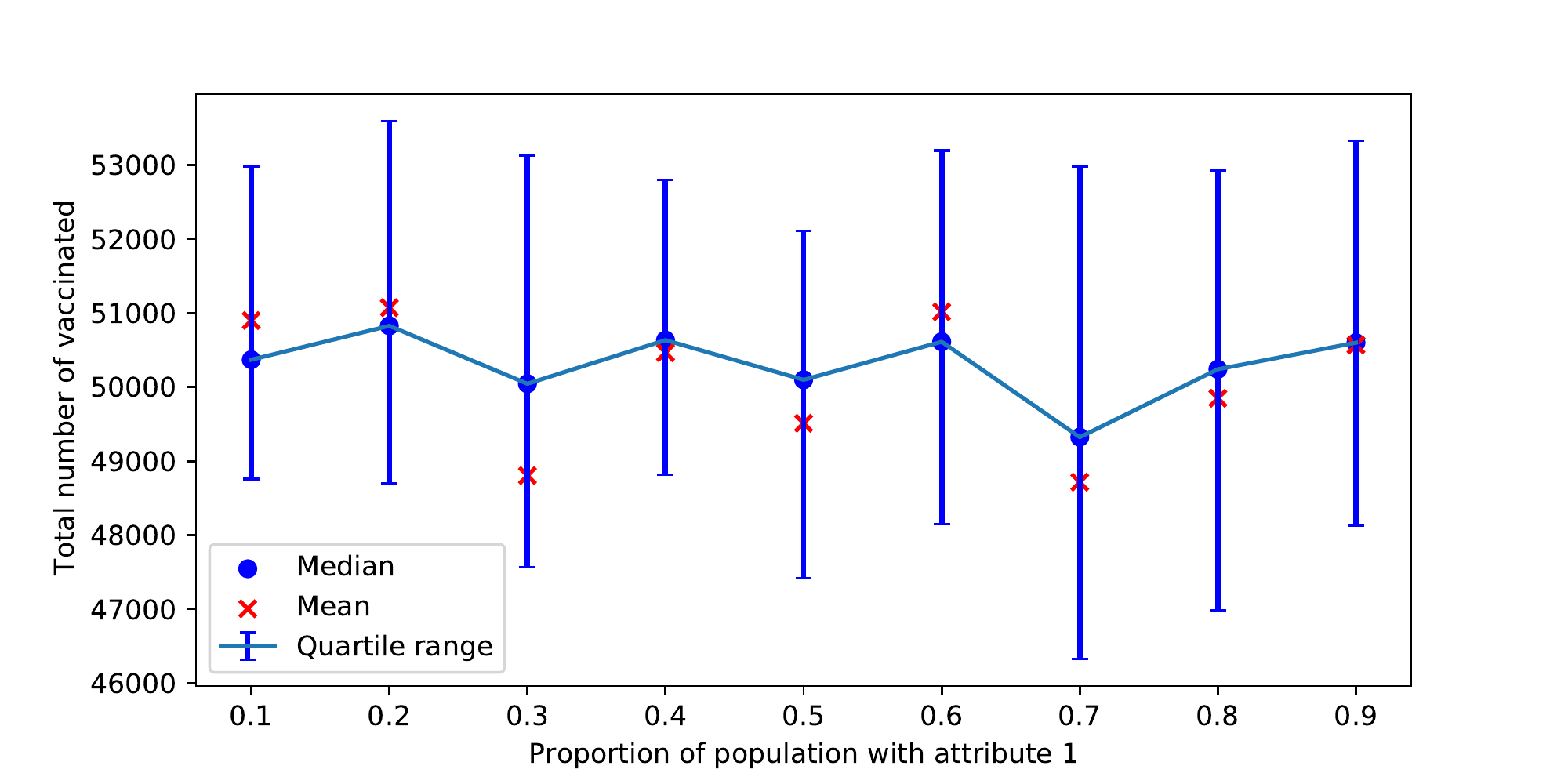}}
\subfigure[]{\includegraphics[width=5.5cm]{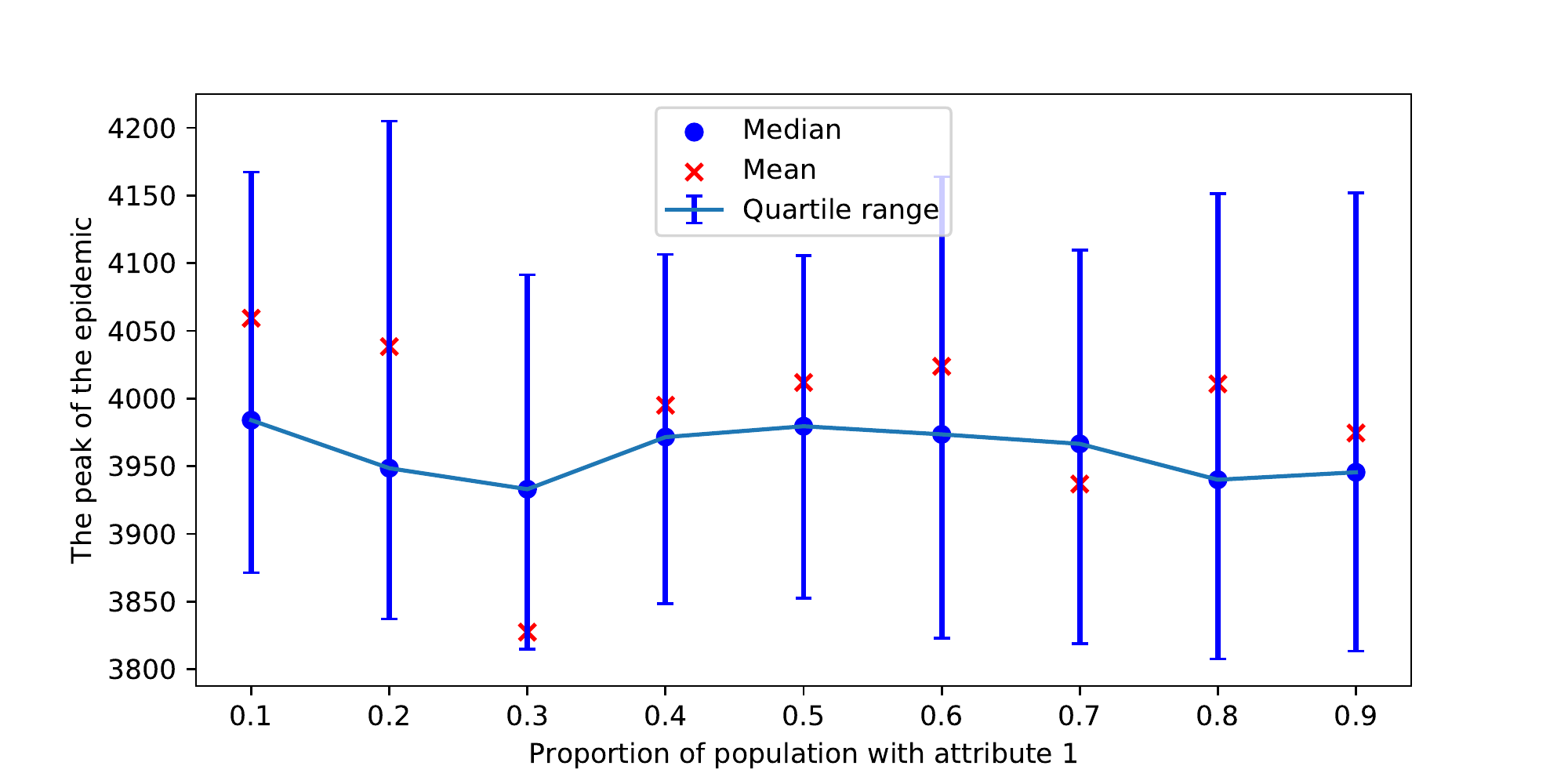}}
\caption{{\bf Fig. 10. Simulations of sizes of the epidemic, the total number of vaccinated children, and the peak of the epidemic on Barab\'{a}si-Albert network model (BAN) for different proportions of attribute 1.} Simulations are done using $\rho=.01$ in  (a), (b) and (c), and $\rho=.001$ in (d), (e) and (f). In all of the simulations $P_{adv}=.0001$.}\label{fig9}
\end{figure}

If the proportion of parents with attribute 1 is more than one-half, vaccination becomes a consensus in the case of ERN; see Fig. 11 (a). Vaccination becomes a consensus in all cases for BAN; see Fig. 11 (b).

\begin{figure}[H]
\centering
\subfigure[]{\includegraphics[width=8.5cm]{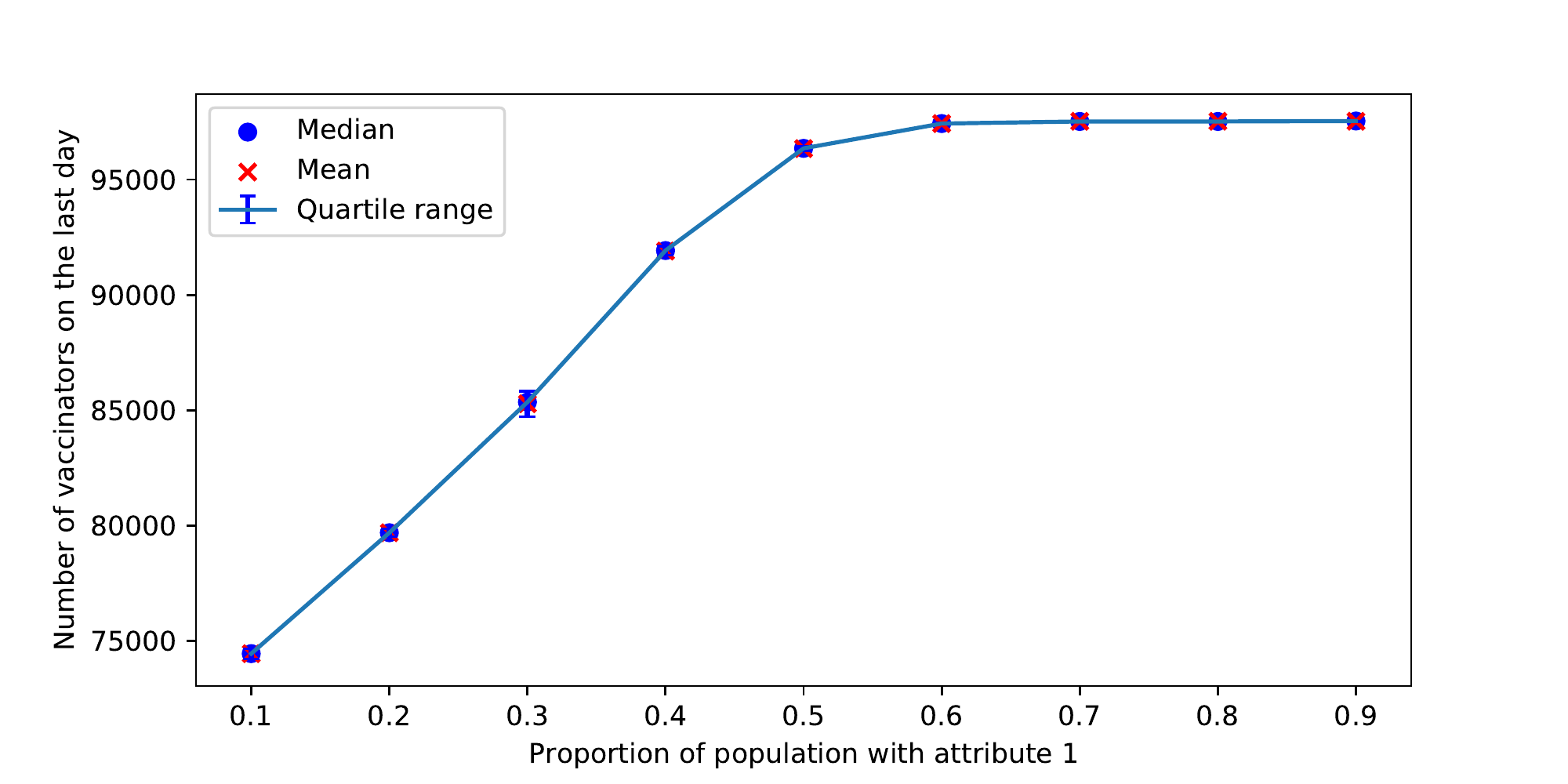}} \subfigure[]{\includegraphics[width=8.5cm]{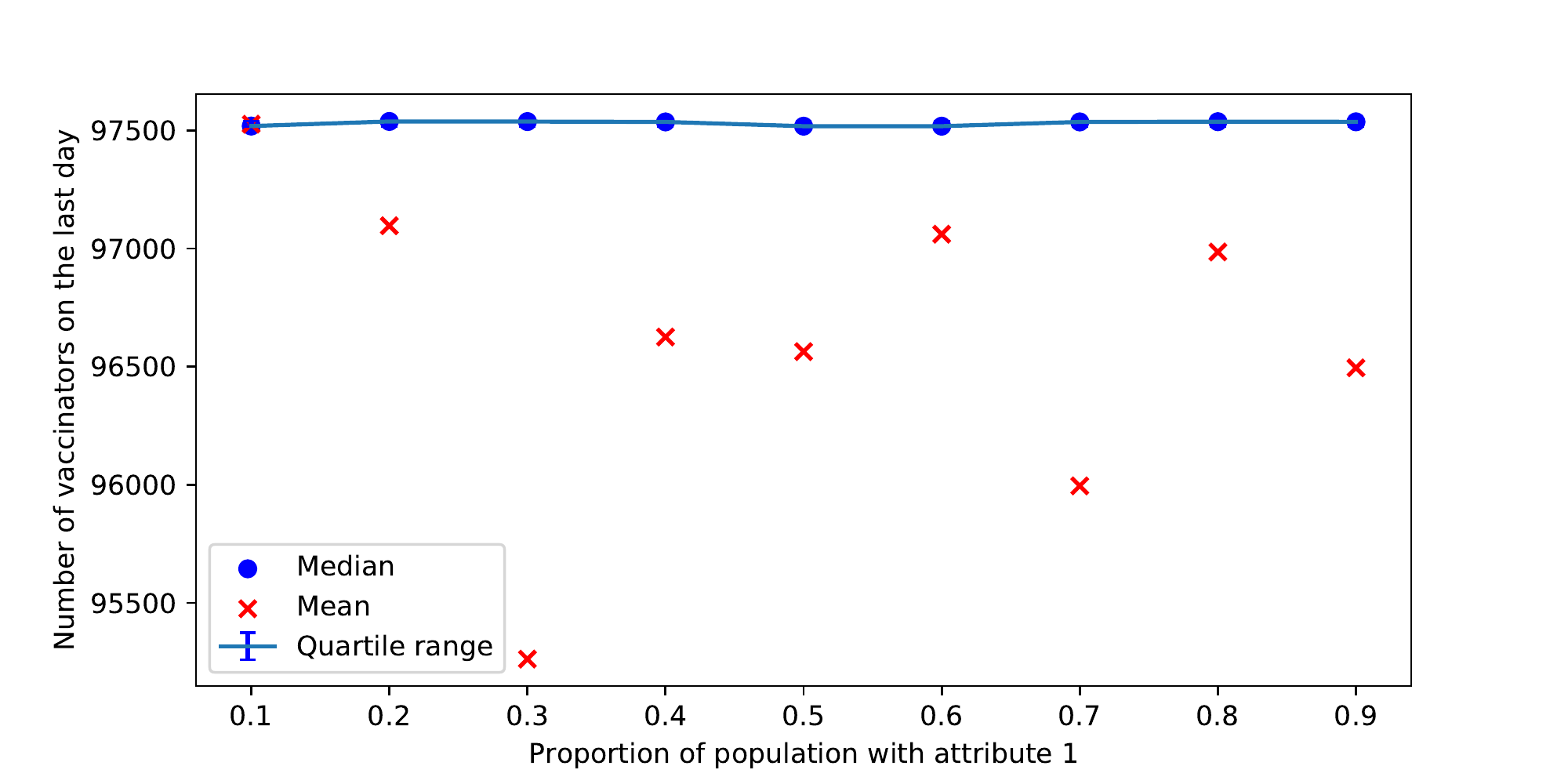}}
\caption{{\bf Fig. 11. Simulations of final total number of vaccinators on Erd\H{o}s-R\'{e}nyi (random) network model (ERN) and Barab\'{a}si-Albert network model (BAN) for different proportions of attribute 1.} Simulations are done on Erd\H{o}s-R\'{e}nyi (random) network model (ERN) in (a) and Barab\'{a}si-Albert network model (BAN) in (b). In all of the simulations are done using $P_{adv}=.0001$ and $\rho=.01$.}\label{fig92}
\end{figure}


\section{Discussions and conclusion}
In this paper, we introduce a Bayesian aggregation model for boundedly rational observational social learning as a method of decision-making concerning vaccination. The learning model is based on social observations of neighbors on the social network through which information is spreading. Other models in the literature are shown to be special cases of our model. Some models, such as the degree of injunctive social norm $\delta$, in contrast, lack the scale that our model enjoys. Using our new model, we explore the influence of social learning on the development of consensus on a network. The model also explores how heterogeneity in the culture of observational social learning affects vaccine uptake.   
We used stochastic simulations of disease and information processes that spread over two overlapping networks. The processes took part in one of two types of networks: Erd\H{o}s-R\'{e}nyi (random) network model (ERN) and Barab\'{a}si-Albert network model (BAN). The results of observational social learning on those different networks and its mutual influence on the disease spread on the overlapping network were dependent on the network type and on the vaccine safety and availability. In ERN, as adverse events become rare due to their small likelihood or inaccessibility to vaccines, the increase in pressure ensued through declaring the adopted strategy with higher probability, especially when $q\geq .5$, will result in an increase in vaccine uptake and eventually vaccination as a social norm. (Notice that $q = .5$ is the rational agent case.) The vaccine uptake in BAN is relatively higher than in ERN, and vice versa for epidemic sizes. Furthermore, the higher the probability of learning when $q\geq .5$ in BAN, the less vaccine uptake will be. 

Based on simulations of epidemic processes on the ERN and BAN, degree distribution plays a role in the levels of vaccine uptake and parental acceptance of vaccination. The uptake of the vaccine in the case of BAN is higher than in ERN. That might be for two reasons: As the disease is known to spread quickly in BAN, the accumulation of the number of cases will be rapid, increasing the subjective probability of vaccinating. The second reason is that the large number of neighbors will increase the inclination toward one of the two opinions even with small values of $q$.  

Vaccine availability and accessibility interact with vaccine adverse events in a way that could influence parental opinion and thus vaccine uptake. In ERN, the more the vaccine is available and the less the chances of its adverse events, it will help to increase the uptake of the vaccine and establish the vaccination as a consensus. In BAN, that paradigm changes as the degree distribution has a heavy tail.

Mixed populations, with two different cultures of sharing and perceiving their opinion about vaccination, could have a significant effect on vaccine uptake. Having a fraction of the population with low learning probability $q_1$ in a population with $q_2>q_1$ could decrease the resulting level of vaccine uptake and increase the size of the epidemic compared to a homogeneous population with the same learning probability $q_2$. 

Social studies using surveys and behavioral game experiments might consider the personal characteristics of each parent that give rise to directional learning probabilities $q_{i,j}$ and $q_{j,i}$. Population surveys can then be used to predict vaccine uptake levels. To elicit an increase in vaccine uptake, it is not enough to consider only the degree of parent link in the information network to spread awareness. Further efforts to promote social information exchange in social norm interventions will induce prosocial decisions about vaccination \cite{Cai2009,Sasaki2018,Goeschl18,Prentice2020}. According to our model, those efforts can lead to a consensus on the vaccination opinion and increase vaccine uptake even in the presence of a fraction of never-vaccinators and amid the vaccine's lack of safety and low availability.

\section{Supplementary Material}
A complete description of the model and methods, parameter values, and auxiliary figures is given in the Supplementary Online Material.
 \section*{References}
 %

\newpage
\setcounter{page}{1}
     \section*{S. Supplementary Material}
\beginsupplement
\subsection*{SI. Model}
In this section, we give a complete description of physical and social networks and the dynamics of disease and information spread.

\subsubsection*{Children physical network}
The first network $\mathcal{N}^C$ connects children to households, so it is the network in which disease transmission occurs from child to child. Links between children represent face-to-face possibilities through school, clubs, communal activities, etc. These links are assumed to be static; that is, they do not change over time once they are established. That would be reasonable since the modeled outbreaks span a few weeks. At this level, we only allow transmission through the network, and no other transmission routes (such as environmental ones) are allowed. In that network, we assume that nodes are formed from a number of $N$ households that are linked through physical transmission connections for each child (agent) in the household. The set of neighbors of the household $i$ on this network is denoted by $N_P(i)$ and their size is $n_P(i)$.

\begin{enumerate}
\item \textbf{Formation.}
A number of children $C_i$ (possibly zero) is randomly assigned to each household $i$, for $i=1,\ldots,N$, using a binomial distribution with parameters $n_c$ and $p_c$. Here, we consider two types of networks:

1) Erd\H{o}s-R\'{e}nyi (random) network (ERN): Two households $i$ and $j$ are physically connected with the probability $p\cdot \sqrt{C_i C_j}$, which is proportional to the geometric mean of the number of children in both households. That probability could be interpreted through Newton's law of gravity and deserves future wider investigation. A household with zero children will not have physical connections. A regular ERN is a special case where $C_i=1$ for all $i$. The parameter $p$ is selected to be less than $1/n_c$. 

2) Barab\'{a}si-Albert Network (BAN): In this network, we connect the first two households and then connect each of the following households, iteratively, according to the weighted degree distribution of the preceding households. That is, the household, say number $j$ ($j=3,4,\ldots,N$), is connected to one of the households $1,2,\ldots,j-1$ according to their weighted degree distribution $\dfrac{C_k d(k)}{\sum_{i=1}^{j-1} C_i d(i)}$, $k=1,2,\ldots,j-1$, where $d(k)$ is the degree of the household $k$. A household with zero children will not have physical connections. A regular BAN is a special case where $C_k=1$ for all $k$. 

See Fig. S1, for simulation instances of the overlapping ERNs and overlapping BANs. 

\item \textbf{Birth process.}
We assume that families decide to have a new baby (start a new pregnancy) with a probability dependent on the number of children per household. We assume that the probability of a new birth in a household $i$, if not already in pregnancy, is given by $$\dfrac{\sigma}{1+\exp(k(C_i-C^*))}.$$ After 280 days $C_i$ is updated to $C_i+1$. The parameter $\sigma$ is the population-level birth rate. The parameter $k$ measures the sensitivity of the probability to number of children in the household and $C^*$ is the median for that probability. We estimated $k$ so as to have that probability approaching zero as $C_i$ approaches $n_C$.

Humans' gestation period is assumed to be 280 days. After delivery, parents either vaccinate or do not vaccinate their new babies. The first 280 days of the simulation of the model are discarded as a burn-in period. 

\item \textbf{Disease transmission within and between households.}
Initially, a number of children $I_0$ are randomly and uniformly selected to be infected. A new infection happens in household, say number $i$, on any single day due to transmission within the household or between households with probability $1-(1-\beta_h)^{I(i)} \cdot(1-\beta)^{n_I(i)/C_i}$; where $\beta$ is the probability to infect a child in another household (through the physical network), $\beta_h$ is the probability to infect a sibling within the same household, and $I(i)=\sum_{k=1}^{\ell}I(i,k)$ is the total number of infected siblings in the same household where $I(i,k)$ is the number of infected siblings at day/stage $k$ of the incubation period. The number $n_I(i)=\sum_{j\in N_P(i)} I(j)$ is the number of infected children connected to that household $i$ through the network. The number $n_I(i)$ is divided by $C_i$ in the probability of infection formula to reflect the approximating assumption that the children in a household have the same number of friends on average. Multiple infections can happen on the same day in the network and the numbers $n_I(i)$ and $I(i)$ are updated every day for all $i$, $i=1,2,\ldots,N$.

\item \textbf{Disease progression.}
The incubation period is assumed to be distributed according to a discretized exponential distribution with mean length of $m_p$ days and maximum $\ell$ days. An infected child on day $j$ after infection moves either to the end of the incubation period (recovers) or stays infected into the following day with probabilities $Q_j$ and $1-Q_j$, respectively, such that 
$$Q_j=\dfrac{\int_j^{j+1} f(x) dx}{\int_j^{\infty} f(x) dx}$$
for $j=0,1,\ldots,\ell-1$, and $Q_{\ell}=1$, where $f(x)$ is the probability density function of the exponential distribution.
\end{enumerate}

\subsubsection*{Parental social network}
The second network $\mathcal{N}^P$ overlaps with the children's physical network and assumes larger social/information links between parents in households. Through social networks, parents share information and perceive opinions to and from network neighbors who are not necessarily physical neighbors but possibly friends, relatives, colleagues, online friends, etc. The set of neighbors of household $i$ on this network is denoted by $N_S(i)$ and their size is $n_S(i)$. To form an overlapping social network with the children's network; links in the children's physical network are retained randomly with a probability of $p_{re}$. Other links are added to any one of the updated non-connected households with probability $p_{ad}$. 

\subsection*{SII. Parameter Values}
The model simulation is run for a large number of times for different sets of values of selected parameters. A parameterization of the model is given in Table 1. The parameters were assigned using the literature, calibration and guesstimation. The time unit is given in days. 

\begin{table}[ht]
\caption{Model parameters, their description and base values.}\label{tabel1}
\centering
\resizebox{\textwidth}{!}{ \begin{tabular}{||c | l| c||} 
 \hline
  \textbf{Parameter} & \textbf{Description} & \textbf{Base values} \\ [0.5ex] 
 \hline\hline
 
$N$  & Number of households (nodes) & $100,000$ \\
\hline

$\sigma$  & Birth rate  & $.005 $ \\
\hline

$n_c$  & Maximum number of children per household & $7$ \\
\hline

$p_c $  & Initial probability of having a child in a household & $.4$ \\
\hline

$k $  & Sensitivity of pregnancy probability to number of children & $2.5$ \\
\hline

$C^* $  & Median number of children in pregnancy probability & $2$ \\
\hline

$p $  & Physical connection probability & $.00013$ \\
\hline



$p_{re} $  & Probability of retaining a physical connection to form the social network & $.6$ \\
\hline 

$p_{ad} $  & Probability of adding a connection to the social network & $.0004$ \\
\hline 

$I_0 $  & Initial number of infected children & $10$ \\
\hline
 
$\beta $  & Human to human transmission probability through network \cite{Phillips2020} & $.06$ \\
\hline

$h $  & Human to human factor of transmission probability within household & $1.5$ \\
\hline

$m_p$, $i_p $  & Mean and maximum length of infectious period & $11$, $16$ days\\
\hline

$NV_0$  & Initial number of all time non-vaccinators & $5\%\cdot N$\\
\hline

$MS_0$  & Initial number of mover-stayer parents & $N-NV_0$\\
\hline

$P_S $  & Probability to vaccinate with social learning & $-$ \\
\hline

$e $  & Vaccine efficacy & $.95$ \\
\hline

$p_{adv} $  & Probability of vaccine adverse event & $.0001-.01$ \\
\hline

$\alpha $ & Degree of relevance of infectiousness to the decision to vaccinate & $.001$ \\
\hline

$\gamma $  & Degree of relevance of vaccine adverse events to the decision to vaccinate & $.01$ \\
\hline

$q$  & Probability to give a correct perception of vaccination  & $0-1$ \\
\hline

$\delta$  & Injunctive social norm  & $.025-.225$ \\
\hline

$\rho$  & Probability to get access to vaccination  & $.001-.01$ \\
\hline

\end{tabular}}

\end{table}

\newpage
\vskip2cm
\subsection*{SIII. Supplementary Figures}

\begin{figure}[H]
\subfigure[]{\includegraphics[width=8.5cm]{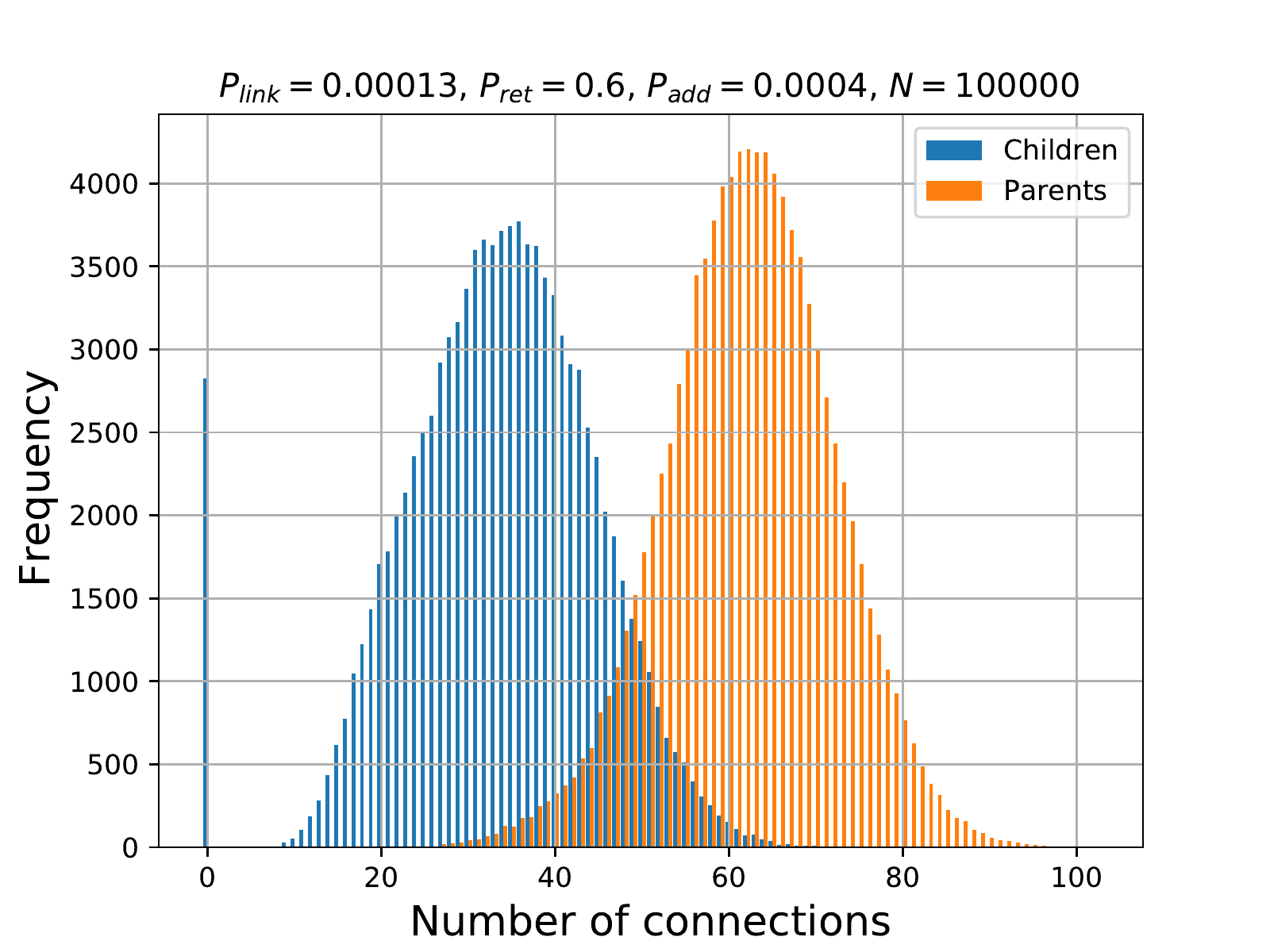}} 
\subfigure[]{\includegraphics[width=8.5cm]{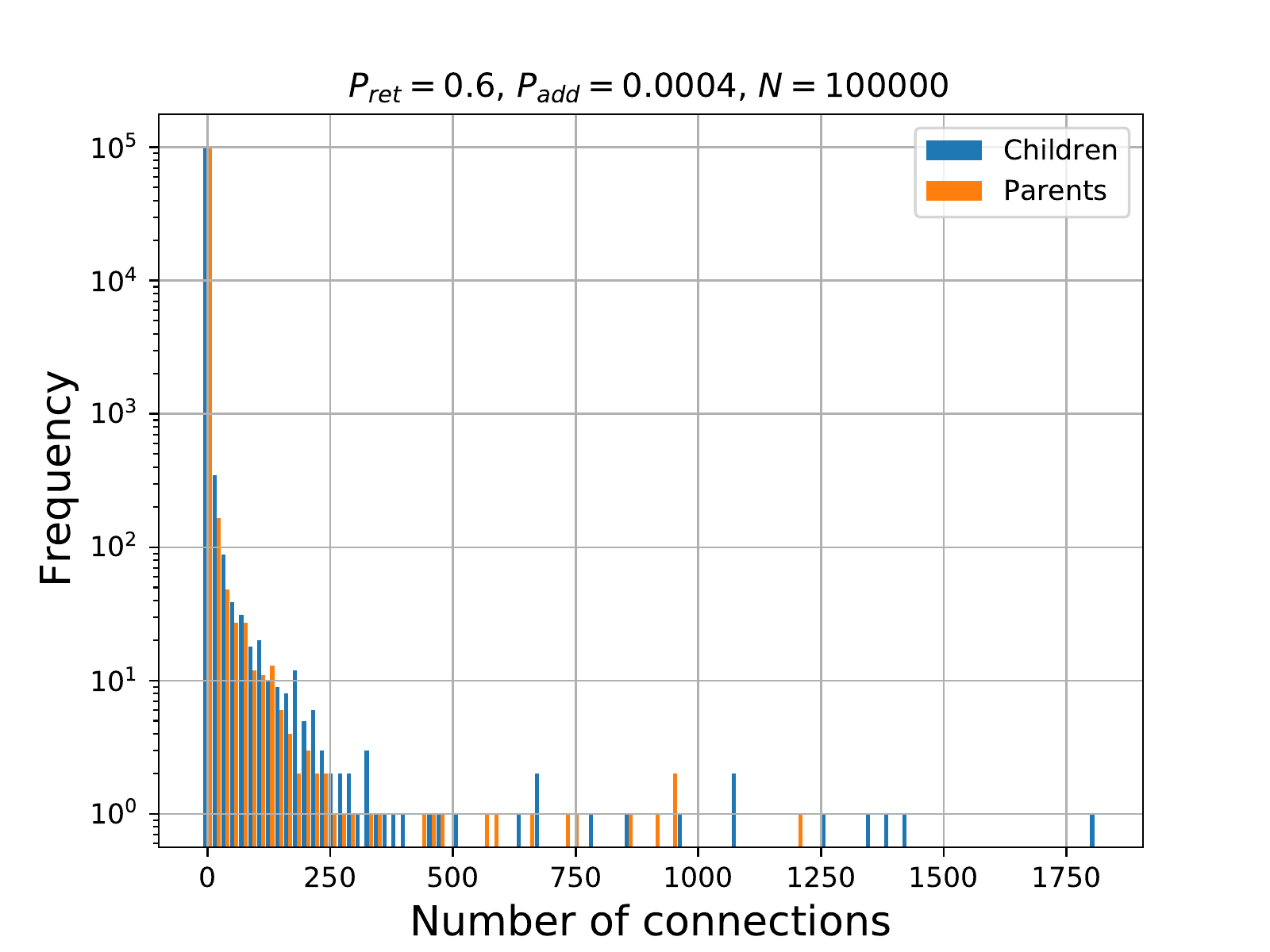}}

\caption{{\bf Fig. S1. Degree histogram for a simulation of Erd\H{o}s-R\'{e}nyi (random) network model (ERN) in (a) and Barab\'{a}si-Albert network model (BAN) in (b).} }\label{figS0}
\end{figure}

\begin{figure}[H]
\subfigure[]{\includegraphics[width=5.5cm]{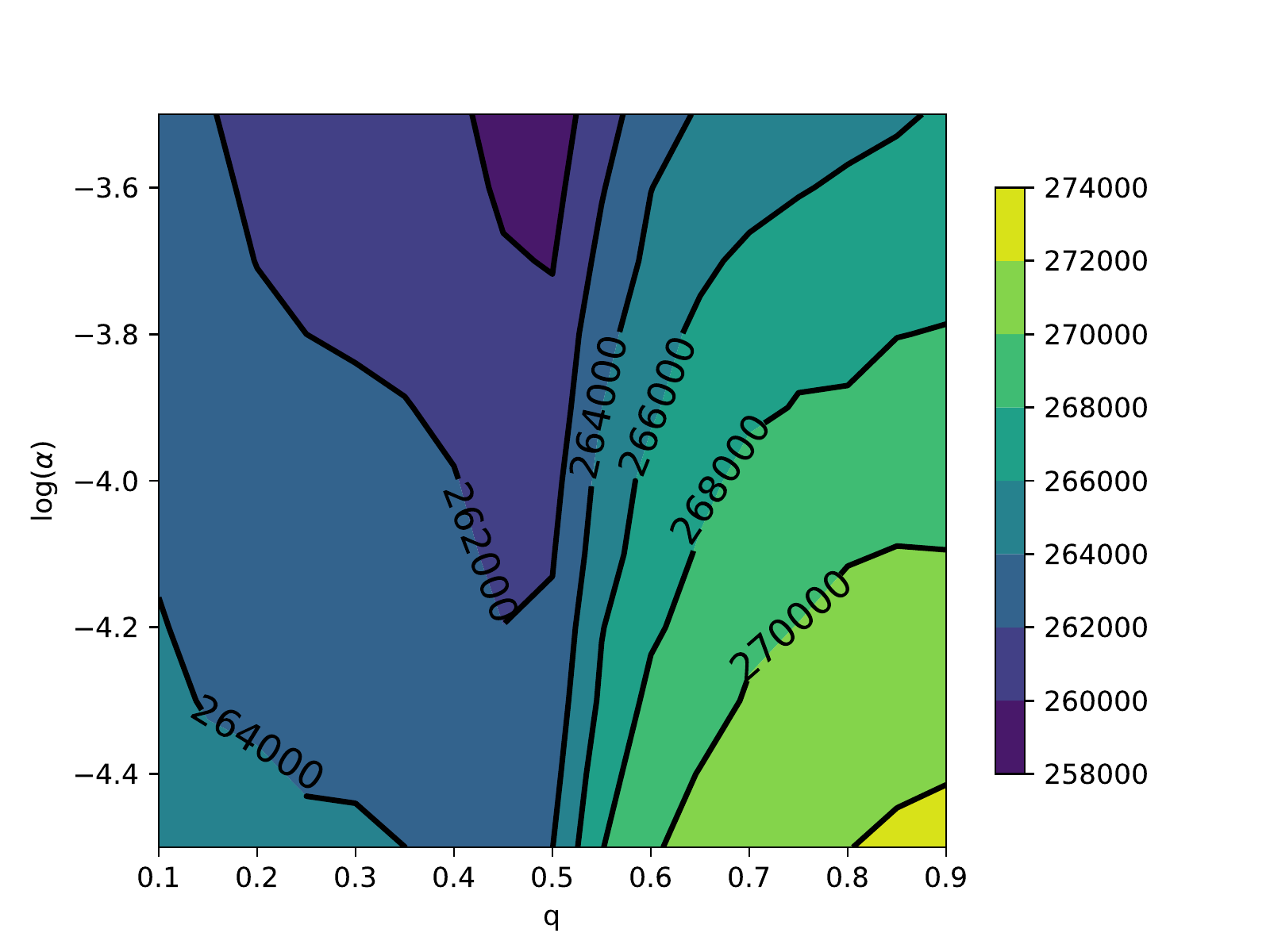}} 
\subfigure[]{\includegraphics[width=5.5cm]{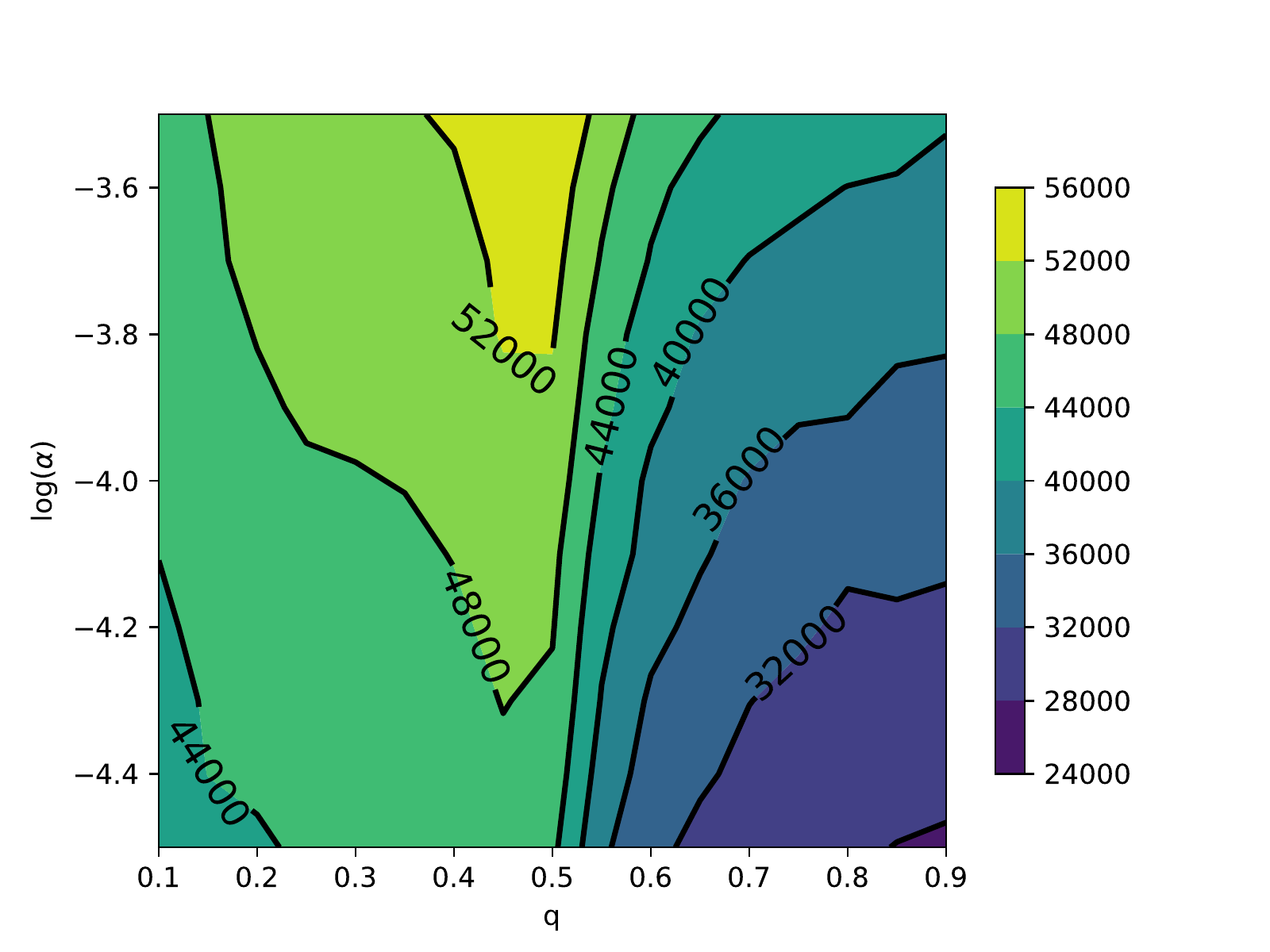}}
\subfigure[]{\includegraphics[width=5.5cm]{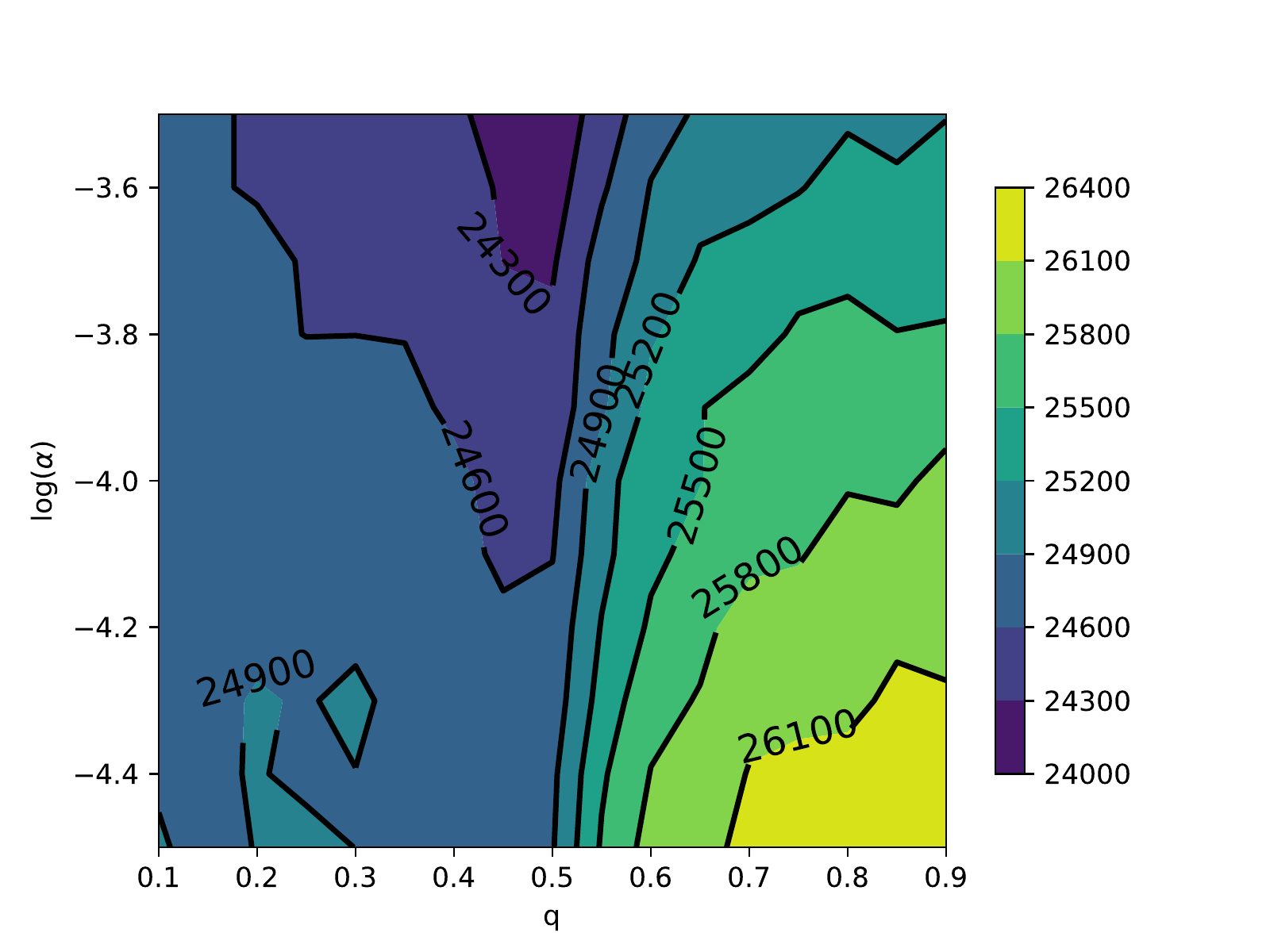}}
\subfigure[]{\includegraphics[width=5.5cm]{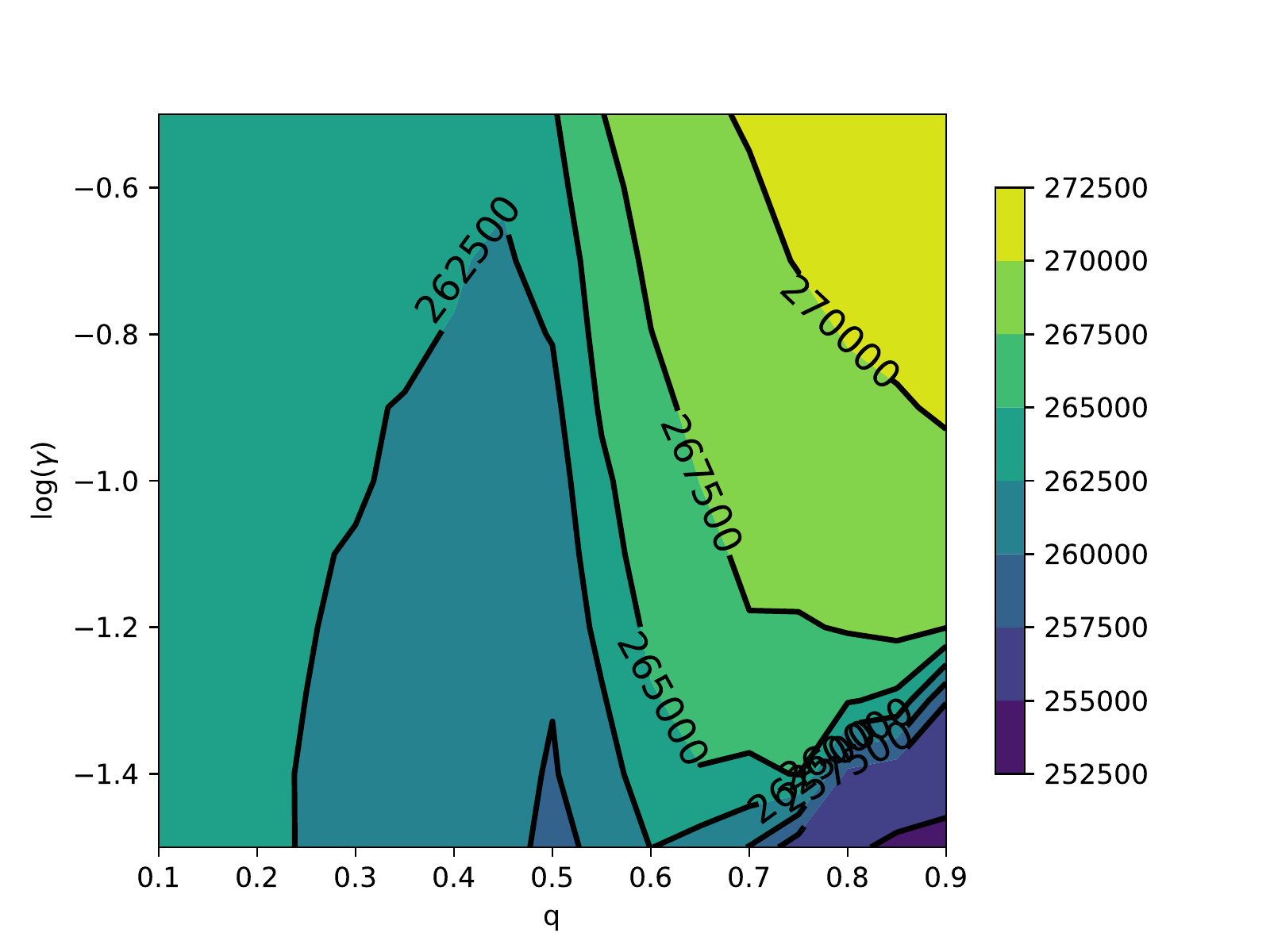}}
\subfigure[]{\includegraphics[width=5.5cm]{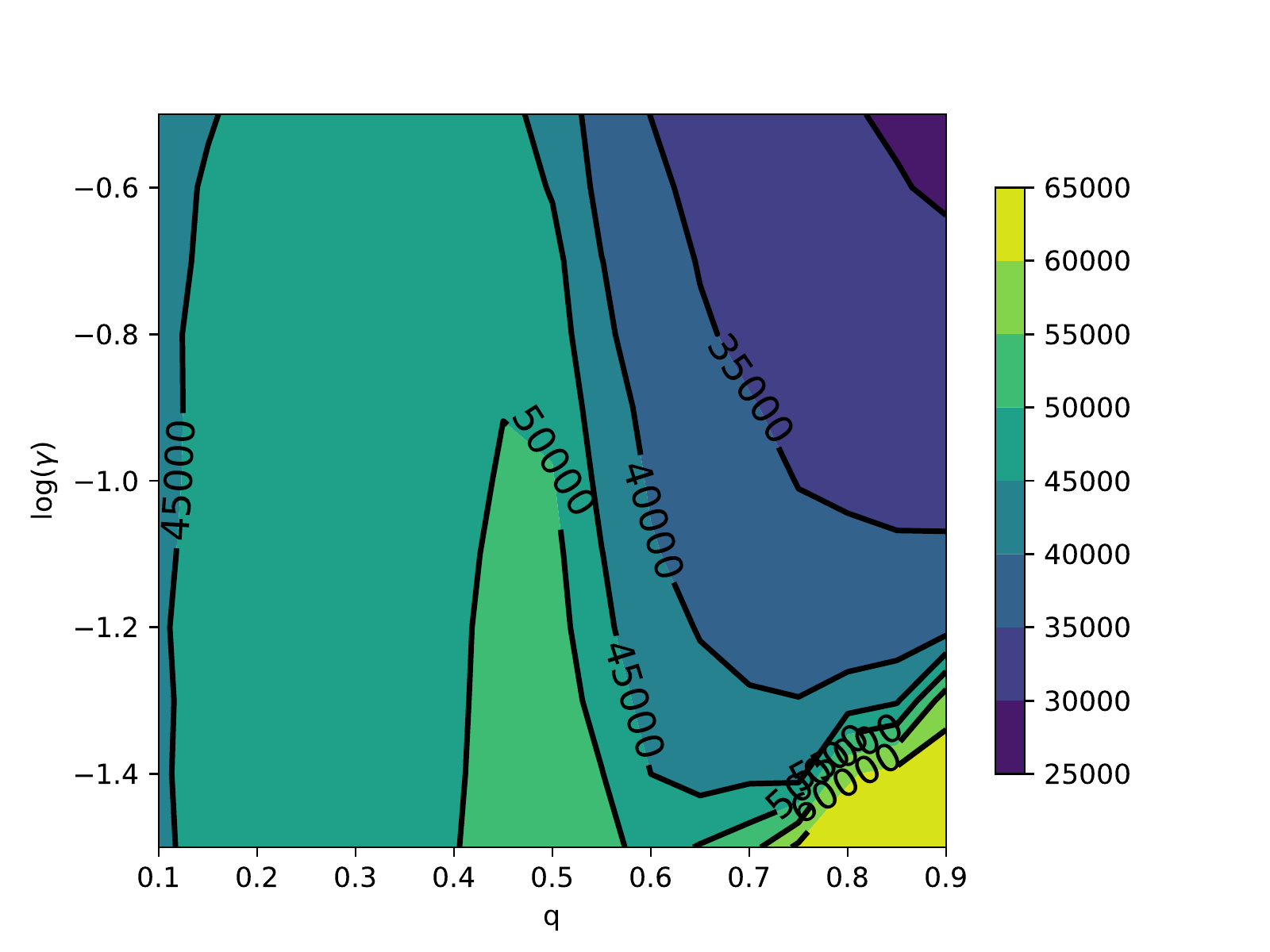}} \subfigure[]{\includegraphics[width=5.5cm]{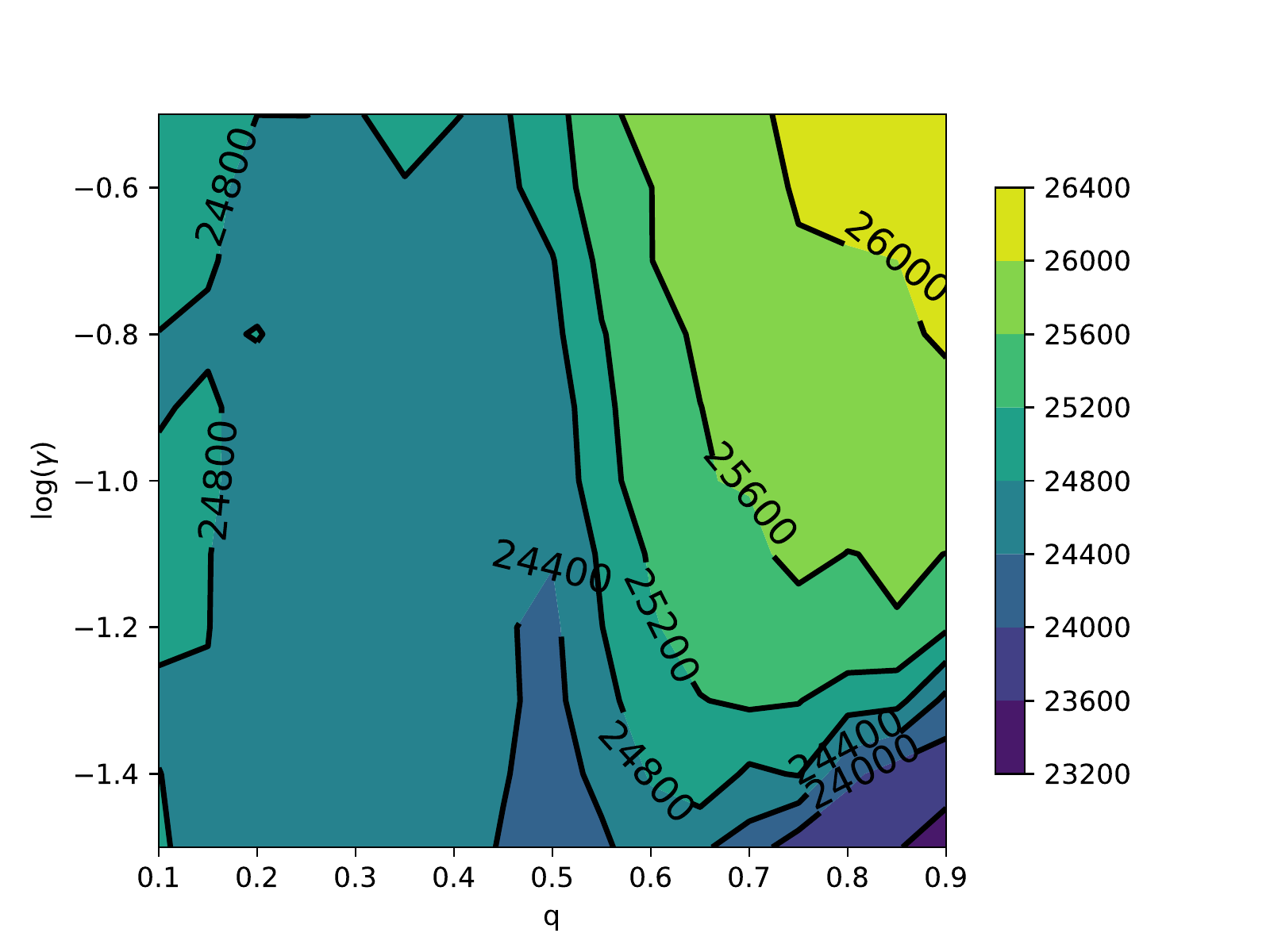}}
\caption{{\bf Fig. S2. Parameter planes of $\alpha$ against $q$ and $\gamma$ against $q$ for sizes of the epidemic, the total number of vaccinated children, and the peak of the epidemic on Erd\H{o}s-R\'{e}nyi (random) network model (ERN).} In all of the simulations, the median value of the simulations are used to plot the parameter planes that are performed at $P_{adv}=.001$ and $\rho=.01$.}\label{figS1}
\end{figure}

\begin{figure}[H]
\subfigure[]{\includegraphics[width=5.5cm]{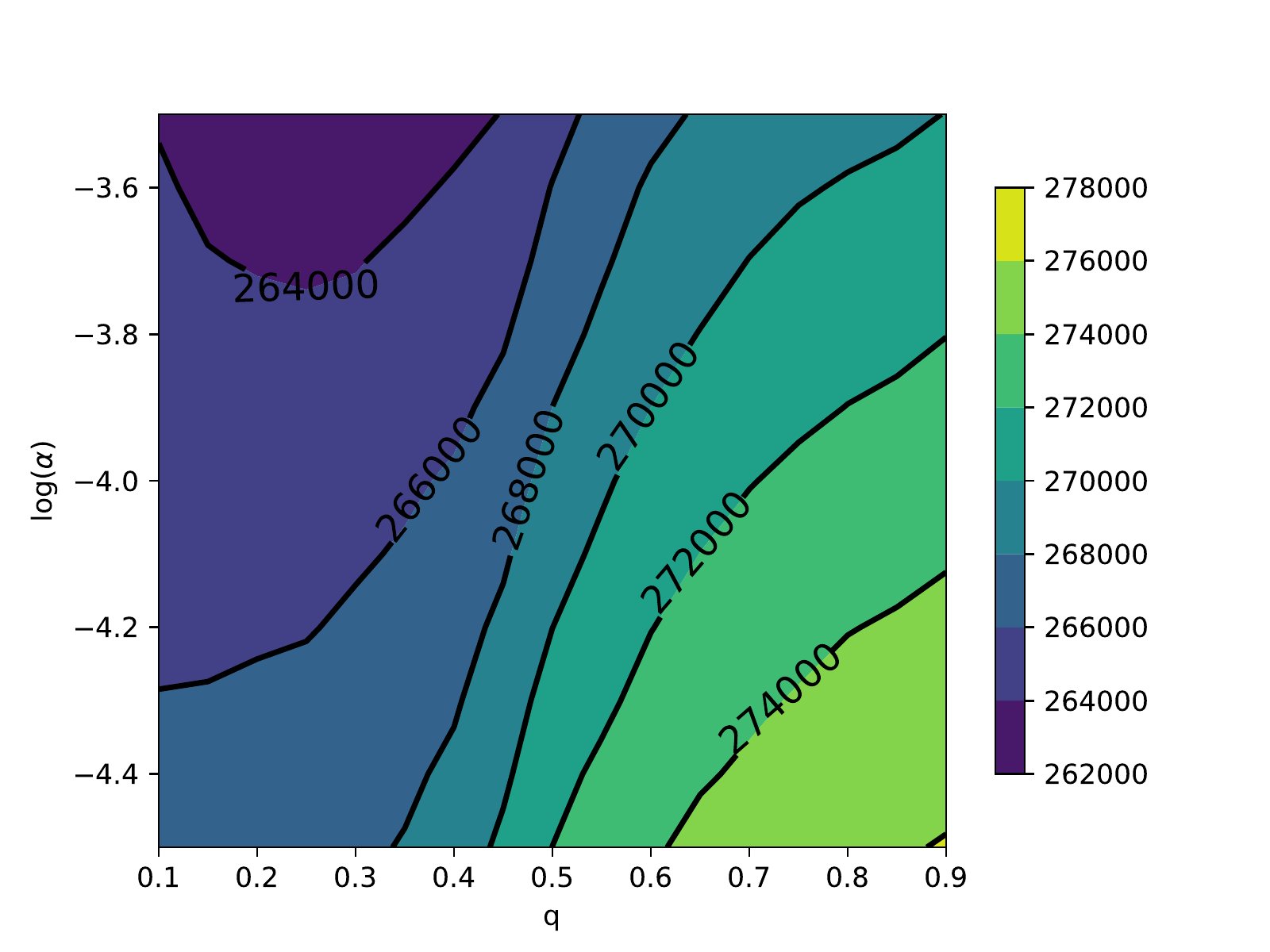}} 
\subfigure[]{\includegraphics[width=5.5cm]{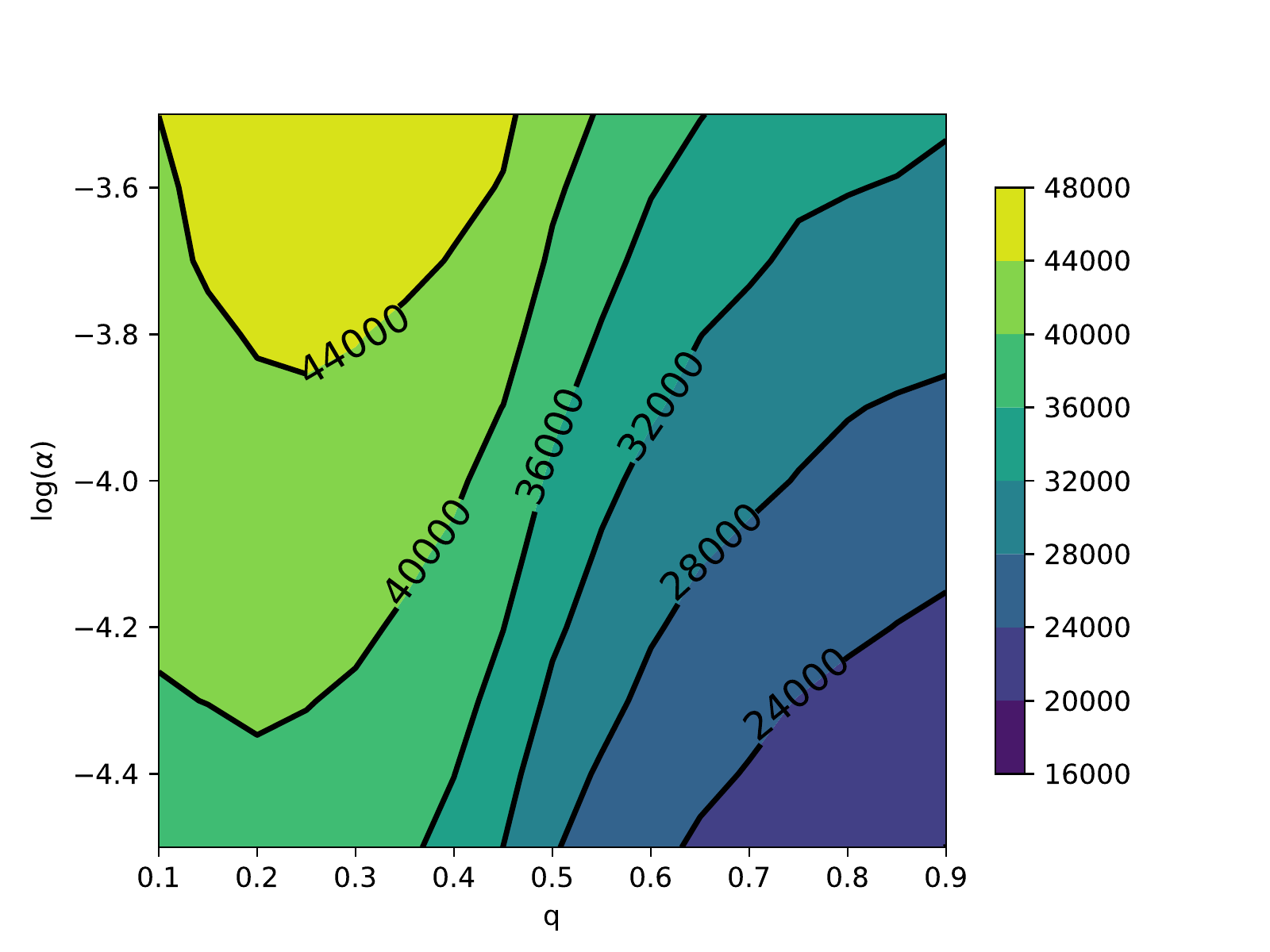}}
\subfigure[]{\includegraphics[width=5.5cm]{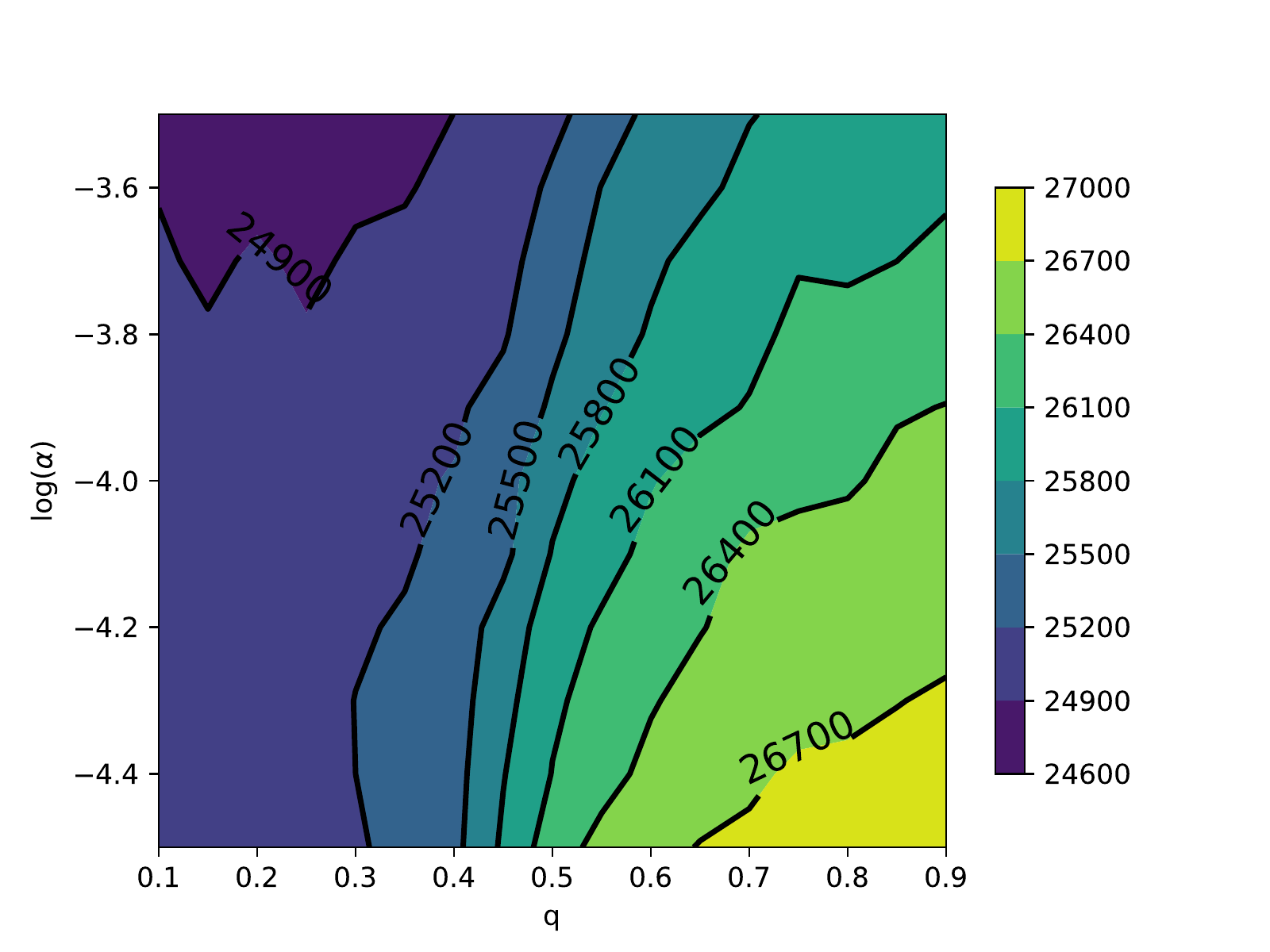}}
\subfigure[]{\includegraphics[width=5.5cm]{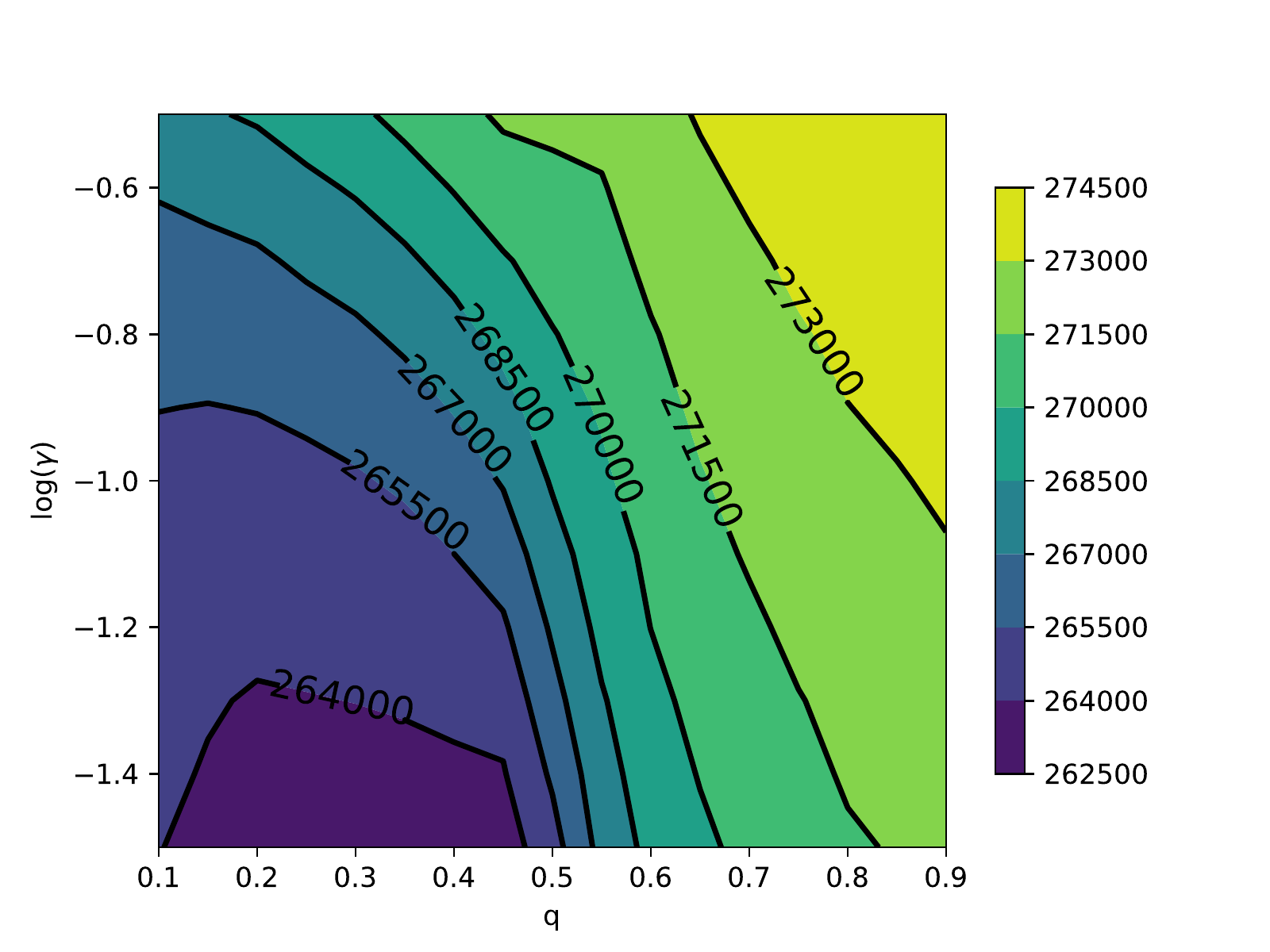}}
\subfigure[]{\includegraphics[width=5.5cm]{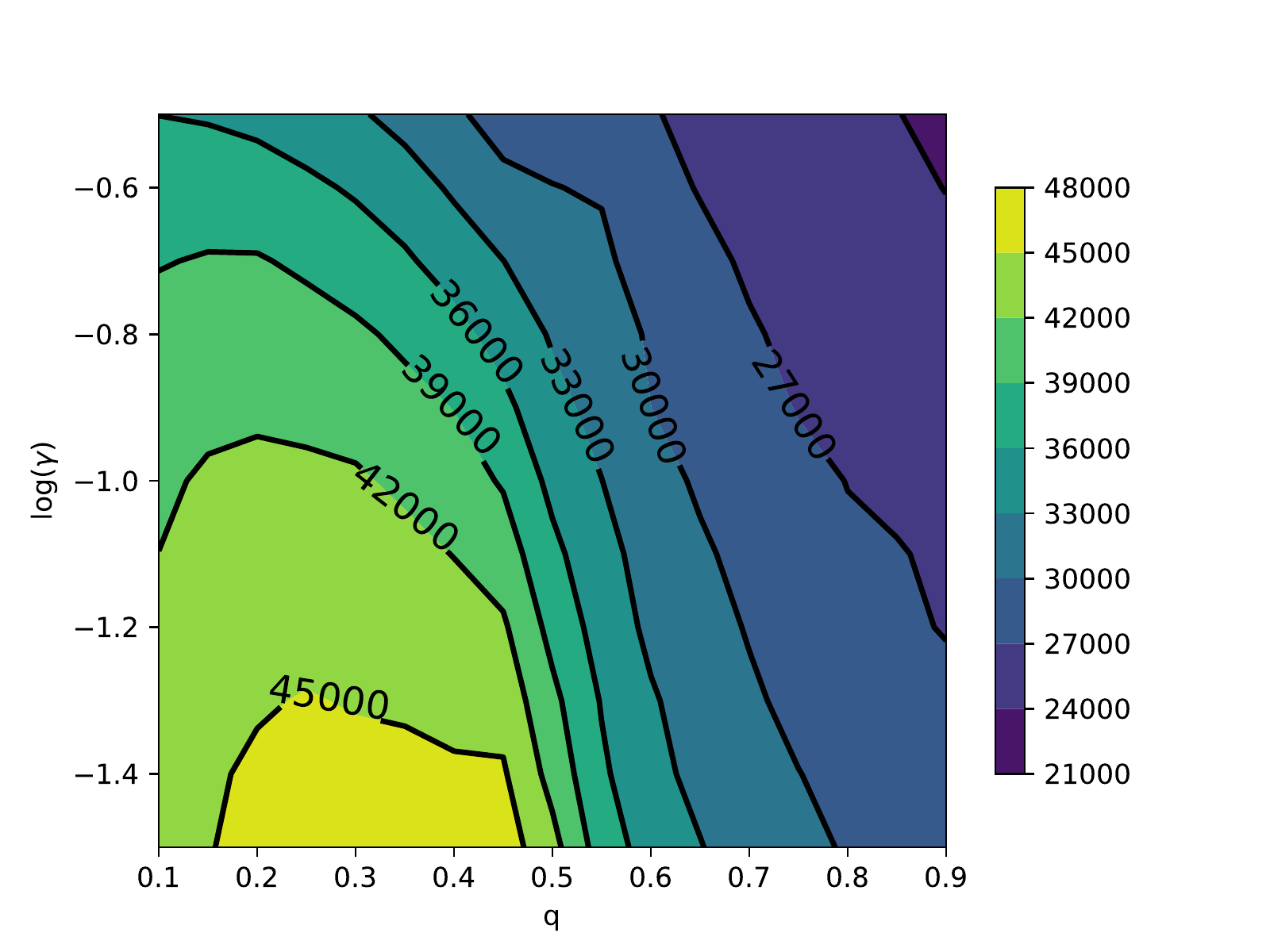}} \subfigure[]{\includegraphics[width=5.5cm]{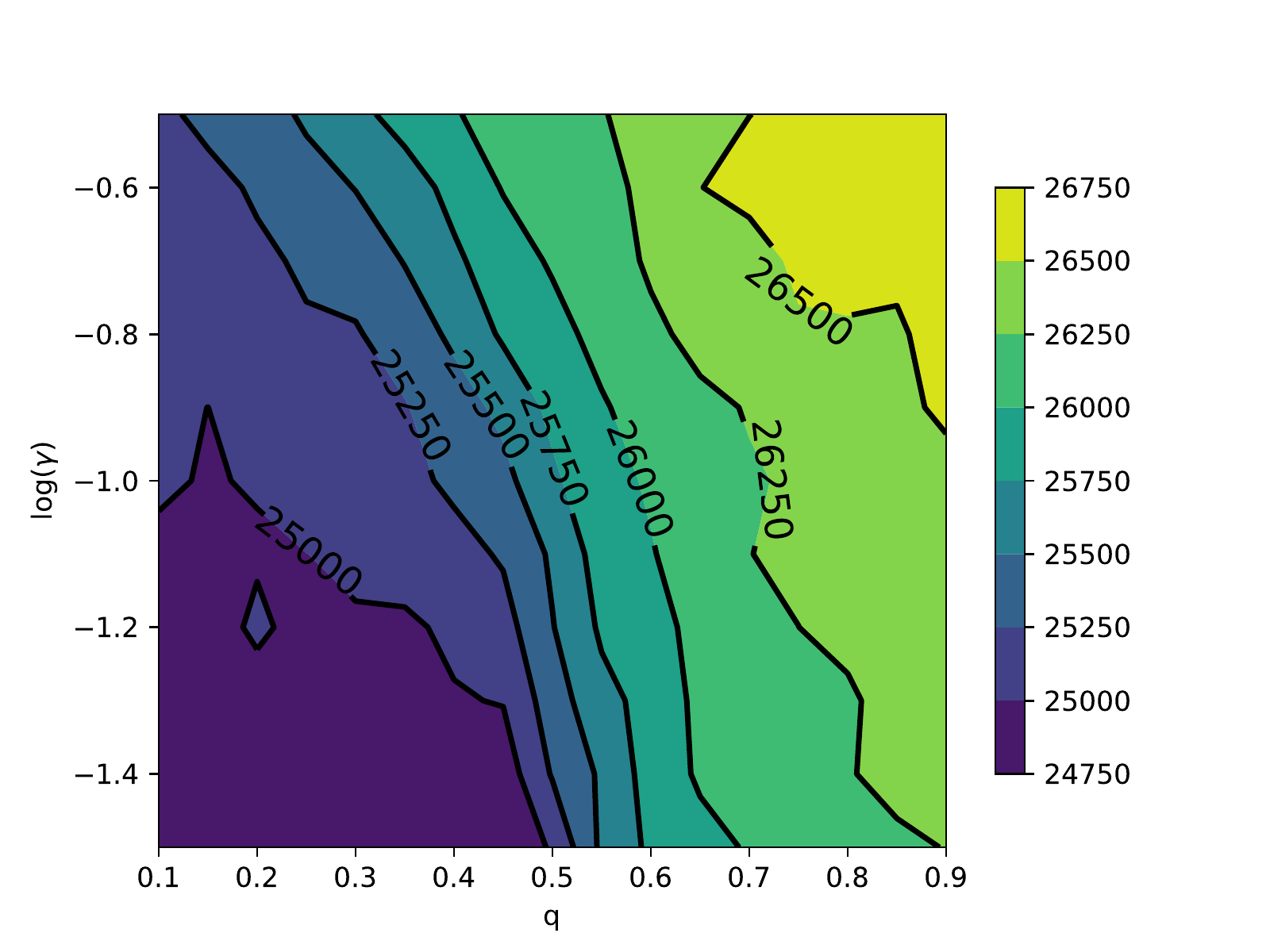}}
\caption{{\bf Fig. S3. Parameter planes of $\alpha$ against $q$ and $\gamma$ against $q$ for sizes of the epidemic, the total number of vaccinated children, and the peak of the epidemic on Erd\H{o}s-R\'{e}nyi (random) network model (ERN).} In all of the simulations, the median value of the simulations are used to plot the parameter planes that are performed at $P_{adv}=.01$ and $\rho=.01$.}\label{figS2}
\end{figure}

\begin{figure}[H]
\subfigure[]{\includegraphics[width=5.5cm]{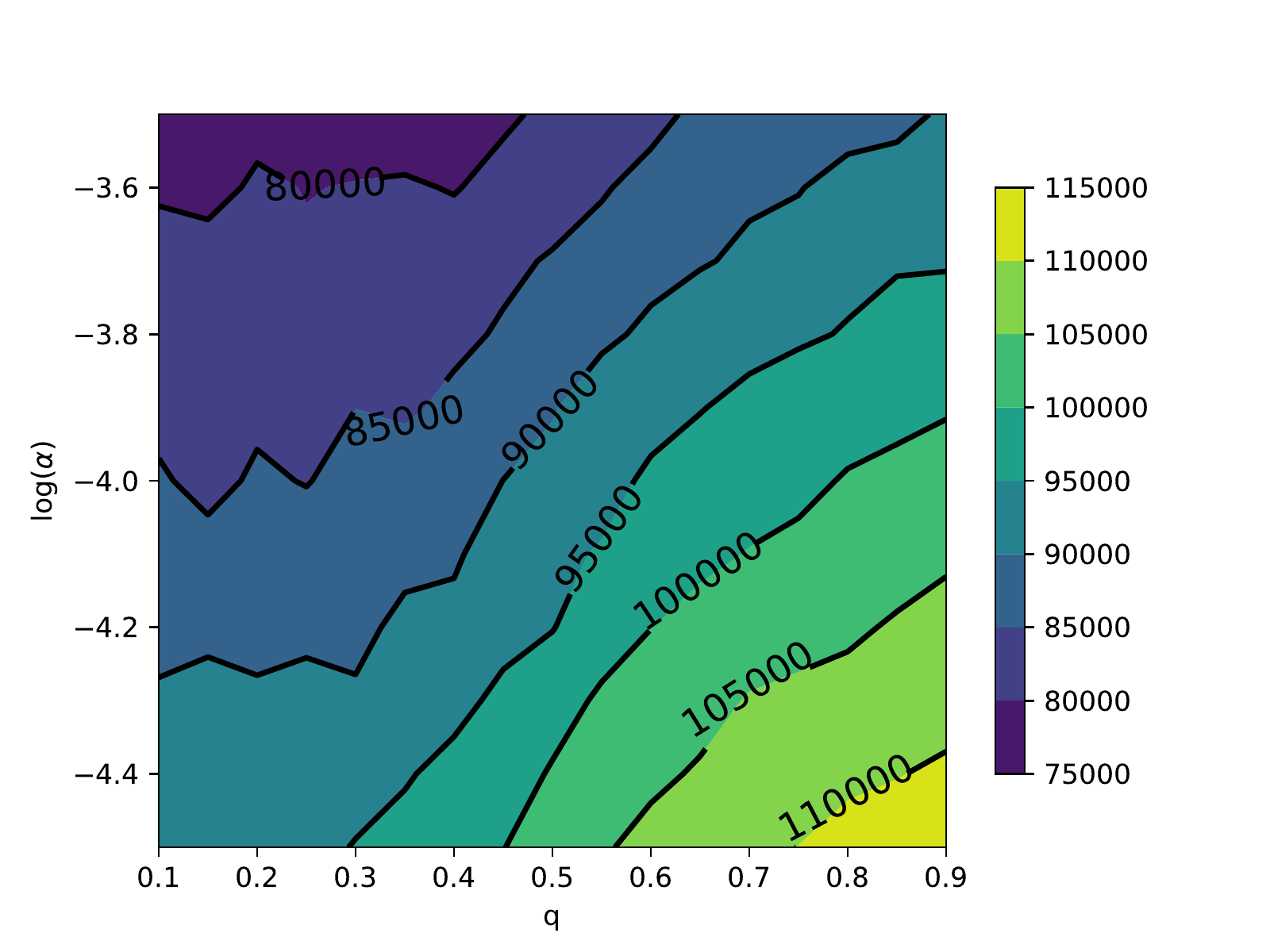}} 
\subfigure[]{\includegraphics[width=5.5cm]{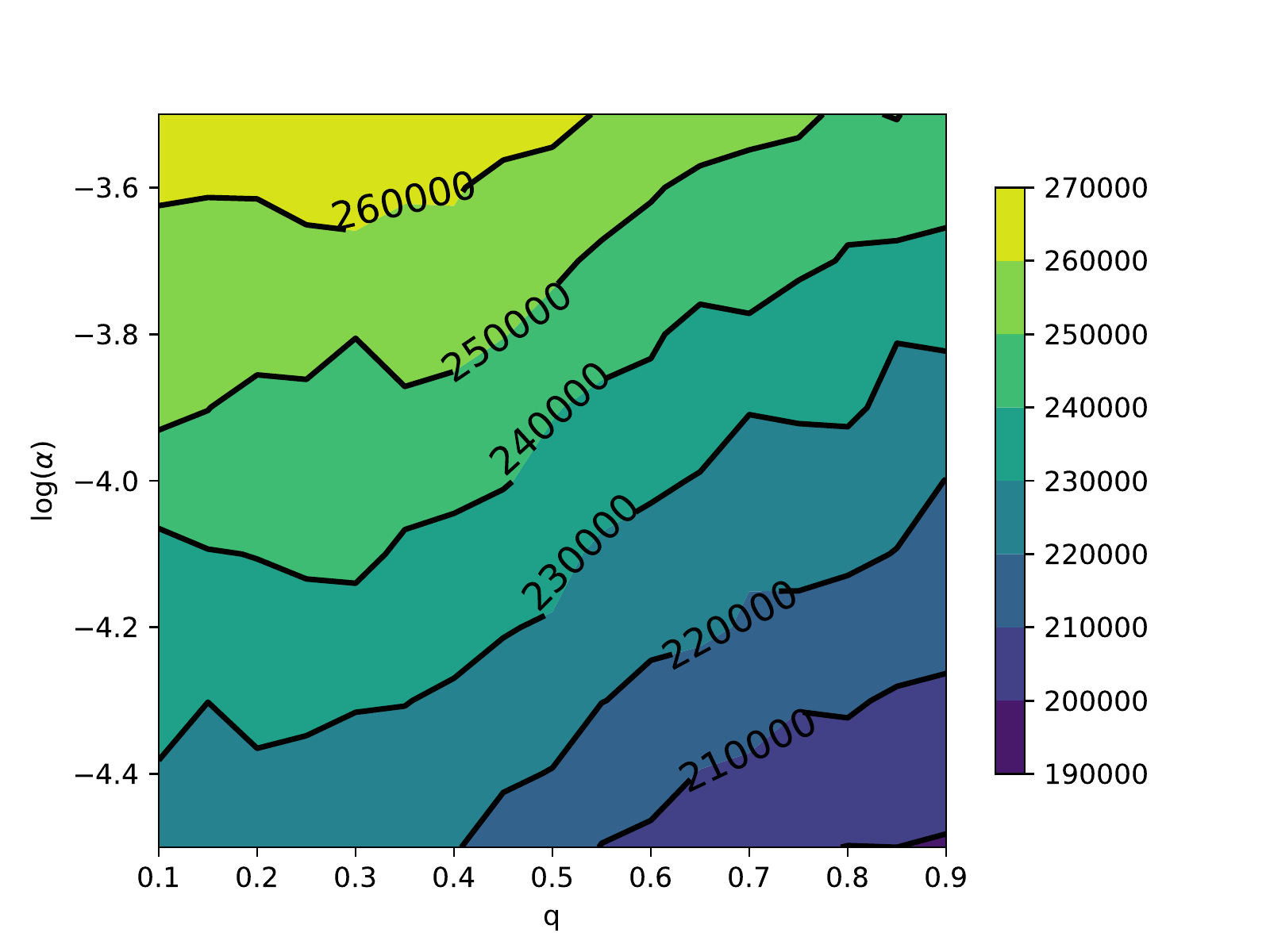}}
\subfigure[]{\includegraphics[width=5.5cm]{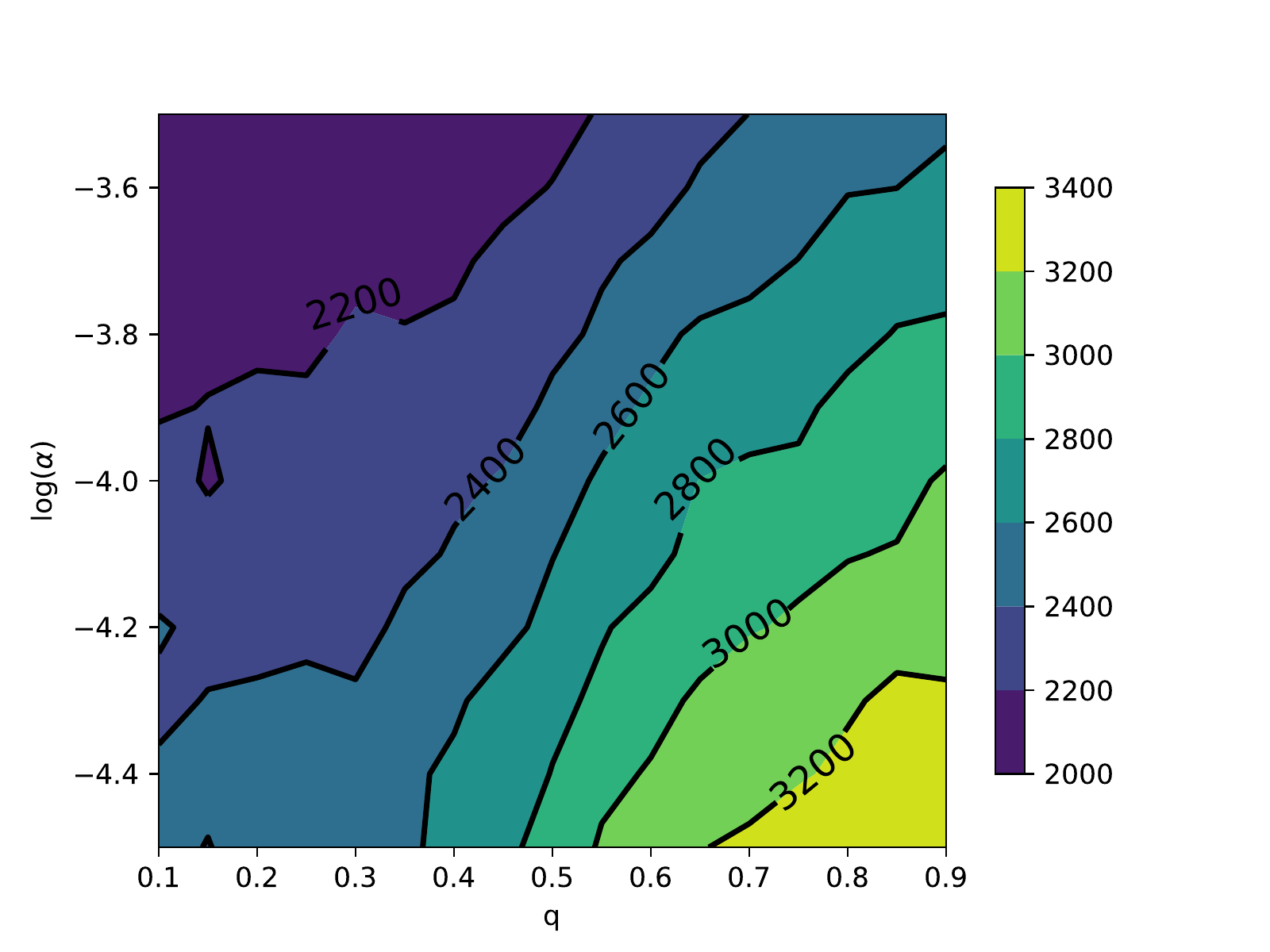}}
\subfigure[]{\includegraphics[width=5.5cm]{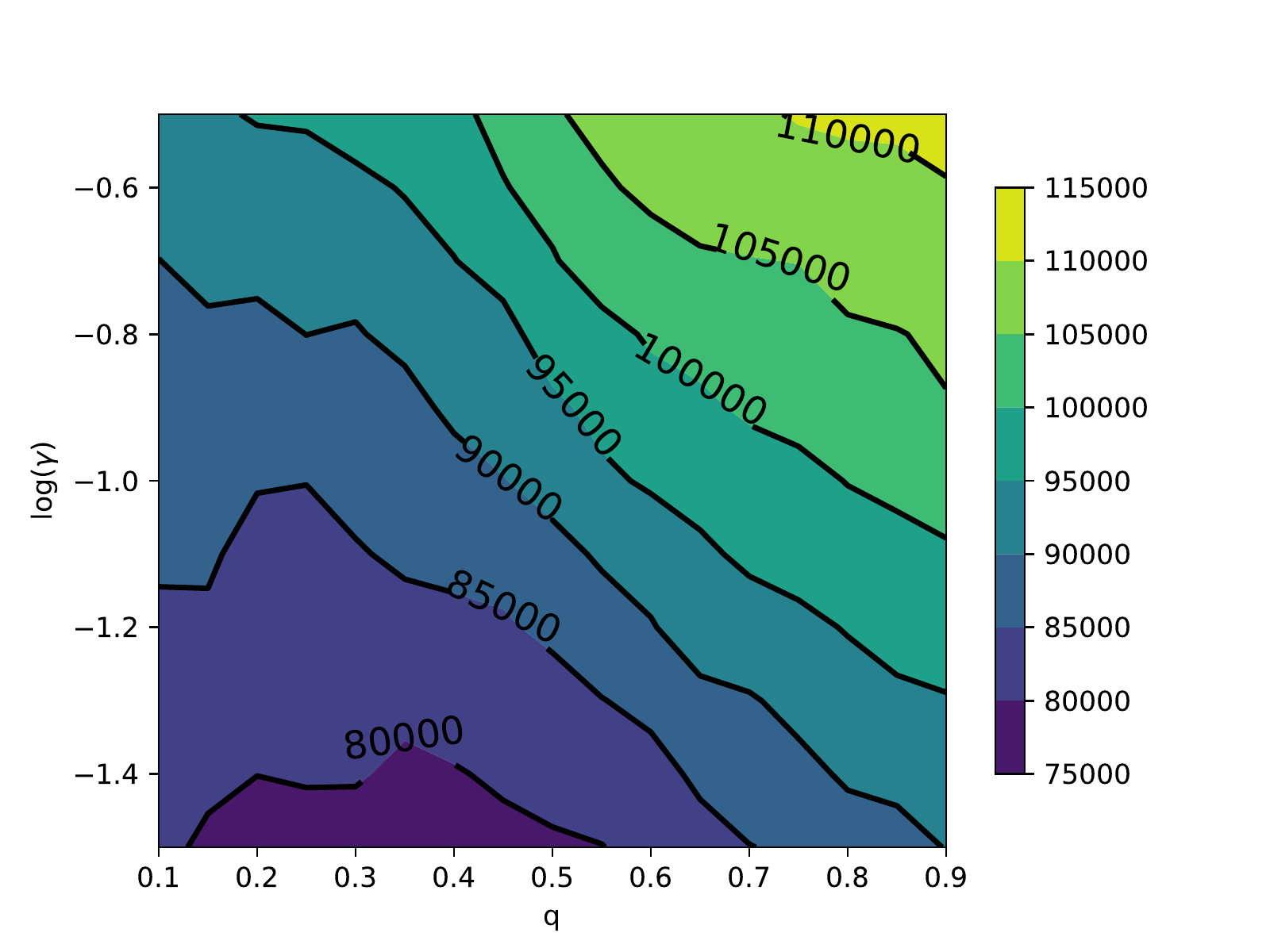}}
\subfigure[]{\includegraphics[width=5.5cm]{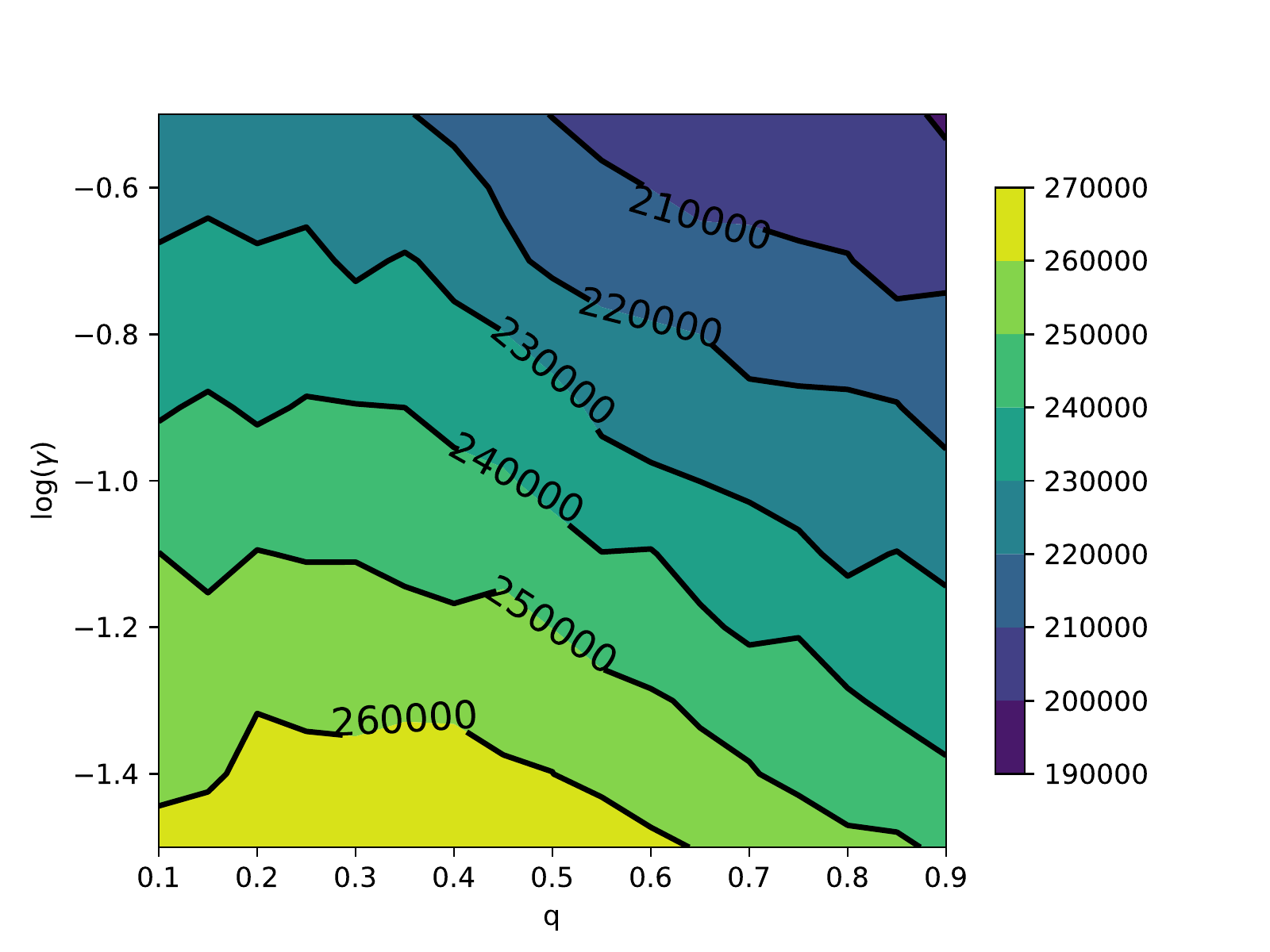}} \subfigure[]{\includegraphics[width=5.5cm]{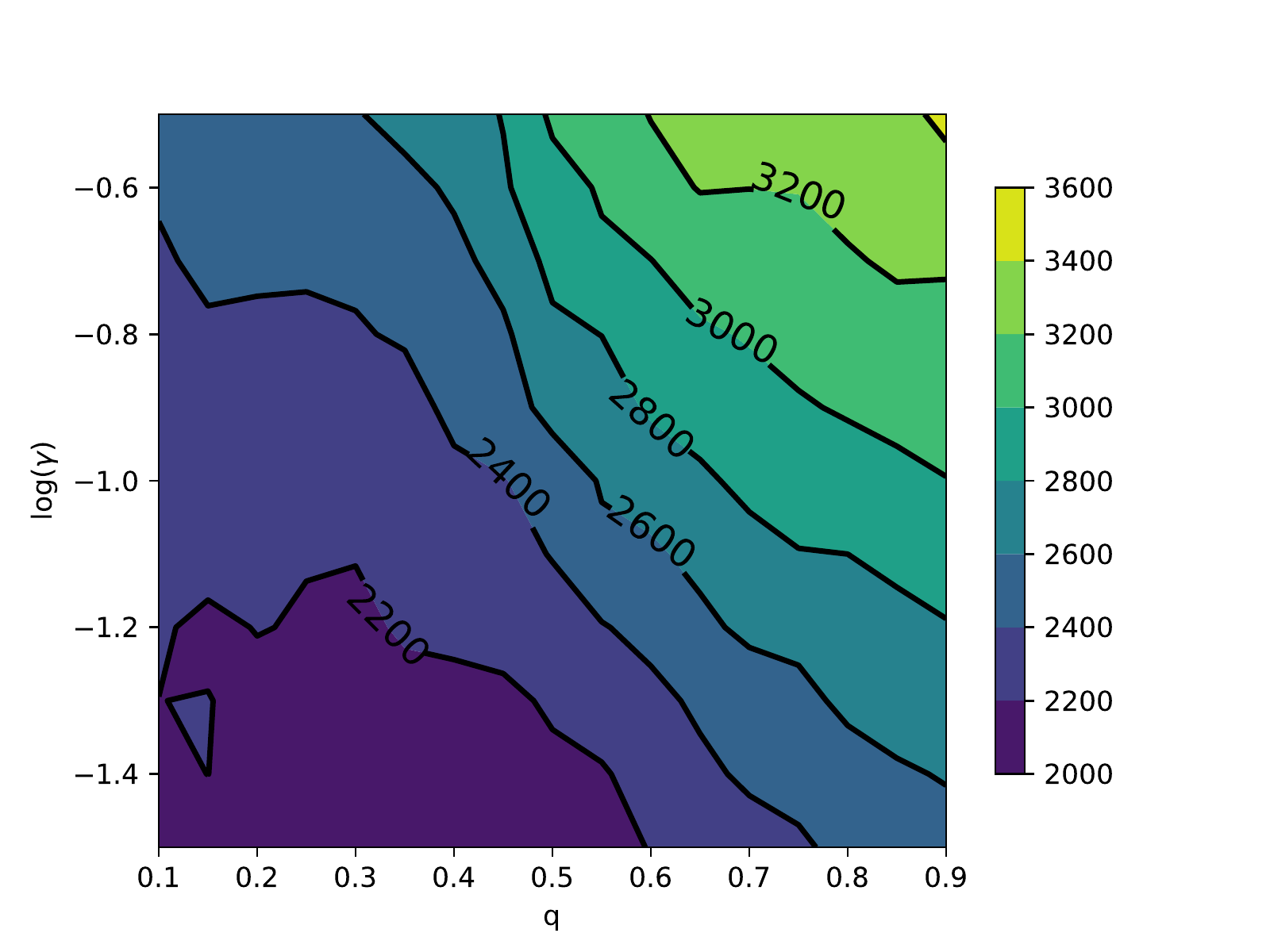}}
\caption{{\bf Fig. S4. Parameter planes of $\alpha$ against $q$ and $\gamma$ against $q$ for sizes of the epidemic, the total number of vaccinated children, and the peak of the epidemic on Barab\'{a}si-Albert network model (BAN).} In all of the simulations, the median value of the simulations are used to plot the parameter planes that are performed at $P_{adv}=.001$ and $\rho=.01$.}\label{figS3}
\end{figure}

\begin{figure}[H]
\subfigure[]{\includegraphics[width=5.5cm]{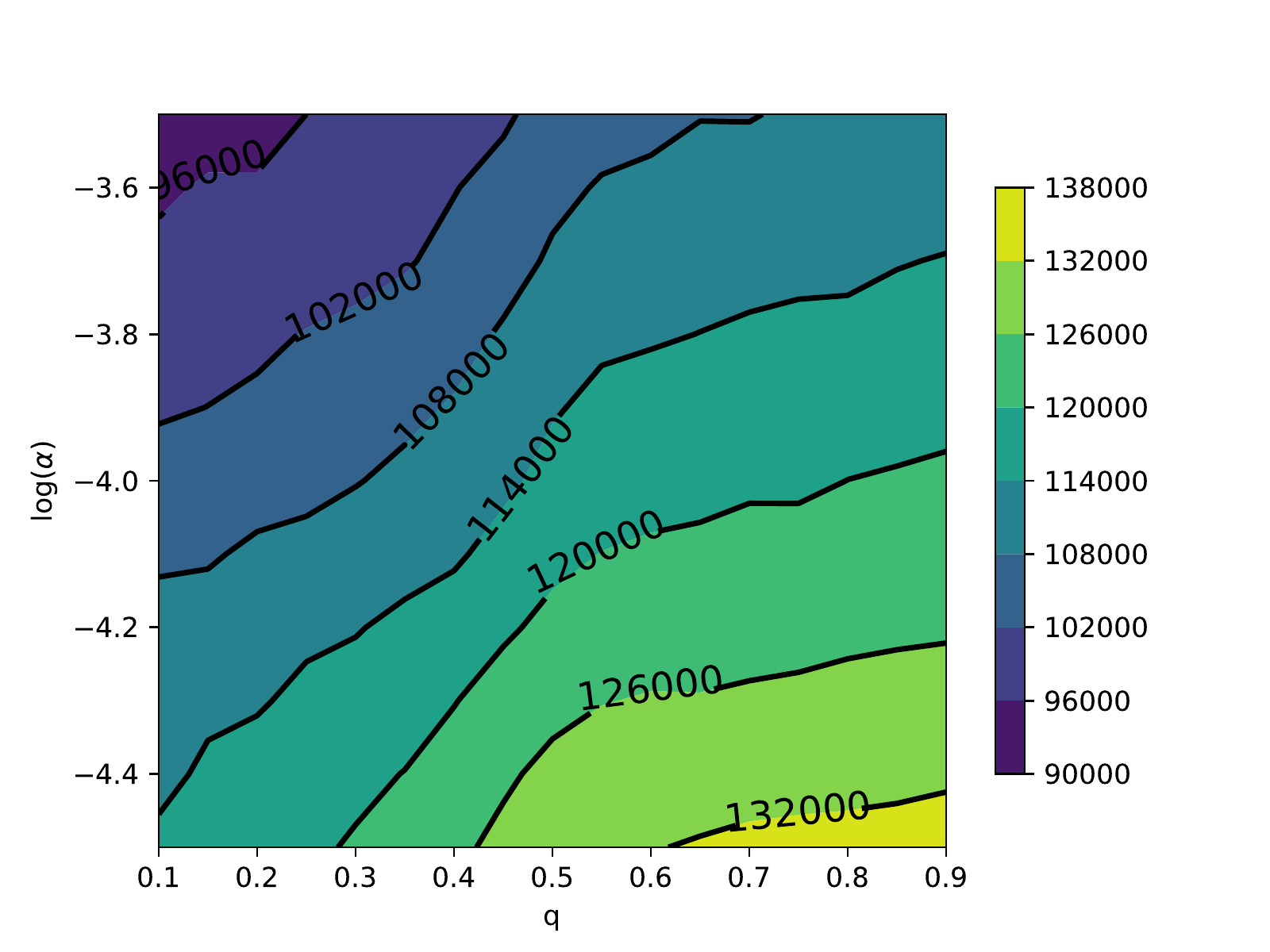}} 
\subfigure[]{\includegraphics[width=5.5cm]{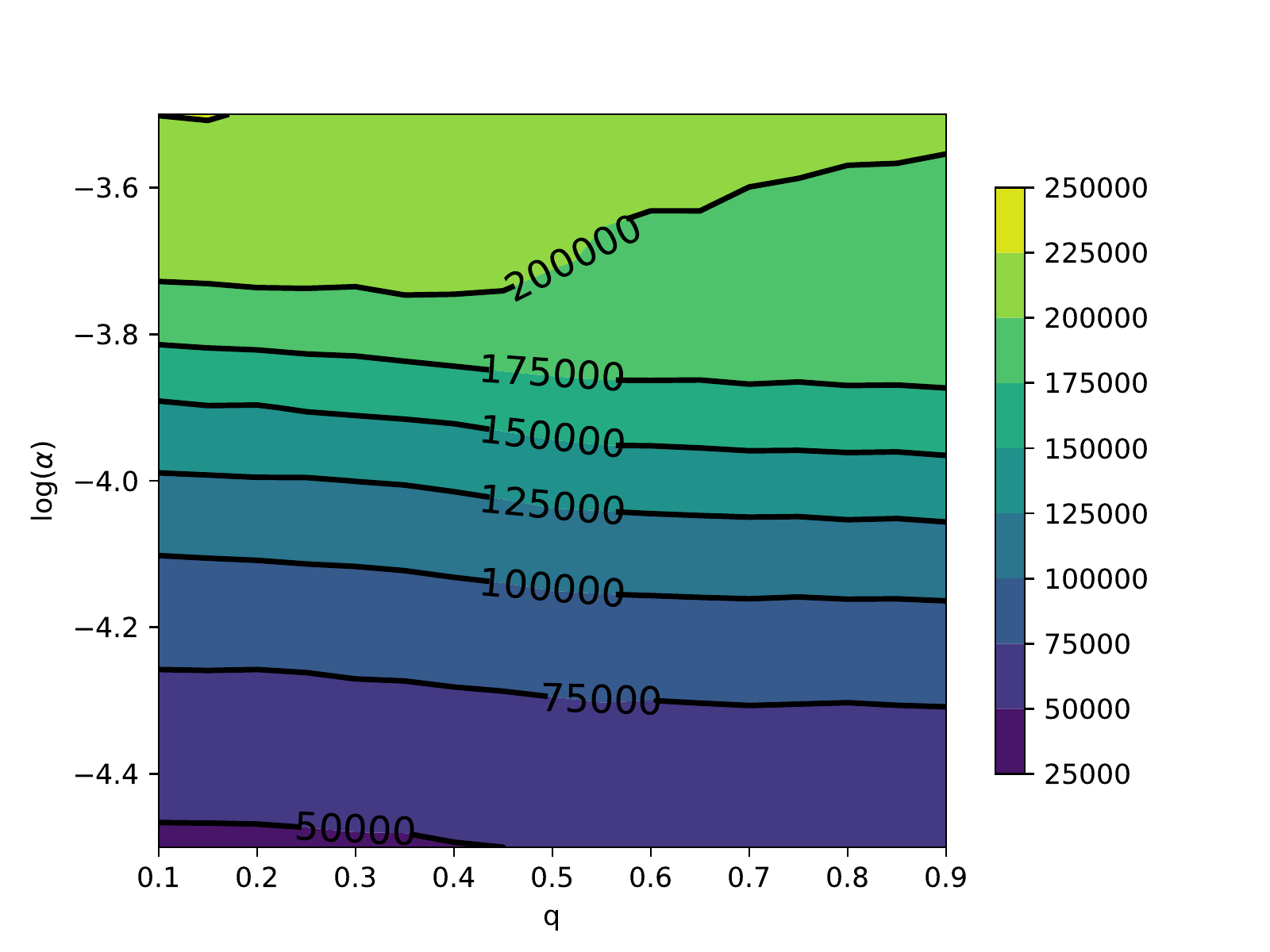}}
\subfigure[]{\includegraphics[width=5.5cm]{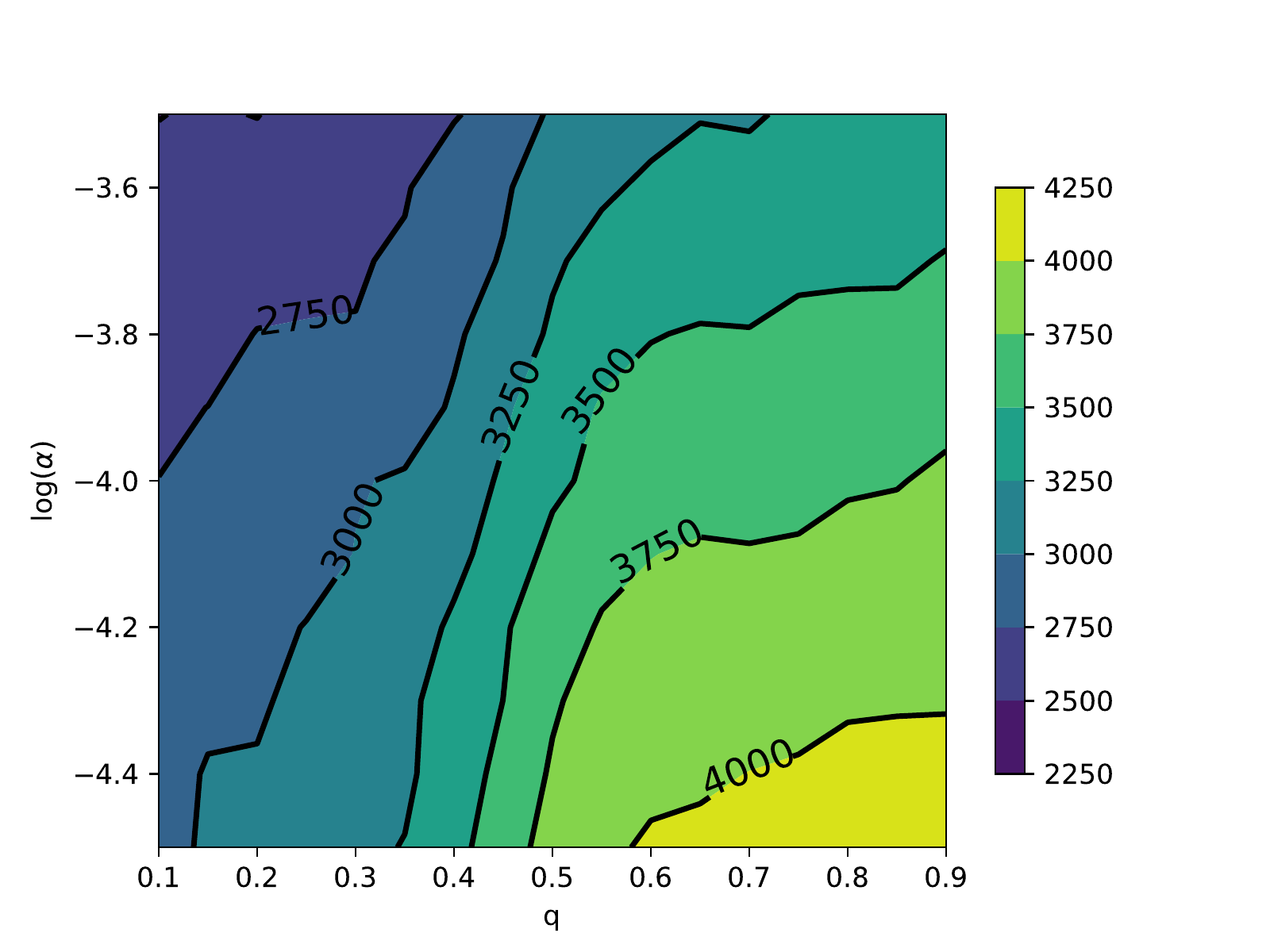}}
\subfigure[]{\includegraphics[width=5.5cm]{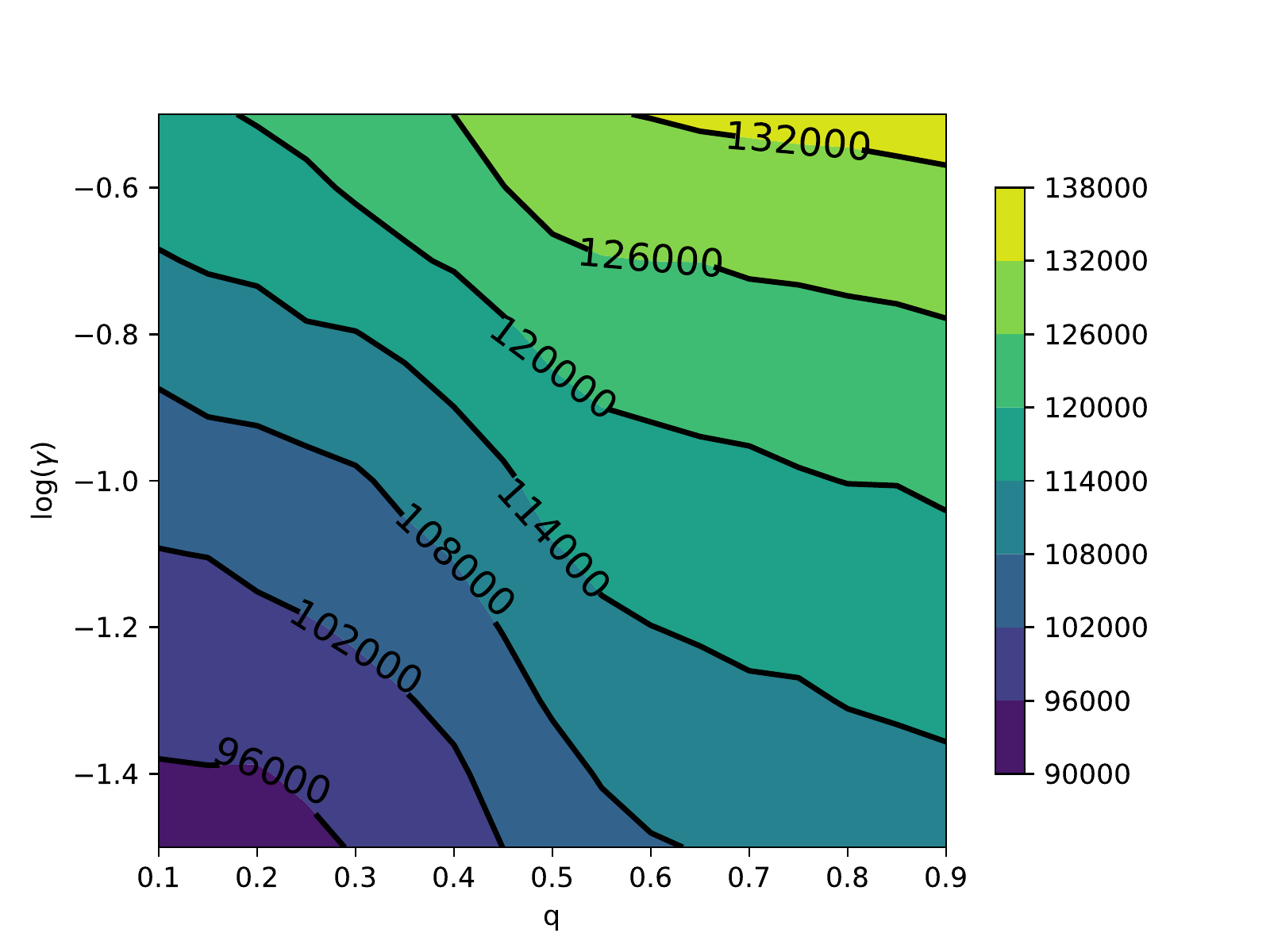}}
\subfigure[]{\includegraphics[width=5.5cm]{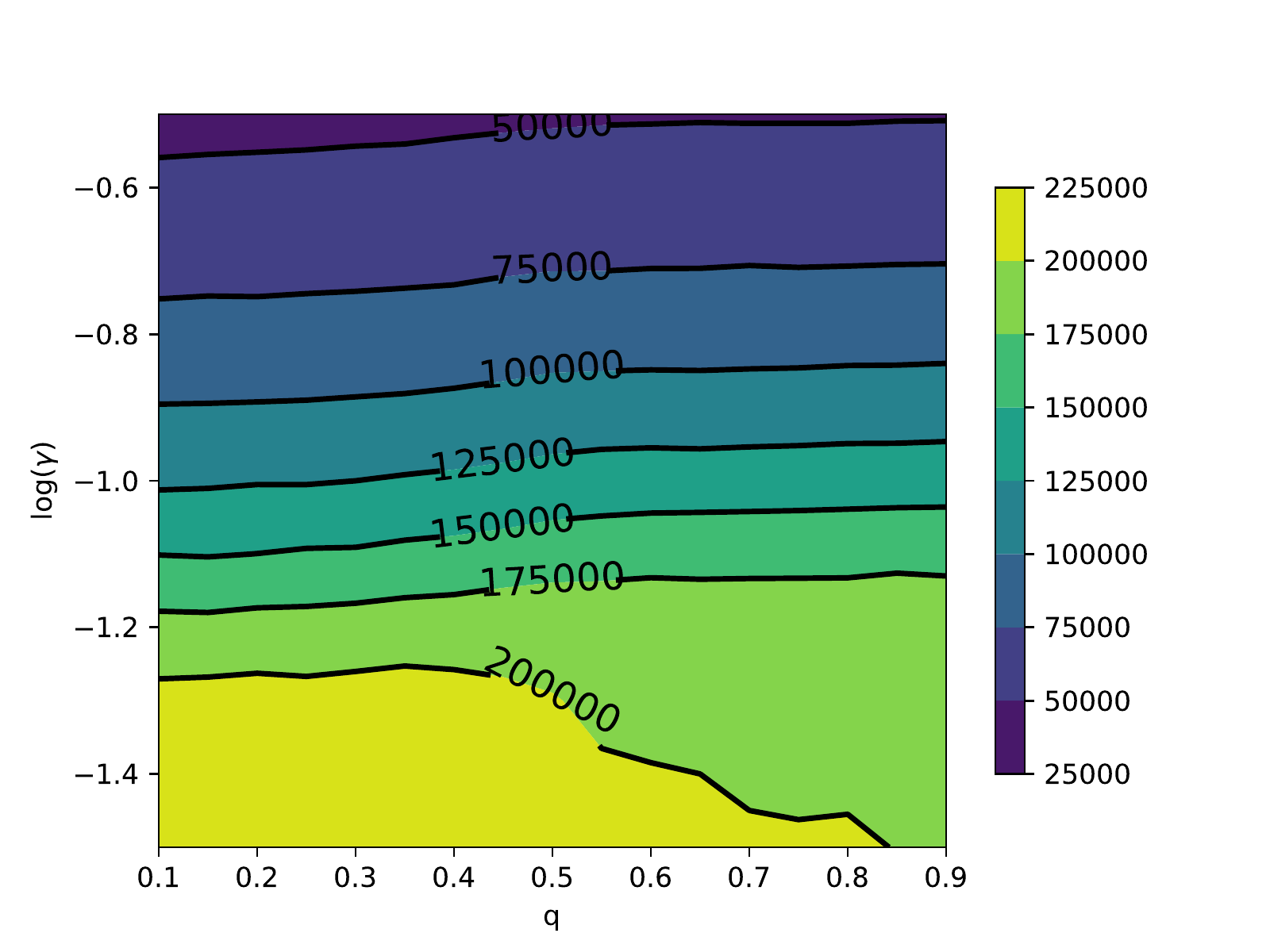}} \subfigure[]{\includegraphics[width=5.5cm]{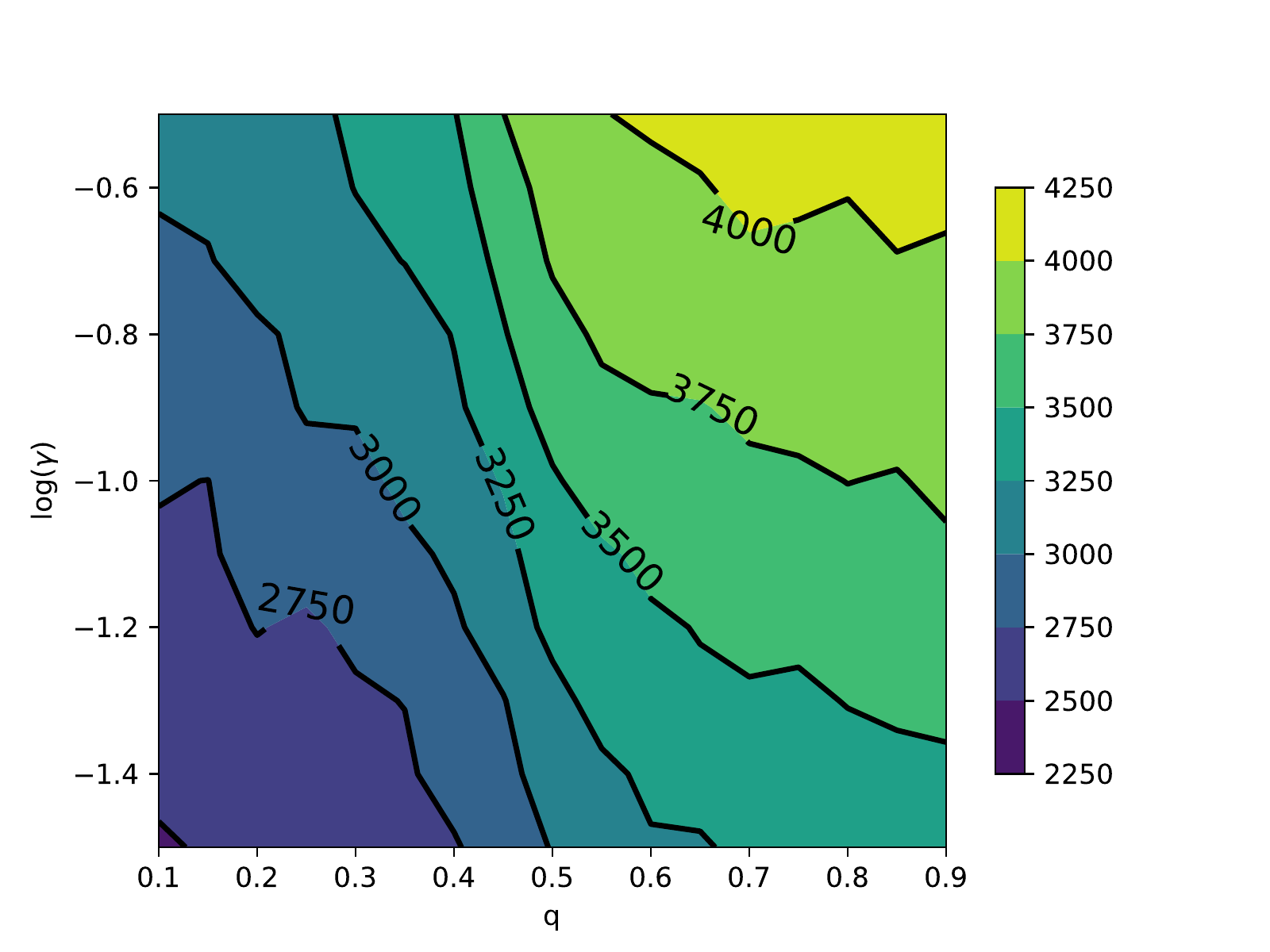}}
\caption{{\bf Fig. S5. Parameter planes of $\alpha$ against $q$ and $\gamma$ against $q$ for sizes of the epidemic, the total number of vaccinated children, and the peak of the epidemic on Barab\'{a}si-Albert network model (BAN).} In all of the simulations, the median value of the simulations are used to plot the parameter planes that are performed at $P_{adv}=.01$ and $\rho=.01$.}\label{figS4}
\end{figure}

\subsection*{SIV. Estimating the Basic Reproduction Number $R_0$}
The basic reproduction number is defined to be the average number of secondary infections caused by an index case introduced to a completely susceptible population till its recovery. To calculate the basic reproduction number as it is defined for the disease process over a network $\mathcal{N}^C$ of households, we carry out algorithm \ref{alg:alg1}. It is based on running the disease process for many number of repetition for an index case in a household that is selected uniformly from the network. Then the expected number of cases caused by the index case is estimated by the mean of binomial distributions whose probabilities are found using Bayes' theorem.

\begin{algorithm}[h!]
\flushleft Input: number of simulation runs  of the disease process $L$, number of simulated networks $W$, inputs for process in table \ref{tabel1}

Output: $R_0$

\textbf{begin}

$\ $\textbf{for} $w= 1,2,\ldots,W$

$\ $ $\vartriangleright$ Generate a network $\mathcal{N}^C_w$ of $N$ households and number of children in households $\{C_i(0):i=1,\ldots,N\}$. Also, generate the starting stand of households about vaccination.

$\ \ $\textbf{for} $i= 1,2,\ldots,N$, such that $C_i(0)>0$

$\ \ \ $\textbf{for $\ell= 1,2,\ldots,L$}

$\ \ \ $ $\vartriangleright$ Select one child in household $i$ to be the index case

$\ \ \ $ $\vartriangleright$ Run the disease process till recovery of the index case, say till day $T$, which will change from one run to another.

$\ \ \ $ $\vartriangleright$ Record the number of new cases in household $j$, $IC_{j}(t)$, and the total number of prevalent children in household $j$ less the index case, $I_{j}(t)$, for all $j\in \{i\} \cup N_P(i)$. Record the total number of prevalent children in the outer neighbors $k$ of the household $j$, $I_{j,k}(t)$, for $t=1,2,\ldots,T$ for all $j\in N_P(i)$, and $k\in N_P(j) -\{i\}$

$\ \ \ $ $\vartriangleright$ Calculate:

The probability that an infection in household $i$ happened due to the index case in household $i$

$$P_i(t)=\dfrac{h\beta}{h\beta+(1-(1-h\beta)^{I_i(t)})+\sum_{j\in N_P(i)}(1-(1-\beta)^{I_j(t)/C_j(t)})}$$
and the probability that an infection in household $j$ happened due to the index case in household $i$

$$P_j(t)=\dfrac{\beta}{\beta+(1-(1-\beta)^{I_i(t)})+(1-(1-h\beta)^{I_j(t)})+\sum_{k\in N_P(j) -\{i\}}(1-(1-\beta)^{I_{j,k}(t)/C_{j,k}(t)})}$$

where $C_{j,k}(t)$ are the number of children in the outer neighbors $k$ of the household $j$. 

$\ \ \ $ $\vartriangleright$ Calculate: $$R_0(w,i,\ell)=\sum_{t=1}^T \sum_{j\in \{i\} \cup N_P(i)}  IC_{j}(t) P_{j}(t)$$

$\ \ \ $\textbf{end}

$\ \ $ $ \vartriangleright$ Return $R_0(w,i)=\frac{1}{L}\sum_{\ell=1}^L R_0(w,i,\ell)$ 

$\ \ $\textbf{end}

$\ $ $ \vartriangleright$ Return $R_0(w)=\frac{1}{N}\sum_{i=1}^N R_0(w,i)$

$\ $\textbf{end}

$ \vartriangleright$ Return $R_0=\frac{1}{W}\sum_{w=1}^W R_0(w)$

\textbf{end}

\caption{To estimate the basic reproduction number using simulations of disease processes on networks.}\label{alg:alg1}
\end{algorithm}

In particular, the probabilities $P_i(t)$ and $P_j(t)$ are the probabilities that the index case is the cause of infection of a new case within the same household $i$ and in the neighbors $j$, respectively. Then the mean number of infections among $IC_i(t)$ and $IC_j(t)$ that are caused by the index case are given by $IC_i(t)P_i(t)$ and $IC_j(t) P_j(t)$, respectively. 

\subsection*{SV. Coding of Simulation.}
Our numerical simulations rely on the NumPy-comptatible Python library CuPy \cite{cupy},  accelerated with NVIDIA CUDA \cite{cuda} for parallel calculations on Graphical Processing Units (GPUs). Most of the calculations were done on a CentOS workstation with $8$ NVIDIA Tesla (Kepler) K80 GPU cards, each having $2496$ CUDA Cores and 12GB memory.  

We introduce a sparse storage format for the adjacency matrices of the networks, where we record only the position of the nonzero elements in the upper triangular part by a single index number (see Fig. S6). With $k$ denoting the index of the upper triangular entries in column-major order and  $(i,j)$ denoting the corresponding row and column indices the following relationships hold:
\begin{align}
k & =i+\frac{\left(j-2\right)\left(j-1\right)}{2}, \label{eq:k} \\
j & =\left \lfloor \frac{3+\sqrt{8k-7}}{2}\right \rfloor, \label{eq:j}\\
i & =k-\frac{\left(j-2\right)\left(j-1\right)}{2}. \label{eq:i}
\end{align}
The indexing formulas \eqref{eq:k}--\eqref{eq:i} provide a fast and efficient storage and calculations through parallel implementations on the GPUs.

\begin{figure}
    \centering
\begin{tabular}{c c c}
 \begin{tabular}{c}
 \begin{tikzpicture}   
[scale=.5,auto=left,every node/.style={circle,fill=blue!20}]   
\node (n1) at (8,9) {1};   
\node (n2) at (10,7)  {2};   
\node (n3) at (8,5)  {3};   
\node (n4) at (5,5)  {4};   
\node (n5) at (4,8)  {5};   
\foreach \from/\to in {n4/n1,n5/n2,n3/n4}     
\draw (\from) -- (\to); 
\end{tikzpicture} 
 \end{tabular}
     & 
     \begin{tabular}{c}
$
A=\begin{bmatrix}{\color{blue}0} & 0 & 0 & {\color{red}1} & 0 \\
0 & {\color{blue}0} & 0 & 0 & {\color{red}1} \\
0 & 0 & {\color{blue}0} & {\color{red}1} & 0 \\
1 & 0 & 1 & {\color{blue}0} & 0 \\
0 & 1 & 0 & 0 & {\color{blue}0}  
\end{bmatrix}
$
     \end{tabular}
     & 
     \begin{tabular}{c}
$
\begin{bmatrix} . & 1 & 2 & {\color{red}4} & 7\\
. & . & 3 & 5 & {\color{red}8}\\
. & . & . & {\color{red}6} & 9 \\
. & . & . & . & 10 \\
. & . & . & . & . 
\end{bmatrix} 
$
     \end{tabular}
\end{tabular}

    {{\bf Fig. S6. One-index representation of network models.}}
    \label{fig:k_index}
\end{figure}

\end{document}